# Geometry Sets the Neutral MLIP–DFT Gap in Non-Covalent Binding, and Charged Systems Fail

Rui Ding,[‡ac] Zixin Ding,[‡b] Rodrigo Pires Ferreira,[‡ac] Diantong Li,[b] Junhong Chen[*ac] and Yuxin Chen[*b]

Pretrained universal machine-learning interatomic potentials (MLIPs) are widely expected to substitute for explicit quantum chemistry when screening molecular interactions. On non-covalent binding we find that this expectation holds for neutral chemistry and fails for charged chemistry. We audit the FAIRChem UMA OMol head, without task-specific fine-tuning, on 5,567 probe–target dimers from 18 benchmarks across nine fidelity arms, using RAPIDS (Rapid Atomistic Probe–target Interaction Discovery Scaffold), a training-free inference-time scaffold. First, on neutral chemistry RAPIDS reaches a mean absolute error (MAE) of 1.37 kcal/mol, matching or beating every explicit DFT single-point fidelity (1.42–1.50 kcal/mol), so the remaining MLIP–DFT gap is geometric rather than a learned-surface error. Second, charged systems form a qualitatively different regime: MAE jumps to 8.5–22.9 kcal/mol across three charged benchmarks; DFT single-point rescoring does not restore accuracy, geometry reoptimization restores it for ionic hydrogen bonds and ion–molecule dimers, and the failure persists across all three MLIP backbones. Third, the residual failures are structured: signed biases cluster by chemistry family, and whether a fidelity fails catastrophically transfers across families more reliably than error magnitude, enabling cost-aware fidelity selection. RAPIDS-Select, a closed-form rule on cheap precomputed features, uses about 37% less compute than a strong static DFT baseline at matched accuracy, and on a newly computed charged benchmark cuts the catastrophe rate from 68% to 5.7%. On WetSenseBench, a wet-lab-grounded host–guest selectivity benchmark from eight sensing studies, RAPIDS without DFT follow-up recovers 82.9% pair-level agreement on the seven studies above its ~1 kcal/mol precision floor.

## 1 Introduction

Pretrained universal machine-learning interatomic potentials (MLIPs) such as UMA (Universal Models for Atoms) and MACE (a higher-order equivariant message-passing interatomic potential) deliver order-of-magnitude speedups over explicit quantum chemistry on broad molecular inputs.[1–3] Low pretraining loss on large molecular corpora is, however, a statement about average force and energy regression on training-distribution conformers, and gives little direct evidence on whether a downstream user can rank non-covalent complexes within 1–2 kcal/mol across chemically distinct regimes. That distinction matters in chemosensing, host–guest recognition, lead optimization, and porous-material screening, where a 1–2 kcal/mol misorder reassigns which probe, ligand, or host is carried forward to experiments.[4–7]

Existing accurate workflows answer this question one system at a time but not at screening scale. CREST conformer exploration on the GFN2-xTB surface with composite density functional theory (DFT) refinement remains a strong reference for small molecular dimers, yet its wall-clock is dominated by repeated semiempirical gradients and DFT follow-ups.[8–10] Docking engines and learned pose generators are orders of magnitude faster, but their scores are calibrated for protein–ligand or geometric plausibility settings and are not general-purpose electronic binding quantities for chemically diverse non-covalent dimers.[11–14] Pretrained universal MLIPs sit between these regimes in cost, but their evaluation resources are organized around standalone molecules, conformers, and general-purpose energy/force supervision rather than downstream binding tasks.[1,15] What is missing is a controlled measurement of where a pretrained universal MLIP is reliable as an error-controlled surrogate simulator of interaction physics inside a realistic screening stack, where it fails, and which intermediate fidelities recover accuracy when it does.

Non-covalent interaction (NCI) ranking suits this measurement. It has a natural multi-fidelity ladder spanning MLIP screening, explicit DFT single points, DFT geometry reoptimization, and CREST sampling, paired with published CCSD(T)-quality references and a clear neutral-versus-charged regime split. Relative energies often differ by only 1 to 2 kcal/mol, so small systematic errors flip rankings, while the cost gap between the cheapest and most expensive arms spans roughly an order of magnitude, so each escalation decision has a real cost.

An alternative to fine-tuning a foundation MLIP for downstream tasks[16–18] is to leave the weights untouched and ask how far a structured inference-time scaffold can carry the pretrained model, paralleling how prompt and tool scaffolds adapt large language models to downstream tasks without fine-tuning. We instantiate this route in RAPIDS (Rapid Atomistic Probe–target Interaction Discovery Scaffold): a three-tier orientation search, limited-memory BFGS (LBFGS) relaxation, and a four-guard validation stack wrapped around the pretrained FAIRChem UMA OMol head. We benchmark RAPIDS against eight additional fidelities on 5,567 dimers from 18 interaction datasets, producing a 5,567 × 9 evaluation log that we use as both a benchmark and an offline replay substrate.

Because non-covalent ranking is ultimately used to decide which probes, receptors, or hosts advance to experiment, we also test whether the computational audit supports downstream sensing decisions. We pair the 5,567 × 9 reference-backed audit with a downstream check against experimental selectivity: WetSenseBench, a reusable wet-lab-grounded selectivity benchmark curated from published host–guest sensing studies.

[a] Pritzker School of Molecular Engineering, University of Chicago, Chicago, IL 60637, USA. E-mail: junhongchen@uchicago.edu
[b] Department of Computer Science, University of Chicago, Chicago, IL 60637, USA. E-mail: chenyuxin@uchicago.edu
[c] Chemical Sciences and Engineering Division, Argonne National Laboratory, Lemont, IL 60439, USA
† Supplementary information (ESI) available: extended background and methods, Fig. S1–S19 and Tables S1–S13.
‡ These authors contributed equally to this work.

**Contributions.** We make five contributions.

1. **Geometry sets the neutral MLIP–DFT gap.** On covered neutral chemistry, RAPIDS reaches an aggregate mean absolute error (MAE) of 1.37 kcal/mol, matching or beating every explicit DFT single-point (SP) fidelity (1.42–1.50 kcal/mol) and beating two of three DFT geometry-reoptimized single-point (GeoSP) arms; only ωB97M-V GeoSP at 1.30 kcal/mol (at ∼4× the RAPIDS cost) edges RAPIDS by ∼5%. A cross-backbone diagnostic on the same RAPIDS-committed geometries with MACE-OMol and ORB-OMol confirms that all three MLIPs reach DFT-SP parity on neutral chemistry once charged motifs are removed, so the residual gap is geometric rather than a learned-surface deficit.
2. **Charged systems form a qualitatively distinct regime.** MAE jumps to 8.5–22.9 kcal/mol on three charged benchmarks, with 36–68% of predictions off by more than 10 kcal/mol. ωB97M-V single points on the MLIP geometry recover at most 0.5 kcal/mol, and on BFDb SSI charged RAPIDS beats them by 0.7 kcal/mol, whereas DFT geometry reoptimization recovers ionic hydrogen bonds and ion–molecule dimers to 2.6–4.9 kcal/mol. The failure persists across all three MLIP backbones we test, ruling out a UMA-specific artifact and pointing at a property of how short-range message-passing potentials currently handle charged binding.
3. **Failure occurrence transfers across chemistry families.** Signed biases cluster by chemistry family across all three backbones, and across held-out families *whether* a fidelity fails catastrophically (|error| > 10 kcal/mol) is predictable out-of-fold (occurrence AUC 0.85–0.95) even though the error *magnitude* is less so (Spearman ρ 0.51–0.65). This transfer asymmetry lets a single closed-form cost-aware selector, RAPIDS-Select, commit the most cost-effective fidelity per system from cheap features alone: ∼37% less compute than a strong static DFT baseline at matched accuracy on covered chemistry, and on a charged benchmark it was never tuned for, near-DFT accuracy at one-third the CREST cost with the cheap-method catastrophe rate cut from 68% to 5.7%.
4. **A controlled audit and a reusable benchmark family.** We deliver these findings through a 5,567-dimer, nine-fidelity audit (MLIP screening, three explicit DFT single-points on the MLIP geometry, three DFT geometry-reoptimization rungs, and two CREST-based references) of the FAIRChem UMA OMol head with no task-specific fine-tuning, and release the resulting 5,567 × 9 evaluation log, the reference and RAPIDS-committed geometries, the cross-backbone MACE-OMol/ORB-OMol diagnostics, a frozen-geometry method-only benchmark, and a newly computed external charged benchmark (981 DES370K-derived[19] systems with CCSD(T)/CBS references) as reusable substrates.
5. **A wet-lab sensing check.** We introduce WetSenseBench, a wet-lab-grounded host–guest selectivity benchmark from eight published sensing studies, on which RAPIDS with no DFT follow-up recovers 82.9% pair-level agreement on the seven studies above its ∼1 kcal/mol precision floor and the wet-lab design target in 6 of 8 studies. The scaffold delivering the audit (RAPIDS) is also exposed as a Skill behind a Model Context Protocol (MCP) interface for agent use.

Figure 1 summarizes the audit and its two central findings: a trust boundary between neutral and charged chemistry, beyond which the cheap MLIP score cannot be used without escalation, and the geometric origin of that boundary.

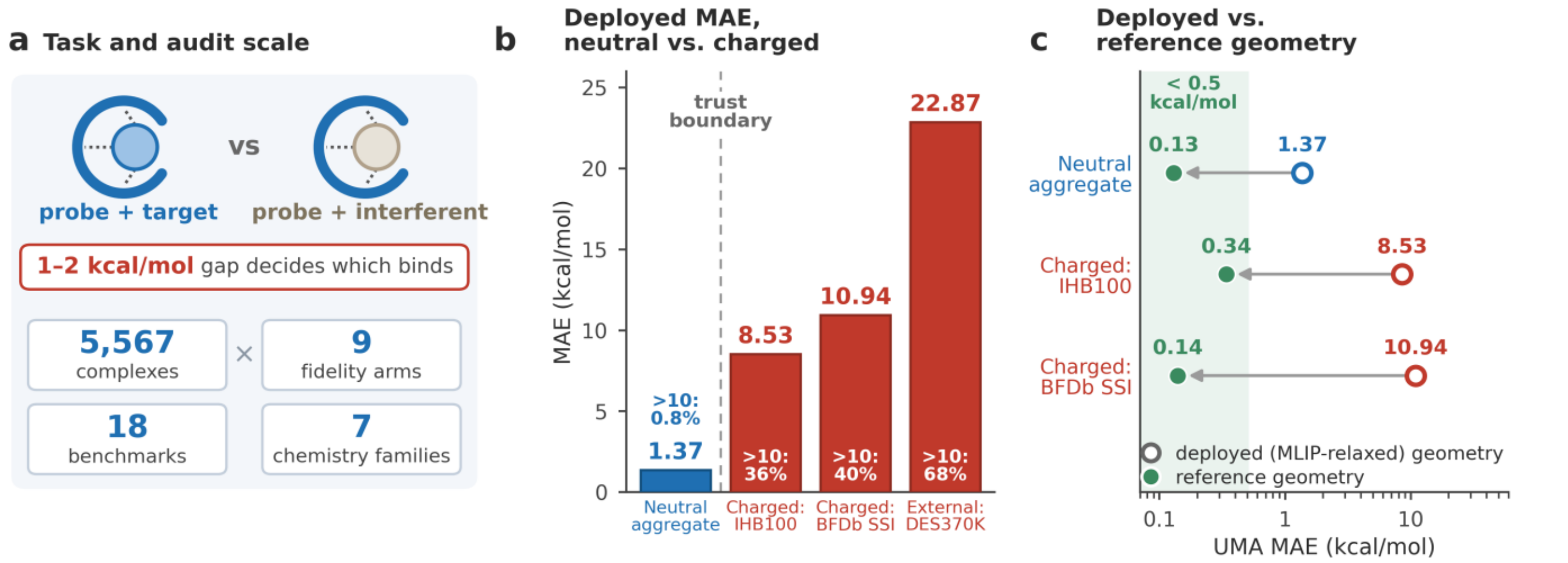

**Fig. 1** A pretrained universal MLIP as a low-cost surrogate for ranking non-covalent binding. **a**, The screening task: ranking probe–target binding where a 1–2 kcal/mol difference decides the experimental call, audited across 5,567 complexes, nine fidelity arms, 18 benchmarks, and seven chemistry families. **b**, A sharp trust boundary: on neutral chemistry the deployed MLIP reaches mean absolute error (MAE) 1.37 kcal/mol, whereas on charged benchmarks MAE rises to 8.53–22.87 kcal/mol; bar labels give the share of predictions off by more than 10 kcal/mol. **c**, The same UMA model evaluated at the benchmark reference geometry (filled) instead of its own relaxed geometry (open): the charged error falls by 25× on IHB100 (to 0.34 kcal/mol) and 75× on BFDb SSI charged (to 0.14 kcal/mol), matching explicit DFT single-point accuracy, so the deployment failure lies in the relaxed geometry rather than in the learned energy surface. The shaded band marks MAE below 0.5 kcal/mol.

## 2 Related Work

**Interaction screening and conformer search.** Ranking non-covalent dimers has been a generate-then-score problem. CREST on the GFN2-xTB surface with composite-DFT refinement sets the accuracy reference but takes hours per dimer.[8–10] Docking engines and diffusion-based pose models are orders of magnitude faster, but their scores target rigid protein-ligand pockets and are not designed for electronic binding across chemically diverse dimers.[11–14] No existing line provides a calibrated low-cost electronic-binding scorer benchmarked on the same task. ESI† Section A.1 covers application background and ESI Section A.2 the broader survey.

**Universal MLIPs for molecular interaction scoring.** Equivariant neural interatomic potentials such as NequIP, MACE, Allegro, and UMA have pushed molecular surrogate modeling well beyond narrow force-field fitting, especially once coupled to large datasets such as SPICE, OMol25, and related cross-domain corpora.[1–3,15,20,21] What is less developed is a full interaction-screening workflow built around such a model, with explicit pose search, validation, fallback logic, and end-to-end benchmark evaluation on chemically diverse non-covalent dimers. The closest precedent is AdsorbML,[22] which wraps pretrained MLIPs in a placement, relaxation, and DFT-verification workflow for adsorbate–surface binding. RAPIDS carries that pattern over to molecular probe–target binding. The workflow itself is therefore not our contribution; we contribute the controlled nine-fidelity audit, the diagnosis of where the surrogate fails and why, and the operating envelope that follows from it. ESI Section A.3 gives the longer universal-MLIP background that motivated this framing.

**Multi-fidelity allocation and offline replay.** Adaptive allocation across cheap and expensive fidelities has a long literature in multi-armed bandits, contextual bandits, multi-fidelity Bayesian optimization, meta-Bayesian optimization, and offline policy evaluation.[23–29] Our nine-fidelity matrix poses such a decision problem, with nontrivial cost–accuracy tradeoffs. The one-shot rule we use, a per-arm regression of each fidelity's error followed by a cost-penalized arg min, is an instance of cost-aware algorithm selection[30,31] and of regression with multi-expert deferral;[32] we do not claim a new routing formulation, only its instantiation on an executed atomistic log and the operating-envelope study it enables. ESI Section A.4 gives the broader literature.

**Scientific agents, chemistry tools, and tool interfaces.** Chemistry and materials agents have shown that large language models can orchestrate literature search, property prediction, cheminformatics, and simulation tools in longer-horizon workflows.[33–38] The Model Context Protocol and progressive-disclosure Skills pattern make it easier to expose such tools to agents without filling the prompt with raw schemas.[39–42] Here these interfaces matter as application-layer packaging for RAPIDS; ESI Sections A.5 and A.6 give the longer discussion.

## 3 Methods

We study a real multi-fidelity screening stack in which each of 5,567 retained probe–target dimers from 18 benchmark datasets is logged across nine fidelity arms; the resulting attempted 5,567 × 9 log records a relaxed binding score, a wall-clock cost, and a status flag per cell, and method-level missing cells (dominated by self-consistent-field (SCF) convergence failures and a small unsupported-system count) are handled at replay time by the fallback rule of §3.4. The main comparison asks when the cheapest learned surrogate is already useful and when explicit higher-fidelity physics changes the answer. Every fidelity emits a relaxed vacuum electronic binding score $\Delta E_{\mathrm{bind}} = E_{\mathrm{complex}} - E_{\mathrm{probe}} - E_{\mathrm{target}}$ at each pipeline's relaxed geometries, where $E_{\mathrm{probe}}$ and $E_{\mathrm{target}}$ are energies of the separately relaxed isolated monomers, each at its own formal charge, rather than of monomers cut out of the complex, so the score includes monomer deformation and more negative values indicate more favorable binding; errors are computed against the published benchmark interaction-energy references $\Delta E_{\mathrm{int}}^*$ evaluated at the dataset-provided geometries. An error is called catastrophic when $|\Delta E_{\mathrm{bind}} - \Delta E_{\mathrm{int}}^*| > 10$ kcal/mol; all catastrophic rates in the paper use this threshold. Full quantity definitions, benchmark-initialization rules, unsupported-system bookkeeping, and thresholds are in ESI Section B.

### 3.1 Benchmark Setup and Nine-Fidelity Ladder

For benchmark evaluation, PubChem retrieval[43] is bypassed. Monomer coordinates, fragment charges, and benchmark reference geometries are taken from the dataset records, the monomers are separated from those records, and each pipeline rebuilds its search start from the same monomer pair. For a fixed system and fidelity, isolated probe and target monomers are evaluated or optimized once and then reused across candidate poses of that system. The benchmark families span canonical small-molecule dimers,[44–46] the Non-Covalent Interactions Atlas H-bond, ionic-H-bond, σ-hole, and London-dispersion benchmarks,[47–50] and the BioFragment Database protein-fragment subsets.[51] Their published reference values are treated as ground truth throughout: CCSD(T)/CBS-style benchmark interaction energies for the small-molecule and Non-Covalent Interactions Atlas sets, and CCSD(T)-F12 silver-standard interaction energies for BFDb.[44–51] The detailed benchmark map with counts, chemistry labels, and source references is given in ESI Table S1.

The nine fidelity arms are RAPIDS; the three fixed-geometry explicit DFT single-point (SP) checks PBE-D3BJ/def2-TZVP, ωB97X-D3BJ/def2-TZVP, and ωB97M-V/def2-TZVP; the three geometry-reoptimized single-point (GeoSP) checks PBE-D3BJ, ωB97X-D3BJ, and ωB97M-V; and the two CREST baselines CREST xTB and CREST xTB+DFT.[52–56] The SP arms apply explicit def2-TZVP single points on RAPIDS geometries. The GeoSP arms reoptimize the RAPIDS geometry at the same functional with def2-TZVP and then evaluate a def2-TZVPD single point. This ladder is designed so that the RAPIDS score, fixed-geometry explicit DFT checks, geometry-corrected explicit DFT checks, and the full CREST baseline are all evaluated on the same retained system set and can be replayed by downstream allocation policies without launching expensive calculations.

### 3.2 RAPIDS Scaffold

RAPIDS (Rapid Atomistic Probe–target Interaction Discovery Scaffold) wraps the pretrained FAIRChem Universal Models for Atoms (UMA) OMol head[1,3] in a search-and-validation workflow; we use the UMA weights as released. The scaffold has four parts: a three-tier orientation search (fixed-anchor Tier 1, random-perturbation refinement in Tiers 2–3), LBFGS relaxation with smart continuation, a validation stack that emits confidence labels and fallback recommendations, and an optional xTB plus analytical linearized Poisson–Boltzmann (ALPB) post-correction. Replay and agent-facing interfaces sit on top of the resulting log. Figure 2 compares the guarded workflow to the CREST composite baseline stage by stage.

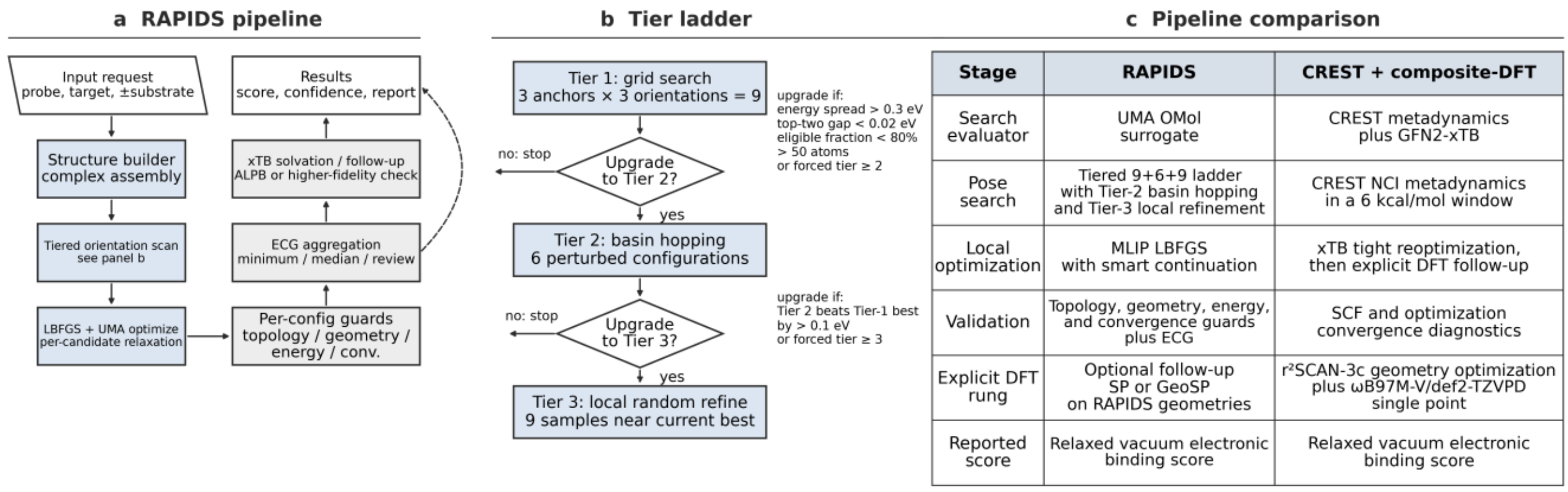

**Fig. 2** RAPIDS architecture and pipeline comparison. **a**, The guarded RAPIDS path builds a probe–target(-substrate) complex, runs a tiered orientation scan, relaxes each candidate on the UMA OMol surface, applies per-configuration guards and a scan-level Energy Consistency Guard (ECG), and returns confidence-annotated outputs with optional xTB solvation or higher-fidelity follow-up. **b**, Tier-upgrade logic. **c**, Stage-by-stage comparison with the CREST plus composite-DFT baseline.

At the workflow level, RAPIDS starts from the benchmark monomers, evaluates a structured 9 + 6 + 9 orientation ladder on the UMA surface, relaxes each candidate with smart continuation, and commits a single pose-ranked relaxed binding score together with confidence metadata. Tier 1 enumerates nine anchor-and-rotation placements, Tier 2 performs six basin-hopping perturbations around the current best candidate, and Tier 3 adds nine local random refinements. The workflow applies four guards, topology, geometry, energy, and convergence, followed by a scan-level Energy Consistency Guard (ECG); all benchmark RAPIDS runs were executed via the MCP `scan_orientations` path so that ECG was active throughout. A key screening rule is the absolute energy guard

$$g_{\mathrm{en}}(r) = \mathbf{1}[\ |E_{\mathrm{bind}}(r)|/N_{\mathrm{small}}(r) > 0.65\ \mathrm{eV/atom} \\ \lor\ |E_{\mathrm{bind}}(r)| > 2.2\ \mathrm{eV}\ ],$$

where the 2.2 eV ceiling (≈50 kcal/mol) is used as an empirical catastrophic-failure sentinel on the same scale as the 50 kcal/mol evaluation clip. These checks decide when RAPIDS can be trusted as a cheap screening oracle and when a system should be escalated to a higher-fidelity arm; they are not intended to rank close neutral candidates. Exact thresholds, guard logic, ECG formulas, pseudocode, and justification of cutoff are given in ESI Section B.1.

### 3.3 High-Fidelity Baseline and Fidelity Ladder

The CREST xTB+DFT benchmark baseline is a cost-controlled CREST plus composite-DFT workflow inspired by Bursch et al.[10] Starting from the same benchmark monomers, it performs CREST non-covalent interaction metadynamics, tight GFN2-xTB reoptimization of the ten lowest xTB candidates, and then forwards the lowest xTB conformer to explicit DFT, using r²SCAN-3c geometry optimization followed by an ωB97M-V/def2-TZVPD single-point evaluation. The isolated monomers are optimized separately with the same method and reused when forming the relaxed binding score. This baseline is intentionally cost-controlled rather than a fully exhaustive DFT reranking protocol. All explicit DFT work uses ORCA 6.1.1. The six intermediate DFT arms (3 SP + 3 GeoSP) on top of RAPIDS-committed geometries support the surface/geometry/sampling decomposition of §4.3; full functional, basis, convergence, charge/spin, and monomer-reuse details are in ESI Section B.2.

### 3.4 Offline Replay, Agent-Facing Packaging, and Downstream Validation

Because the same retained systems are evaluated across all nine fidelities, the resulting 5,567 × 9 matrix can be reused as an offline replay log for cost–accuracy studies. Each row is one probe–target system, each column is one fidelity arm, and each populated cell stores a relaxed binding score, a wall-clock cost, and a status flag. Replaying allocation heuristics and retrospective replay policies over this common log lets us compare fixed fidelities, rule-based escalation strategies, chemistry-aware selectors, and meta-learning selectors without launching new DFT jobs.[23,24,57] The main paper reports 40 replay policies. Rule-based policies and frozen meta-models are the closest to direct deployment; selectors that use retrospective all-arm feedback or online updates should be read as offline replay studies rather than as strict online bandit simulations. A policy $\pi$ on system $s_t$ returns an ordered sequence $M_t = (a_{t,1}, \ldots, a_{t,m_t})$, the committed arm is $a_t^{\dagger} = \mathrm{last}_{\mathrm{n}}\mathrm{onNaN}(M_t)$, and the per-system reward is

$$r_t^{\pi} = -\min(|\Delta\hat{E}_{\mathrm{bind}}(s_t, a_t^{\dagger}) - \Delta E_{\mathrm{int}}^{\star}(s_t)|,\ 50\ \mathrm{kcal/mol}).$$

Cost accounting, feedback regimes, and the full policy taxonomy are given in ESI Section B.6. The replay families include fixed ALWAYS-*X* baselines, disagreement and progressive-ladder rules, cheap-ensemble selectors, UCB/THOMPSON-style methods, chemistry-aware selectors such as CHEMUCB-8, and stacked or frozen meta-learning models.[58,59] The full policy taxonomy, update rules, and deployability breakdown are given in ESI Section B.6.

**One-shot cost-aware fidelity selection.** The replay policies above may probe several arms per system with online feedback. Deployment, by contrast, is usually *one-shot*: a single fidelity is committed per system with no retries, decided in advance from cheap information. For this regime we build RAPIDS-SELECT, a closed-form cost-aware rule that, for each system *x*, commits the arm $\arg\min_a [\, \mathrm{cost}(a) + \lambda\, \widehat{\mathrm{err}}_a(x) \,]$, where $\widehat{\mathrm{err}}_a(x)$ is a per-arm gradient-boosted estimate[59] of arm *a*'s absolute error from ~156 cheap precomputed features (three-backbone MLIP scan summaries and cross-backbone disagreement, relaxation and geometry descriptors, per-fragment chemistry, and charge), and λ (fixed at 800) trades cost against predicted error. The regressors are fitted on the execution log to absolute errors capped at $E_{\max}$ = 50 kcal/mol, the evaluation clip used throughout, so $\widehat{\mathrm{err}}_a(x)$ predicts, before arm *a* runs, the quantity whose negative is the replay reward $r_t^{\pi}$. Unlike the training-free RAPIDS scaffold, RAPIDS-Select is a trained component. The pool is a five-arm subset of the ladder (RAPIDS, PBE-D3BJ SP, PBE-D3BJ GeoSP, CREST xTB, CREST xTB+DFT); the regressors are trained with charged systems included and use no hard charge short-circuit, so the cost rule itself escalates charged systems by their high predicted cheap-arm error. To stress-test the selector on chemistry outside the neutral-heavy suite, we additionally computed an external charged benchmark of 981 DES370K-derived charged dimers with CCSD(T)/CBS-quality references, evaluated across the same fidelity arms. Selector construction, the cheap-feature list, literature baselines, and the charged benchmark are detailed in ESI Section C.7.

**Agent-facing packaging.** The scaffold is also exposed as a progressively disclosed Skill backed by a Model Context Protocol (MCP) server,[39–42] so that an LLM agent can invoke RAPIDS, the follow-up DFT/CREST arms, run inspection, and replay-policy queries through one tool interface. This packaging does not change the underlying model or thresholds; interface details and a worked agent case study are documented in the RAPIDS GitHub repository (Data availability).

**Wet-lab-grounded selectivity evaluation.** To test whether RAPIDS rankings transfer beyond computational references, we curate WETSENSEBENCH from eight published host–guest sensing studies, each contributing (probe, target, interferent) records with a qualitative selectivity call read from the original paper. RAPIDS is run in its cheapest mode (no DFT follow-up, training, calibration, solvent, counterion, or ionic-strength model), and for each record we take the sign of the RAPIDS interaction-energy difference between the probe–target and probe–interferent complexes as the predicted preference, scored against the reported call (pair-level agreement plus a study-level top-binder check). Corpus construction, the 14 → 8 scope filter, exclusions, and the per-study failure-mode analysis are in ESI Section C.10.

# 4 Results

### 4.1 Benchmark setup and the three questions

The benchmark results below use the common 5,567-dimer log defined in §3.2–§3.3. The set comprises A24, S66, and X40 small-molecule dimers,[44–46] the Non-Covalent Interactions Atlas H-bond, ionic-H-bond, σ-hole, and London-dispersion benchmarks HB375/IHB100, HB300SPX, SH250, and D1200,[47–50] and the BioFragment Database (BFDb) protein-fragment subsets.[51] Each retained system is evaluated with all nine fidelity arms of §3.1; missing cells are handled as in §3.4. Primary metrics are the MAE and the Spearman rank correlation ρ between each method's relaxed binding score and the published benchmark interaction-energy reference. Following the statistics protocol in §3.4, per-system absolute errors are clipped at 50 kcal/mol to limit the influence of rare outliers. Across the nine-fidelity matrix only 0.4% of per-system errors exceed this clip, concentrated on charged systems where the cheap surrogate fails and on a small number of CREST or SCF-convergence failures, so the neutral aggregates below are essentially unaffected by the clip. On the neutral aggregate, RAPIDS costs 14.8 h against 125.3 h for a strong CREST xTB+DFT benchmark baseline, Always-CREST xTB+DFT, so the cheapest end of the ladder already costs 12% of that baseline's cumulative total. The results answer three questions: **(Q1)** on neutral non-covalent interaction (NCI) dimers and at a fixed geometry, does the UMA OMol head usefully replace the explicit ωB97M-V/def2-TZVP single-point rung used in the matrix, and what is the dominant source of the residual RAPIDS↔CREST+DFT benchmark-relative gap; **(Q2)** on neutral chemistry, how does the RAPIDS pipeline compare with each DFT and CREST fidelity in cost–accuracy terms; and **(Q3)** can calibration, allocation heuristics, and replay policies on the offline log improve the neutral cost–accuracy frontier relative to fixed single-fidelity baselines. Charged-system failure and recovery are collected separately in §4.5. A downstream check then moves beyond the static audit: a test of RAPIDS-alone selectivity calls against WETSENSEBENCH (§4.9).

### 4.2 Neutral per-method landscape

We first identify the neutral benchmarks on which RAPIDS is competitive and those on which explicit DFT or CREST-based sampling materially changes the ranking or error profile. Across the 16 neutral benchmarks, no single fidelity uniformly wins; the full per-benchmark MAE/ρ heatmaps are deferred to ESI Fig. S2. On H-bond sets, GeoSP variants typically take the best MAE, while RAPIDS falls in the 1.4–1.8 kcal/mol range and remains well below CREST xTB on A24/HB375. On halogen, σ-hole, and D1200 dispersion subsets, the winner is almost always one of the three GeoSP variants: CREST xTB is rarely the per-benchmark winner on a neutral set (best only on the crystal-structure BFDb BBI and HSG subsets), and CREST xTB+DFT tops only S66 among the 16 neutral benchmarks. RAPIDS comes within ∼20% of the per-benchmark best on D1200_Halogens and SH250, and is essentially tied with the best CREST variant on D1200_HBCNO. The strongest neutral result for RAPIDS is the BFDb SSI family, where it is the best-MAE method on all four neutral SSI subsets. The BFDb SSI outcome is benchmark-relative. Those reference structures are crystal-constrained fragments rather than in-vacuo minima, so a DFT geometry optimization can move the complex away from the benchmark reference while still improving the underlying electronic description. This also explains the recurring split between MAE and ρ: on several neutral benchmarks, especially the SSI subsets, RAPIDS is closer to the crystal-constrained reference by MAE while DFT-GeoSP variants preserve the ordering of binding strengths more faithfully by ρ. Best-MAE and best-ρ therefore answer different screening questions.

#### 4.3 Neutral decomposition and cost–accuracy frontier

Q1 asks whether replacing the cheap learned arm by explicit physics changes the answer once geometry and sampling are controlled; the intermediate fidelities are designed for this by holding two of three axes (evaluator, geometry, sampler) fixed at a time. Here ωB97M-V/def2-TZVP SP is an explicit single-point on the RAPIDS-committed geometry, while ωB97M-V GeoSP adds a same-functional geometry reoptimization plus a def2-TZVPD single-point. With $\hat{y}_R$, $\hat{y}_{SP}$, $\hat{y}_{Geo}$, $\hat{y}_C$ denoting RAPIDS, ωB97M-V SP, ωB97M-V GeoSP, and CREST xTB+DFT scores respectively, the RAPIDS-versus-CREST+DFT gap admits the telescoping decomposition

$$\hat{y}_R - \hat{y}_C = \delta_{surf} + \delta_{geom} + \delta_{samp},$$

with $\delta_{surf} = \hat{y}_R - \hat{y}_{SP}$, $\delta_{geom} = \hat{y}_{SP} - \hat{y}_{Geo}$, and $\delta_{samp} = \hat{y}_{Geo} - \hat{y}_C$ (per-system definitions and method-string mappings in ESI Section B.4).

First, $\delta_{geom}$ also changes the basis set, from def2-TZVP at the SP stage to def2-TZVPD at the GeoSP stage (§3.3). A control with ωB97M-V/def2-TZVPD single points on the RAPIDS-committed geometries of all 16 neutral benchmarks (4,686 systems) separates the two: at fixed geometry the larger basis moves the neutral MAE from 1.44 to 1.53 kcal/mol, and reoptimizing the geometry at fixed basis then lowers it to 1.30 kcal/mol. The geometry contribution at matched basis is therefore −0.23 kcal/mol, larger than the −0.14 kcal/mol of the combined step. Second, $\delta_{samp}$ folds in the optimization-functional change from ωB97M-V at GeoSP to r²SCAN-3c at CREST Stage 3, so the "sampling" label refers to the full stack-difference and does not isolate sampling. We summarize via adjacent MAE-against-reference shifts; cancellation against the reference means these do *not* directly bound MAE($|\delta_{surf}|$), MAE($|\delta_{geom}|$), or MAE($|\delta_{samp}|$). On the neutral aggregate, Always-RAPIDS reaches 1.37 kcal/mol MAE and Always-ωB97M-V/def2-TZVP SP at 1.44 kcal/mol; swapping the UMA surface for an explicit ωB97M-V single-point at the same (RAPIDS) geometry yields no aggregate benchmark-relative improvement and in fact a marginal +0.07 kcal/mol regression. Moving to Always-ωB97M-V GeoSP at the same functional drops aggregate MAE to 1.30 kcal/mol, and Always-CREST xTB+DFT reaches 1.80 kcal/mol. The three adjacent ladder shifts on the neutral aggregate are: RAPIDS→ ωB97M-V/def2-TZVP SP +0.07 kcal/mol, ωB97M-V SP→ ωB97M-V GeoSP −0.14 kcal/mol, and ωB97M-V GeoSP→CREST xTB+DFT +0.50 kcal/mol. The first shift is consistent with a negligible fixed-geometry surrogate gap on covered neutral chemistry; if anything, RAPIDS is marginally below explicit ωB97M-V SP at the same geometry. The second corresponds to geometry reoptimization at fixed ωB97M-V functional, and remains the larger contributor in magnitude. The third mixes CREST sampling, r²SCAN-3c optimization, and the final SP stack, so we do not read it as a pure sampling contribution. This answers Q1 for neutral chemistry: within the UMA OMol coverage and at a fixed geometry, the benchmark-relative ladder is consistent with a negligible fixed-geometry surrogate contribution and a larger geometry contribution. The charged ladders behave differently and are treated separately in §4.5.

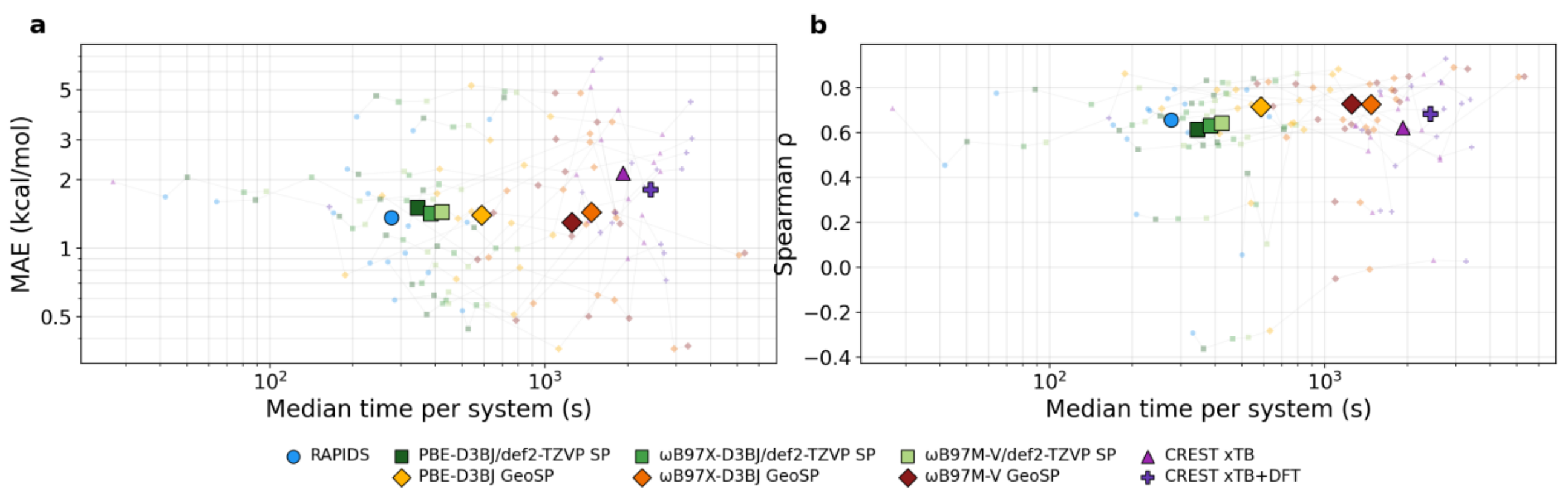

**Fig. 3** Neutral static-fidelity overview across the 16 neutral benchmarks. **a**, MAE and **b**, Spearman ρ versus median wall-clock time per system (seconds, log scale). Faint markers and light gray connectors show benchmark-level values for the nine static fidelities; the larger colored markers show the median across the 16 neutral benchmarks for each fidelity. Costs here are per-system medians; Fig. 5 and Table 1 report per-system means. RAPIDS is at the low-cost end of the static ladder, and GeoSP and CREST xTB+DFT occupy progressively more expensive rungs.

The static ladder of Fig. 3 has three cost zones (aggregate cost range 14.8–125.3 h across the nine methods). RAPIDS occupies the low-cost corner at 14.8 h (12% of Always-CREST xTB+DFT, ∼10% below the fastest DFT-SP baseline), with aggregate MAE (1.37 kcal/mol) at or below the entire DFT-SP cluster (1.42–1.50 kcal/mol) on the neutral aggregate, and within ~5% of the strongest GeoSP arm. DFT-SP methods cluster at intermediate cost and moderate MAE, with the range-separated hybrids (ωB97X-D3BJ and ωB97M-V) slightly ahead of the pure functional. GeoSP costs 2–4× as much as DFT-SP; only ωB97M-V GeoSP, at 64.8 h and aggregate MAE 1.30 kcal/mol, strictly beats Always-RAPIDS on the neutral aggregate (PBE-D3BJ GeoSP 1.40 and ωB97X-D3BJ GeoSP 1.44 are both above it). CREST xTB+DFT has the highest cost and is Pareto-dominated on the neutral aggregate. This agrees with the decomposition: RAPIDS is already a viable low-cost inner-loop predictor for neutral NCI chemistry, and the only fidelity that materially improves on it (at ~4× cost) is ωB97M-V GeoSP.

#### 4.4 Offline replay over the neutral error structure

The residual neutral error has structure: the low-cost arm is competitive but not uniformly best, signed biases cluster by chemistry family (§4.6), and the static ladder spans a large cost range (RAPIDS at 14.8 h to Always-CREST xTB+DFT at 125.3 h). This structure can be exploited without new calculations by replaying allocation policies over the fully observed 5,567 × 9 log (§3.4). Across 40 replayed policies the strongest, ChemUCB-8, reaches 1.01 kcal/mol at 65.1 h, about half the Always-CREST xTB+DFT cost, improving MAE below even the strongest static arm (ωB97M-V GeoSP, 1.30 kcal/mol) at essentially the same cost (64.8 h); an online per-benchmark bias correction, updated from delayed reference feedback, also moves RAPIDS from 1.37 to 1.30 kcal/mol at no added inference cost (ESI Fig. S1; full inventory in ESI Table S7). These policies may probe several arms per system with retrospective feedback, so they are best read as an offline upper bound on what the log's structure allows rather than a directly deployable procedure; the deployable one-shot regime, in which a single fidelity is committed per system with no retries, is the subject of §4.8. System-level error-distribution diagnostics, aggregate Pareto summaries, and family-level policy rankings are deferred to ESI Figs. S5, S6, S7, and S8.

#### 4.5 Charged systems: a sharp trust boundary

Charged motifs mark the trust boundary of the pretrained MLIP. On both charged benchmarks (IHB100, 100 ionic hydrogen-bond complexes; BFDb_SSI_charged, 776 salt-bridge and charged-sidechain pairs) RAPIDS fails: MAE of 8.53 kcal/mol on IHB100 and 10.94 kcal/mol on BFDb_SSI_charged, with 36% and 40% of predictions off by more than 10 kcal/mol, against 1.37 kcal/mol and under 1% on the neutral aggregate. Adding ωB97M-V/def2-TZVP SP on the RAPIDS geometry does not close the gap (8.04 and 11.66 kcal/mol). At fixed geometry the MLIP and DFT agree (paired MAE 1.2 and 1.8 kcal/mol), and on BFDb_SSI_charged RAPIDS beats the DFT single point by 0.7 kcal/mol, so the error sits in the geometry. On IHB100, DFT geometry reoptimization is the main recovery mechanism: ωB97M-V GeoSP is the best-MAE method (MAE 4.90 kcal/mol), followed by CREST xTB+DFT (5.79 kcal/mol). On BFDb_SSI_charged every relaxing arm, from RAPIDS to CREST xTB+DFT, lands within 10.6–12.2 kcal/mol. These crystal-derived sidechain pairs (salt bridges near −100 kcal/mol and same-charge pairs above +55 kcal/mol) relax in vacuum away from the fixed geometry at which the reference interaction energy is defined, and relaxing from the reference geometry itself on the UMA surface already moves the binding score by 9.6 kcal/mol on average (§4.7); per-benchmark MAEs are in ESI Table S9 and ρ in ESI Fig. S2. This failure pattern is consistent with both missing long-range electrostatics in finite-cutoff MLIPs and thinner pretraining coverage of charged species; we do not disentangle those axes, but the gap between fixed-geometry and geometry-corrected DFT on IHB100, and on the external DES370K set of §4.8 (22.5 versus 2.6 kcal/mol), shows that the RAPIDS committed geometry itself is distorted in the charged regime, and the cross-backbone diagnostic in §4.6 shows the charged error is shared across backbones rather than UMA-specific. Ionic hydrogen bonds and ion–molecule dimers therefore require escalation to a DFT geometry-correction rung or to CREST xTB+DFT; the validation guards catch catastrophic topology/geometry/energy failures, but charged systems should be escalated by charge/regime membership even when no guard fires (trigger-rate and flagged/unflagged separations in ESI Tables S11–S12; charged diagnostics and case studies in ESI Figs. S3, S4, S7, S11, and S12, with neutral comparison cases in ESI Figs. S13–S14).

#### 4.6 Cross-backbone diagnostics and end-to-end replacement

To test whether the failure modes identified above are specific to UMA, we perform a cross-backbone analysis with MACE-OMol[2] and ORB-OMol[60] at three levels: fixed-geometry single-point scoring on RAPIDS-committed geometries (L1), MLIP-only relaxation of those geometries (L2), and end-to-end replacement (L3) in which the full RAPIDS scaffold runs with the alternative backbone as evaluator. L1/L2 isolate surface bias and relaxation drift under controlled geometry; L3 produces a complete alternative-backbone screening workflow.

The diagnostic gives three results (ESI Fig. S15). *(i) Chemistry-family-structured signed bias.* The same hydrogen-bond, σ-hole, halogen, and dispersion families carry consistent signed biases across UMA, MACE-OMol, and ORB-OMol, reinforcing the calibration view of §4.4: the residual surface gap is at least partially *learnable per chemistry family*. *(ii) Paired surface error.* On shared systems, paired MAE against ωB97M-V/def2-TZVP SP is 0.56 kcal/mol for ORB-OMol, 0.62 kcal/mol for UMA and 2.24 kcal/mol for MACE-OMol. UMA and ORB-OMol also stay within about 2 kcal/mol of DFT-SP on charged motifs (1.75 and 2.31 kcal/mol on the 876 charged systems), whereas the MACE-OMol gap is dominated by them (11.59 kcal/mol); once charged motifs are removed all three MLIP backbones track DFT-SP to within 0.5 kcal/mol on neutral chemistry (0.41 UMA, 0.50 MACE-OMol, 0.24 ORB-OMol). *(iii) The charged error is shared across backbones.* At fixed geometry no alternative scorer improves on UMA on the charged family (capped MAE 10.7 kcal/mol for UMA, 13.4 for MACE-OMol and 11.0 for ORB-OMol, against 11.2 for ωB97M-V SP and 10.9 for ωB97M-V GeoSP). MLIP-only relaxation does not recover charged accuracy on either backbone (ΔMAE = +0.48 kcal/mol for MACE-OMol and +0.10 kcal/mol for ORB-OMol), whereas DFT geometry reoptimization recovers ionic hydrogen bonds (IHB100: 8.04 → 4.90 kcal/mol for ωB97M-V, against +0.86 and

+0.47 kcal/mol for MACE-OMol and ORB-OMol relaxation). Run end-to-end, the three backbones reach 10.7 (UMA), 11.1 (MACE-OMol), and 7.9 kcal/mol (ORB-OMol) on the charged systems where they converge, against 1.3–1.4 kcal/mol on neutral chemistry, so changing the backbone does not remove the charged trust boundary. ORB-OMol relaxation (L2) beats DFT-GeoSP accuracy on covered chemistry (on the fair subset ORB-OMol L3 leads at 2.43 kcal/mol, followed by UMA-RAPIDS at 2.46 and ORB-OMol L2 at 2.49, all ahead of ωB97M-V GeoSP at 2.56), but this lead is coverage-conditional: it holds only on the subset where both alternative backbones converge, whereas UMA-RAPIDS commits a value on all 5,567 systems against MACE-OMol's 89% and ORB-OMol's 79% end-to-end convergence, with the failures concentrated in specific chemistry families (MACE-OMol on heavy halogens and σ-holes, ORB-OMol on protein-sidechain dispersion; ESI Fig. S16). UMA-RAPIDS therefore remains the deployed pipeline, and MACE-OMol and ORB-OMol are usable as alternative-backbone arms only with explicit chemistry-family exclusions. Full L3 ranking, BiasCorr cross-backbone behavior, and complementary-coverage breakdown are in ESI Section C.9. Cross-MLIP disagreement also yields a cheap escalation signal: the standard deviation across the three backbones' fixed-geometry predictions correlates with downstream error (Spearman $\rho$ = 0.53 overall and 0.55 on charged systems), and flagging the top 5% most-disagreeing systems for escalation reaches 0.82 precision. The three backbones therefore form a cheap ensemble whose spread flags systems for escalation, and the same spread is one of the cheap features RAPIDS-Select uses (§3.4).

#### 4.7 Isolating surface bias on frozen reference geometries

The decomposition of §4.3 and the cross-backbone diagnostic above both attribute the residual MLIP–DFT gap to geometry rather than the learned surface, but each still carries a geometry source (the RAPIDS-committed pose). A companion frozen-geometry benchmark removes geometry as a variable entirely: on the exact reference structures that define the published CCSD(T)/CBS-style and CCSD(T)-F12 interaction energies, we recompute single-point energies for all 5,567 neutral and charged systems with ten methods (GFN2-xTB, the UMA, MACE-OMol, and ORB-OMol backbones, and six DFT single points), so every method is scored on identical geometries and only method bias remains (protocol in ESI Section B.5, per-method statistics in ESI Table S13). Because released UMA pretraining metadata do not support an exhaustive duplicate audit against every benchmark geometry, we read this frozen-geometry result as a geometry-intervention and method-bias diagnostic rather than a train/test generalization claim. Held to the reference geometry, the UMA surface is already at explicit-DFT accuracy on neutral chemistry (MAE 0.13 kcal/mol as a median across the 16 neutral benchmarks, essentially tied with the best DFT single point, ωB97M-V/def2-TZVPD at 0.12 kcal/mol, and below the other five). The charged benchmarks give the stronger test. Where the deployed pipeline shows a large charged error on MLIP-committed geometries (MAE 8.53 and 10.94 kcal/mol on IHB100 and BFDb_SSI_charged, §4.5), the same UMA surface evaluated on the reference geometry attains MAE 0.34 and 0.14 kcal/mol on those same two benchmarks ($\rho$ = 0.996 and 0.9998), matching or beating the best DFT single point on both. This collapse of the charged error by 25–75× when geometry is frozen (10.94 → 0.14 kcal/mol on BFDb_SSI_charged) places the charged trust boundary on geometry rather than the learned energy surface: UMA reproduces charged interaction energies almost exactly when handed the benchmark reference structure, and the deployment failure is that the MLIP relaxes charged complexes to distorted geometries. This near-exact frozen-geometry accuracy is UMA-specific among the learned backbones: MACE-OMol retains a non-negligible signed surface under-binding bias (+1.1 kcal/mol, with a much larger charged-subset frozen-geometry MAE of 8.4) and GFN2-xTB a −2.0 kcal/mol bias, so it is a property of the UMA surface rather than of MLIPs in general. Two low-cost diagnostic probes (Fig. 4) localize this failure to the geometry the MLIP commits to, both the pose it selects and the structure it relaxes into. In reference-seeded relaxation, each complex is relaxed on the MLIP surface starting from its benchmark reference geometry, and the rigid-body-aligned RMSD it drifts away predicts deployment error per system (BFDb_SSI_charged Spearman $\rho$ = +0.64). The residual ωB97M-V force at the MLIP-committed geometry is markedly larger on charged than neutral systems (0.123 vs 0.029 eV/Å) and points toward the DFT minimum in ∼97% of cases, so the MLIP relaxes charged complexes to geometries whose local gradient field is inconsistent with higher-fidelity optimization even though its energy at the benchmark geometry is near-exact. The energetics of the reference-seeded relaxations separate the two geometry contributions: relaxing from the reference geometry, with separately relaxed monomers, moves the UMA binding score away from the reference by 0.33 kcal/mol on neutral chemistry (median across the 16 benchmarks), 1.77 kcal/mol on IHB100, and 9.62 kcal/mol on BFDb_SSI_charged, against deployed errors of 1.37, 8.53, and 10.94 kcal/mol. The IHB100 error therefore comes mostly from pose search, and the BFDb_SSI_charged error mostly from relaxation away from the crystal-derived reference structure, which every relaxing method inherits. An ωB97M-V relaxation started from the reference geometry reaches 1.94 kcal/mol on 50 IHB100 systems, against 5.12 kcal/mol for ωB97M-V GeoSP started from the RAPIDS pose on the same systems.

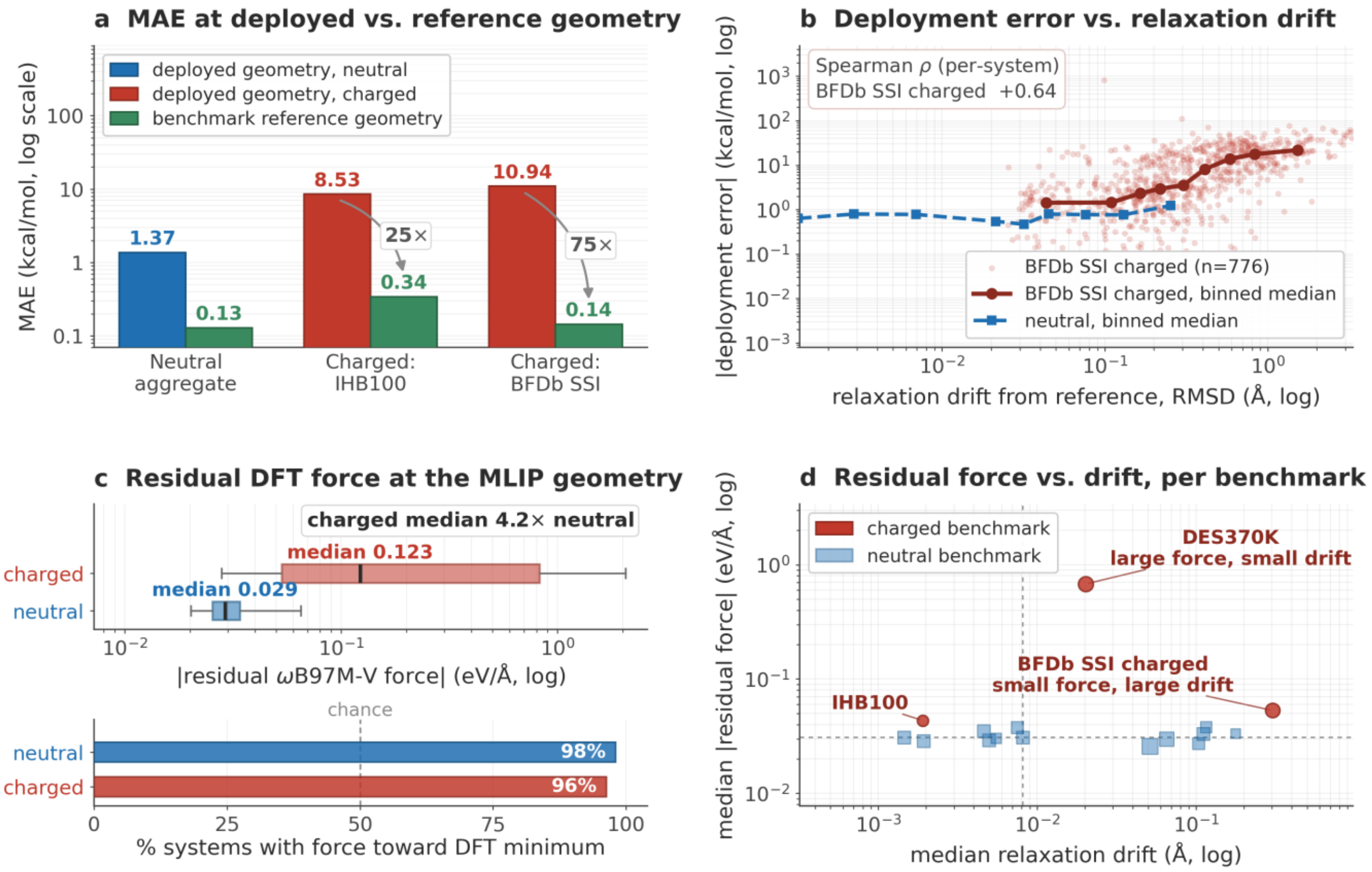

**Fig. 4 The MLIP is accurate at the reference geometry but relaxes charged complexes away from it.** **a**, At the benchmark reference geometry the MLIP reproduces the published interaction energies to 0.13–0.34 kcal/mol, including on both charged benchmarks, whereas the same model evaluated on its own relaxed (deployed) geometry gives 1.37 kcal/mol on the neutral aggregate and 8.53 (IHB100) / 10.94 (BFDb SSI charged) kcal/mol; freezing the geometry collapses the charged error by 25× and 75× respectively. **b**, On BFDb SSI charged the failure is per-system and tracks how far the MLIP relaxes away from the benchmark geometry: the drift correlates with the deployment error (Spearman ρ = +0.64, *n* = 776), while the neutral trend is flat by comparison. **c**, Residual ωB97M-V forces evaluated at the MLIP geometry are larger on charged complexes (median 0.123 vs 0.029 eV/Å, 4.2×) and are systematically directed: for ∼97% of systems the residual force points toward the DFT minimum (98% neutral, 96% charged), far above the 50% chance level. The charged magnitude spread is carried largely by DES370K (median 0.68 eV/Å), where the residual-force magnitude also predicts the per-system error (Spearman ρ = +0.66, *n* = 981). **d**, Per-benchmark medians on the force–drift plane separate the charged failures into two groups (marker area scales with benchmark size; dotted lines mark the neutral medians): DES370K shows large residual forces with little drift (median 0.68 eV/Å, 0.02 Å), a pose-search failure, whereas BFDb SSI charged shows small forces with large drift (0.053 eV/Å, 0.30 Å), where the MLIP minimum lies away from the reference geometry. The two quantities are decorrelated *within* each benchmark (Spearman ρ = −0.07 DES370K, −0.02 BFDb SSI charged); the pooled charged value (−0.48) is an artifact of the two benchmarks occupying opposite regions of this plane. Each group has its own per-system error predictor: residual force on DES370K and drift on BFDb SSI charged. Red denotes charged systems, blue neutral, and green the benchmark reference geometry throughout.

### 4.8 RAPIDS-Select: choosing the most cost-effective fidelity per system

The trust boundary makes blanket policies wasteful: running the cheapest arm everywhere fails catastrophically on charged motifs, while running an accurate DFT arm everywhere spends several times more compute than covered neutral chemistry needs. Deployment therefore asks which single fidelity is most cost-effective for a given system, decided in advance from cheap features (§3.4). RAPIDS-Select answers with one closed-form rule, $\arg\min_a [\,\mathrm{cost}(a) + \lambda\, \widehat{\mathrm{err}}_a(x)\,]$, that relies on a transfer asymmetry (Fig. 5a): across held-out chemical families, *whether* an arm makes a catastrophic error (|error| > 10 kcal/mol) is predictable out-of-fold (occurrence AUC 0.85–0.95), whereas the error *magnitude* is less so (Spearman ρ 0.51–0.65), so the selector can steer away from catastrophic arms without predicting exact errors.

On the 18-benchmark suite RAPIDS-Select commits at 387 s per system for a macro-averaged MAE of 2.49 kcal/mol and a 5.6% catastrophic rate (Fig. 5b). This 2.49 is over the five-arm subset and the 18-benchmark macro-average that *includes* the two charged benchmarks, so it is not comparable to the neutral-only single-method figures of §4.3. Against the strongest static baseline in this pool, Always-PBE-D3BJ GeoSP, RAPIDS-Select cuts per-system cost by ∼37% (paired bootstrap over benchmarks, significant) at statistically indistinguishable error and catastrophic rate, and it matches or beats every published one-shot selection baseline we re-evaluated on the identical cost accounting, with the lowest cost of all six at statistically equal or better MAE (SATzilla-style,[31] FrugalML-style,[61] learning-to-defer,[62,63] ALORS,[58] and a multiclass Rice-style classifier;[30] Table 1).

**Table 1** RAPIDS-Select against the static arms and literature one-shot selectors on identical caches, benchmarks, and per-system cost accounting. In-distribution: macro average over the 18 leave-one-benchmark-out folds; external charged: the 981-system DES370K benchmark. Cost is mean per-system wall-clock seconds, MAE is in kcal/mol, and Cat. is the fraction of systems with |error| > 10 kcal/mol. ORACLE is the per-system best arm and is not deployable.

| | In-distribution | | | External charged | | |
|---|---|---|---|---|---|---|
| Method | Cost (s) | MAE | Cat. (%) | Cost (s) | MAE | Cat. (%) |
| **RAPIDS-Select** (λ = 800) | **387** | **2.49** | **5.6** | **692** | **3.01** | **5.7** |
| Always-RAPIDS | 277 | 2.56 | 5.7 | 230 | 22.87 | 68.4 |
| Always-PBE-D3BJ SP | 283 | 2.75 | 6.3 | 240 | 20.17 | 62.8 |
| Always-PBE-D3BJ GeoSP | 615 | 2.54 | 5.6 | 479 | 3.04 | 6.7 |
| Always-CREST xTB | 1803 | 3.34 | 6.5 | 1550 | 15.44 | 30.9 |
| Always-CREST xTB+DFT | 2256 | 2.97 | 6.5 | 2041 | 2.41 | 5.2 |
| SATzilla-style, cost-blind[31] | 998 | 2.53 | 5.7 | 1859 | 2.47 | 5.0 |
| FrugalML-style cascade[61] | 402 | 2.43 | 4.8 | 739 | 2.68 | 4.1 |
| Learning-to-defer rule[62,63] | 477 | 2.49 | 5.2 | 1760 | 2.49 | 4.9 |
| ALORS with cost term[58] | 799 | 2.50 | 4.8 | 1448 | 2.66 | 5.5 |
| Rice-style multiclass[30] | 1087 | 2.38 | 4.7 | 1887 | 2.32 | 5.0 |
| ORACLE (per-system best) | 1016 | 1.27 | 2.4 | 1494 | 1.04 | 1.0 |

On the external charged benchmark (981 DES370K systems) the cheap arms fail (Always-RAPIDS 22.9 kcal/mol MAE, 68.4% catastrophic), whereas RAPIDS-Select commits at 692 s / 3.01 kcal/mol / 5.7% (Fig. 5c): near the accuracy of Always-CREST xTB+DFT (2.41 kcal/mol) at roughly one-third its cost, with the cheap-method catastrophe rate cut from 68% to 5.7%. It matches the accuracy of the strong mid-fidelity Always-PBE-D3BJ GeoSP arm (3.01 vs 3.04 kcal/mol) at a lower catastrophic rate (5.7% vs 6.7%) but at higher cost (692 vs 479 s). Because this benchmark is entirely charged, the result shows that the selector avoids the cheap-arm failure without a charge rule; it does not show that per-system routing beats a blanket escalation of charged systems to PBE-D3BJ GeoSP. The selector does this in one shot from cheap features, without a hard-coded charge branch: cheap arms carry high *predicted* error on charged systems, so the cost rule self-escalates them (consistent with the charged trust boundary of §4.5). An ablation isolating charged-aware training data as the dominant factor and the λ cost–risk frontier are detailed in ESI Section C.7.

A hand-written CHARGEGATE rule (commit PBE-D3BJ GeoSP on any fragment-charged system, else RAPIDS) reaches 303 s / 2.42 kcal/mol / 4.8% in distribution; RAPIDS-Select (387 s / 2.49 / 5.6%) learns this behavior from cheap features without being given the charge rule. On the external charged benchmark the rule reduces to blanket PBE-D3BJ GeoSP (479 s / 3.04 kcal/mol / 6.7%), and RAPIDS-Select has the lower error and catastrophic rate (3.01 kcal/mol, 5.7%).

The cost weight was fixed at λ = 800 s (kcal/mol)$^{-1}$, where the selector is more than a third cheaper than Always-PBE-D3BJ GeoSP at lower error; Table 2 shows what happens away from it. Every λ in the table is cheaper than Always-PBE-D3BJ GeoSP, matching its error at λ = 200 and beating it from λ = 400 upward. Lowering λ buys further savings (λ = 200: 50% below Always-PBE-D3BJ GeoSP at the same 2.54 kcal/mol), whereas raising it trades the saving for a small accuracy gain (2.47 kcal/mol at λ = 3200, with a 5% saving). The arm-selection pattern shifts gradually: RAPIDS is committed on 62%, 57%, and 52% of in-distribution systems at λ = 200, 800, and 3200, and on the external charged set the dominant arm moves from PBE-D3BJ GeoSP (94% and 83% of systems at λ = 200 and 800) to CREST xTB+DFT (68% at 3200), which lowers external MAE from 3.17 to 2.58 kcal/mol at 3.3× the cost. Retraining the regressors with the error cap set to $E_{\max}$ = 20 or 100 instead of 50 kcal/mol leaves this pattern intact: RAPIDS remains the majority in-distribution arm (53–56% of systems), PBE-D3BJ GeoSP the dominant external charged arm (78–84%), and the in-distribution saving relative to Always-PBE-D3BJ GeoSP stays within 35–39%.

**Table 2** Sensitivity of RAPIDS-Select to the cost weight λ (s per kcal/mol); λ = 800 is the reported operating point. Saving is the in-distribution cost reduction relative to Always-PBE-D3BJ GeoSP (615 s, 2.54 kcal/mol, 5.6%). Protocol and units as in Table 1.

| | In-distribution | | | | External charged | | |
|---|---|---|---|---|---|---|---|
| λ | Cost (s) | MAE | Cat. (%) | Saving (%) | Cost (s) | MAE | Cat. (%) |
| 200 | 309 | 2.54 | 5.7 | 50 | 472 | 3.17 | 6.8 |
| 400 | 349 | 2.52 | 5.7 | 43 | 532 | 3.09 | 6.3 |
| **800** | **387** | **2.49** | **5.6** | **37** | **692** | **3.01** | **5.7** |
| 1600 | 452 | 2.48 | 5.4 | 27 | 1164 | 2.79 | 5.3 |
| 3200 | 585 | 2.47 | 5.4 | 5 | 1581 | 2.58 | 4.9 |

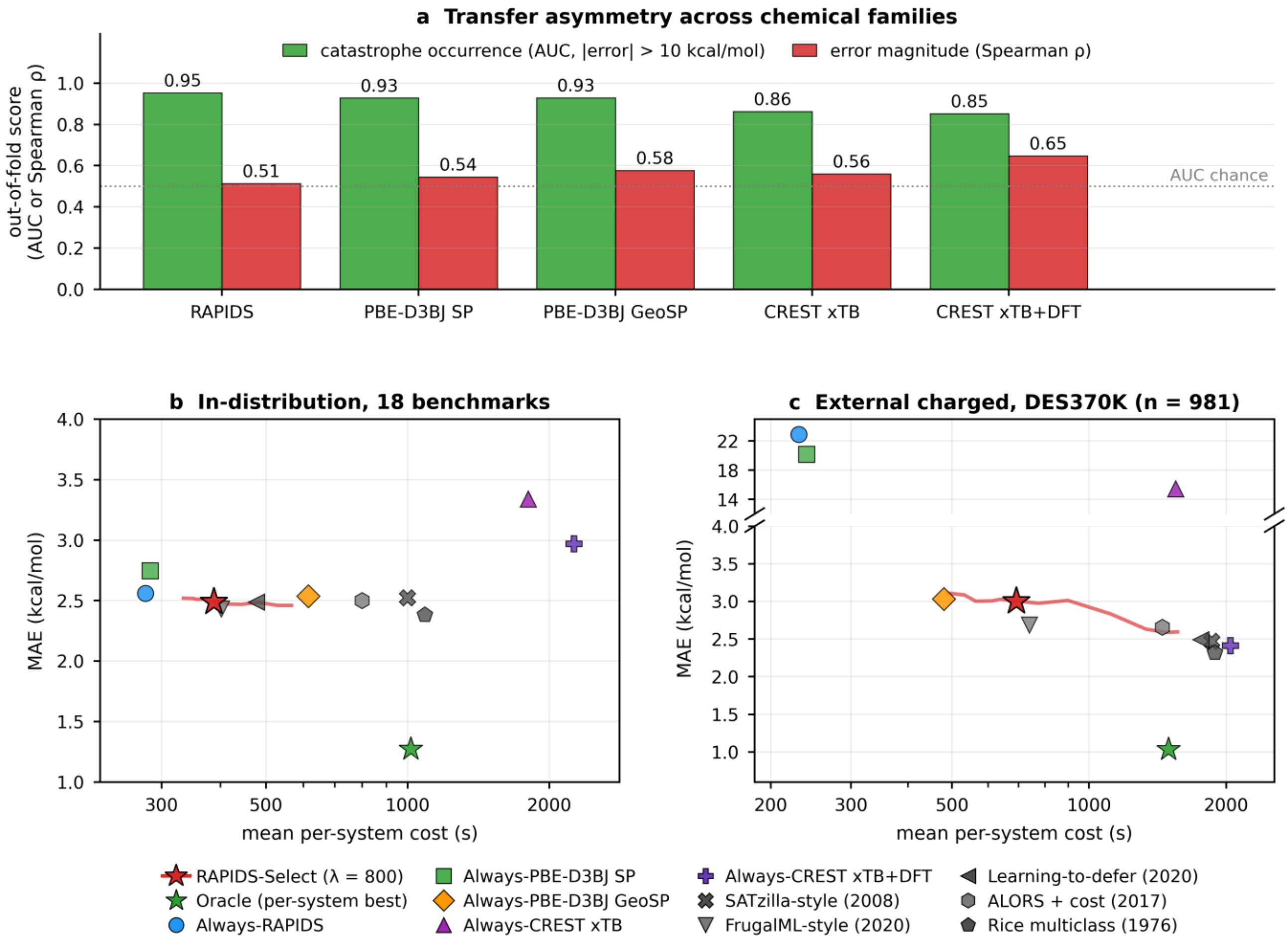

**Fig. 5** RAPIDS-Select: one-shot cost-aware fidelity selection. **a**, Transfer asymmetry across held-out chemical families: for each fidelity arm, catastrophe *occurrence* (|error| > 10 kcal/mol; green) is predictable out-of-fold (AUC 0.85–0.95) whereas error *magnitude* (red) is less so (Spearman ρ 0.51–0.65). **b**, In-distribution cost–accuracy over the 18-benchmark suite and **c**, external charged cost–accuracy on the 981-system DES370K benchmark. The red star is RAPIDS-Select (cost-aware arg min, λ = 800), and the red curve traces its cost–accuracy frontier as λ varies (387 s / 2.49 kcal/mol / 5.6% in-distribution; 692 s / 3.01 / 5.7% on charged). Colored markers are the five static fidelity arms, gray markers are literature one-shot selectors on identical benchmarks and cost accounting, and Oracle is the per-system best arm; per-system costs are wall-clock seconds, and the y-axis of **c** is broken to show the cheap arms that fail on charged systems. Full numbers in Table 1 and interpretation in §4.8.

Holding out whole chemistry groups is a stricter test than leaving out single benchmarks. We partition the 18 benchmarks into nine groups (eight neutral and one charged; ESI Table S4), hold out each group in turn, retrain the regressors on all remaining systems, and route the held-out group. On every neutral group RAPIDS-Select stays cheaper than Always-PBE-D3BJ GeoSP (211–364 versus 351–840 s per system) at comparable error (better, or within 0.25 kcal/mol), and averaged over all nine groups it costs 31% less at lower MAE. The held-out charged group is the hard case: with IHB100 and BFDb_SSI_charged both unseen, the router reaches 11.1 kcal/mol MAE and a 38% catastrophic rate at 1232 s, level with Always-PBE-D3BJ GeoSP on the same systems (11.4 kcal/mol, 39%) at 3.2× its cost (384 s). Held out singly, the router trails Always-PBE-D3BJ GeoSP on IHB100 (6.8 versus 5.0 kcal/mol) and is level on BFDb_SSI_charged (12.0 versus 12.2). On an unseen charged chemistry per-system routing matches blanket escalation in error at 3.2× the cost, so blanket escalation of charged systems to a geometry-changing arm is the cheaper default, consistent with the ablation that identifies charged-aware training data as the selector's dominant factor (ESI Section C.7).

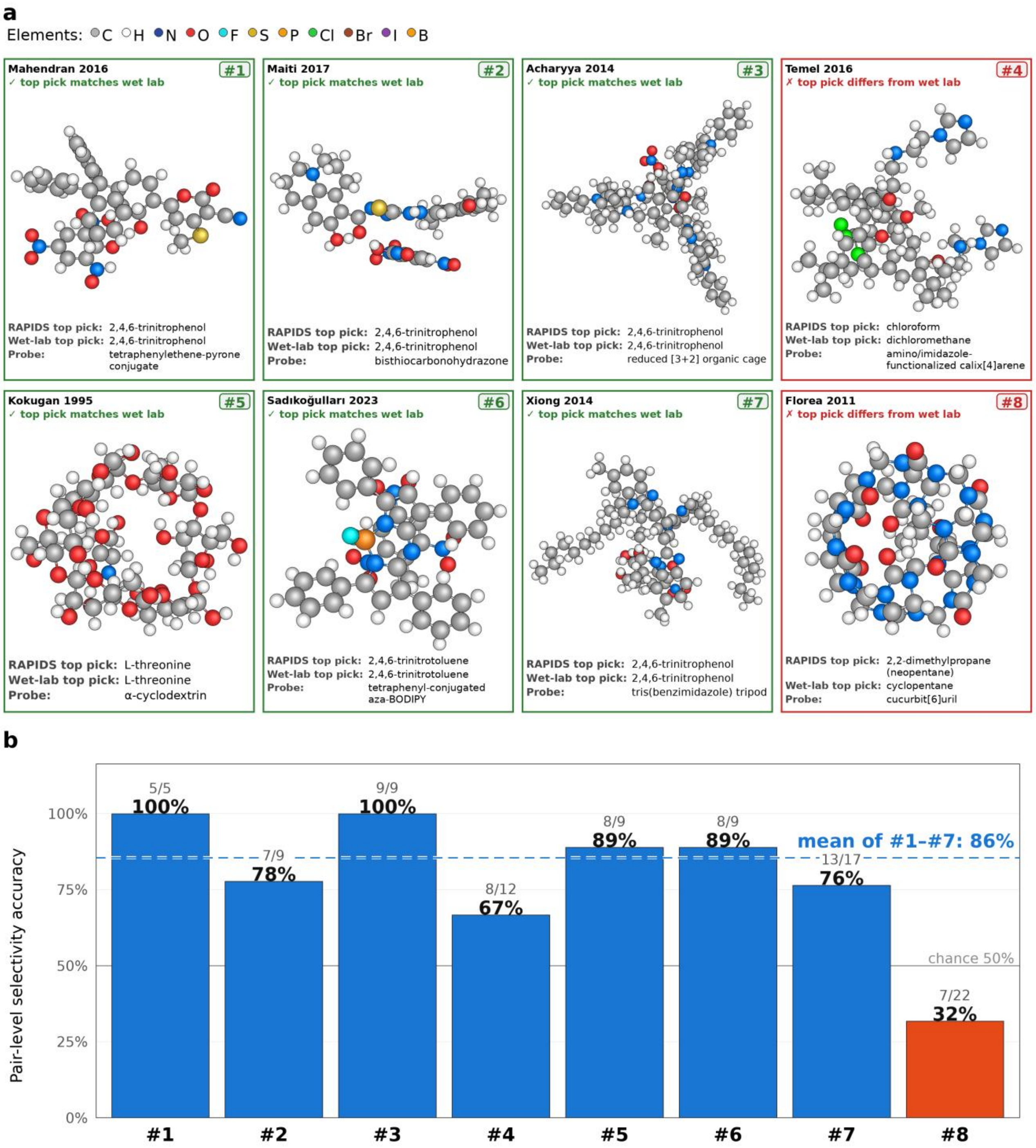

**Fig. 6** WetSenseBench: RAPIDS with no DFT follow-up against 8 wet-lab host–guest sensing studies. **a**, For each study, the RAPIDS-predicted highest-affinity binder rendered as its vacuum-optimized complex; a green border marks agreement with the wet-lab design target (6 of 8), red marks disagreement (#4 calix[4]arene, chloroform vs. dichloromethane; #8 CB[6], a precision-floor case). Picric acid recurs as the pick in #1–#3 and #7 because those four studies are picric-acid explosive-sensing assays. **b**, Pair-level selectivity accuracy per study; the seven studies above the precision floor (blue) average 86%, while CB[6] (orange, #8) drops to 32% and is analyzed separately in ESI Section C.10.

### 4.9 Wet-lab-grounded selectivity validation

The benchmark above evaluates RAPIDS against computational references. As a final test we ask whether the cheapest arm alone, RAPIDS with no DFT follow-up, recovers *experimental* selectivity calls. We assemble WetSenseBench, a wet-lab-grounded host–guest selectivity benchmark of 8 published sensing studies[64–71] (92 probe–target–interferent pairs spanning explosive nitroaromatics, nitrophenols, amino-acid enantiomers, volatile organics, and saturated hydrocarbons). For each pair we predict the selectivity direction from the sign of $\Delta E = E_{int}(\text{probe} + \text{target}) - E_{int}(\text{probe} + \text{interferent})$, where $E_{int}$ is the RAPIDS interaction (binding) energy (negative = bound), and score a binary hit against the paper's reported call, with no training, calibration, or solvent model. Across the 7 studies above the model's ~1 kcal/mol precision floor, RAPIDS reaches 82.9% pair-level agreement (58 of 70 pairs; the mean of the seven per-study accuracies is 86%, the dashed line in Fig. 6b), and on 6 of the 8 studies its top binder is the design target, the analyte the probe was built to detect (Fig. 6a). The one outlier study (cucurbit[6]uril against saturated hydrocarbons, 32%) lies at that ~1 kcal/mol floor: its selectivity-determining gaps are sub-kcal/mol, one wet-lab label flips with ionic strength (outside gas-phase scope), and two guests hit conformer-scan pose artifacts. The benchmark construction, per-pair $\Delta E$ structure, and the failure-mode analysis are in ESI Section C.10.

## 5 Discussion

We audited a pretrained universal MLIP on non-covalent binding across a 5,567-dimer, nine-fidelity ladder. The audit isolates the non-covalent interaction-ranking step that recurs in molecular sensors, supramolecular hosts, porous adsorbents, and functional interfaces (periodic solids are outside its scope), and this focus lets the reliability boundary be measured across thousands of systems and multiple fidelity rungs. On covered neutral chemistry, the MLIP is already a strong low-cost screen: on the neutral aggregate RAPIDS is at or below every DFT single-point fidelity, and the only fidelity that materially improves on it is ωB97M-V GeoSP, by ~5% at roughly 4× the cost. MACE-OMol and ORB-OMol on the same RAPIDS-committed geometries reach DFT-SP parity once charged motifs are removed, so the neutral MLIP–DFT gap is a geometry gap rather than a learned-surface gap. Charged systems behave differently. MAE rises to 8.5–22.9 kcal/mol, with 36–68% of predictions off by more than 10 kcal/mol, and DFT single points at the MLIP geometry do not recover accuracy. DFT geometry reoptimization or CREST xTB+DFT restores it for ionic hydrogen bonds and ion–molecule dimers, while crystal-derived salt bridges stay limited by relaxation away from the reference structure. The same failure repeats across all three backbones, so it is not UMA-specific. The errors that remain on neutral chemistry cluster by chemistry family across backbones. Whether an arm fails catastrophically transfers across chemical families even though the error magnitude transfers less reliably, so a single closed-form cost-aware selector (RAPIDS-Select) commits the most cost-effective fidelity per system from cheap features alone. It uses roughly 37% less compute than a strong static DFT baseline at matched accuracy and, on charged chemistry it was never tuned for, recovers near-DFT accuracy at a fraction of the CREST cost while cutting the cheap-method catastrophe rate from 68% to 5.7%.

Our errors are deviations of method-specific relaxed scores against benchmark interaction-energy references rather than frozen-geometry interaction energies. The accuracies reported here (neutral MAE 1.37 kcal/mol with ρ = 0.66; selector MAE 2.49–3.01 kcal/mol) support coarse screening and catastrophe avoidance; on their own they do not establish reliable ordering of candidates separated by only 1–2 kcal/mol, which would require pairwise accuracy stratified by the reference gap.

WetSenseBench tests the low-cost limit: RAPIDS alone, with no DFT follow-up, is compared against qualitative selectivity calls from eight published sensing studies; its 82.9% above-floor pair-level agreement and 6-of-8 design-target recovery suggest that the neutral non-covalent ranking signal survives into many real host–guest sensing corpora, while the single cucurbit[6]uril precision-floor case marks where sub-kcal/mol gaps, solvent, ionic strength, and pose artifacts remain outside scope. RAPIDS is therefore best read as a high-throughput prioritization and escalation scaffold, not a replacement for DFT, CREST, or experimental validation on final hits.

## Author contributions

R.D. conceived the study and led the work, including the RAPIDS framework, the benchmark-computing DFT and MLIP simulations, the analyses, and the writing. R.P.F. and Z.D. contributed to algorithm development. D.L. contributed to the Bayesian optimization literature review. Y.C. and J.C. supervised the project and provided funding and guidance. All authors reviewed and approved the manuscript.

## Conflicts of interest

There are no conflicts to declare.

## Data availability

The RAPIDS benchmark family supporting this study is openly available at Zenodo: https://doi.org/10.5281/zenodo.20544147. The deposit contains the 5,567 × 9 multi-fidelity evaluation log (per-system interaction energies, wall-clock costs, and status flags for all nine fidelity arms, together with the published reference interaction energies and per-system charge and benchmark-family metadata); the RAPIDS-committed and reference per-system geometries; the cross-backbone MACE-OMol and ORB-OMol L1/L2/L3 datasets; the frozen-geometry ten-method benchmark; the 981-system DES370K external charged benchmark used by RAPIDS-Select, with CCSD(T)/CBS-quality references; and the WetSenseBench wet-lab selectivity benchmark with its per-pair geometries. Analysis, offline-replay, RAPIDS-Select, and figure-generation scripts are included in the same deposit. The RAPIDS scaffold source code is openly available at https://github.com/ruiding-uchicago/RAPIDS, together with a note documenting the agent interface and a worked agent case study. References 72–168 are cited in the ESI.

## Acknowledgements

This project is supported by Schmidt Sciences, LLC, through the Eric and Wendy Schmidt AI in Science Postdoctoral Fellowship, and by the University of Chicago Data Science Institute through the 2024 AI+Science Research Initiative. This work was also supported in part by the National Science Foundation Future Manufacturing project 2037026, Robust Intelligence project 2313131, 2332475, and CAREER award 2543755. An award of computer time was provided by the U.S. Department of Energy's Innovative and Novel Computational Impact on Theory and Experiment (INCITE) Program. This research used resources of the Argonne Leadership Computing Facility and the Carbon Cluster at the Center for Nanoscale Materials, both U.S. DOE Office of Science user facilities at Argonne National Laboratory, supported by the U.S. Department of Energy under Contract No. DE-AC02-06CH11357. Additional computational resources were provided by the University of Chicago Research Computing Center through proposal #68909 and by the ACCESS program (Delta cluster, allocation MAT240097), supported by NSF grants 2138259, 2138286, 2138307, 2137603, and 2138296.

## Notes and references

**Electronic Supplementary Information (ESI)**

# Geometry Sets the Neutral MLIP–DFT Gap in Non-Covalent Binding, and Charged Systems Fail

Rui Ding,[‡ac] Zixin Ding,[‡b] Rodrigo Pires Ferreira,[‡ac] Diantong Li,[b] Junhong Chen[*ac] and Yuxin Chen[*b]

[a] Pritzker School of Molecular Engineering, University of Chicago, Chicago, IL 60637, USA. E-mail: junhongchen@uchicago.edu
[b] Department of Computer Science, University of Chicago, Chicago, IL 60637, USA. E-mail: chenyuxin@uchicago.edu
[c] Chemical Sciences and Engineering Division, Argonne National Laboratory, Lemont, IL 60439, USA
[‡] These authors contributed equally to this work.

This ESI contains Sections A–D, Figs. S1–S19 and Tables S1–S13. Figures, tables and numbered sections (§) without an "S" or a letter refer to the main article; superscript reference numbers follow the main article's reference list, in which references 72–168 are those cited only in this ESI.

# Contents

# A Additional Related-Work Background

This section collects the longer background discussions that support the main-text Introduction and Related Work sections. It retains the fuller application motivation, search-and-score pipeline literature, universal-MLIP background, multi-fidelity decision literature, chemistry-agent survey, and MCP-plus-Skills interface discussion that were compressed in the main paper for readability.

## A.1 Why non-covalent molecular binding matters

RAPIDS targets a concrete computational task: for a probe–target pair, predict a pose-ranked relaxed vacuum electronic binding score

$$\Delta E_{\text{bind}} = E_{PT}(R_{PT}^{\text{opt}}) - E_P(R_P^{\text{opt}}) - E_T(R_T^{\text{opt}})$$

on a FAIRChem UMA OMol surrogate trained at the ωB97M-V/def2-TZVPD level, with an optional Analytical Linearized Poisson–Boltzmann (ALPB) implicit-water single-point correction. The quantity we study is a relative energetic ordering across candidate non-covalent complexes. It sits apart from binding free energies, docking scores, and kinetic observables. Many downstream applications depend on resolving such orderings when competing poses are separated by only 1–2 kcal/mol, at throughputs well beyond first-principles screening.

### A.1.1 Chemical sensors and biosensors as an application anchor

Synthetic chemosensors and biosensors built around macrocyclic hosts (cyclodextrins, calixarenes, cucurbit[n]urils, pillararenes) transduce host–guest recognition into optical or electrochemical signals, and small relative energetic differences among competing host–analyte complexes often track sensor selectivity against structurally similar interferents.[5] The cucurbituril family illustrates how tightly tuned this ordering can be: CB[7] and CB[8] reach monovalent association constants up to $10^{17}$ $M^{-1}$ in water via ion–dipole contacts at the portals, hydrophobic inclusion, and release of high-energy cavity water.[4] Anion sensing is even more constrained because hydration enthalpies exceed most receptor–anion interaction energies, forcing designs that rely on directional C–H···$X^-$, halogen-bond, and cooperative hydrogen-bond contacts.[72]

Device-level platforms inherit the same relative-ordering problem. Luminescent metal–organic frameworks embed host–guest chemistry in a porous scaffold whose emission shifts with guest inclusion, enabling discrimination among volatile organics, heavy-metal ions, and explosives.[73] Broadly responsive chemiresistive elements recover selectivity statistically through electronic-nose arrays whose response patterns are classified by machine learning models trained on large adsorption landscapes.[74] Wearable and point-of-care devices extend the same chemistry to sweat, tears, interstitial fluid, and breath, requiring receptors that separate biomarkers from structurally similar metabolites at sub-micromolar concentration,[75] and probes for nitroaromatic explosives, nerve-agent simulants, and chemical-warfare agents depend on turn-on/turn-off events driven by favorable non-covalent contacts that outcompete environmental interferents.[76] Kinetics, transduction, surface coverage, and matrix effects enter the observed signal in each case; what the computation is asked to supply is a reliable ranking of candidate probe–analyte interaction energies under small gaps and across flexible partners.

### A.1.2 The same pattern across adjacent domains

The same pattern recurs outside sensing. Structure-based drug discovery advances when congeneric ligands are reordered by a few kcal/mol against the target pocket and off-targets; alchemical free-energy protocols tightened this ranking but still inherit the underlying non-covalent energetics,[6] and macrocyclic host–guest blind challenges continue to stress-test the same physics.[77] In porous materials, gas storage, isomer separation, and catalytic selectivity are decided by $\lesssim 2$ kcal/mol differences between competing guest binding modes,[7] and pharmaceutical-cocrystal formation is controlled by relative lattice energies across competing polymorphs.[78] The question handed to the energy model often reduces to the same ranking subproblem: order candidate complexes under small gaps at screening throughput, even when the observable of interest is not itself a dimer free energy.

## A.2 Conformer search and ranking pipelines

Predicting which non-covalent probe–target arrangement dominates requires scanning a high-dimensional pose manifold and ranking candidates whose vacuum electronic binding scores frequently differ by less than 1–2 kcal $mol^{-1}$. Existing pipelines divide into three families, each with a different cost–accuracy tradeoff.

**Physics-based search on approximate potential-energy surfaces (PESs).** The dominant workflow for small, flexible non-covalent interaction (NCI) dimers is CREST,[8,79] widely adopted in the Grimme-group pipeline and community NCI conformer studies. CREST drives RMSD-biased metadynamics and genetic crossover on the semiempirical GFN2-xTB surface,[9] with an –nci wall restraint that keeps the dimer associated; a GFN-FF pre-screen[80] and a CENSO density functional theory (DFT) re-ranking stage[81] book-end the search. The broader enhanced-sampling family (replica exchange,[82] metadynamics[83]) shares the same template but typically runs on classical force fields, which systematically misorder halogen-bond, σ-hole, and ionic contacts. CREST is the strong baseline because its search operates on a quantum-mechanical PES with no ligand/receptor distinction. The bottleneck is cost: GFN2-xTB gradients dominate the inner loop, and layering a DFT single-point or re-optimization on top pushes per-system wall-clock into the regime that forecloses high-throughput screening (a claim we make concrete against our own measured CREST+DFT timings in the results), while ab initio molecular dynamics (MD) is ruled out entirely by its per-step DFT cost.

**Docking and heuristic pose generation.** Docking engines such as AutoDock Vina[11] and Glide[12] search translational, rotational, and torsional degrees of freedom against a grid-interpolated empirical score tuned on protein–ligand training sets, with deep-learning rescorers[84] and diffusion-based pose samplers[13] inheriting the same generate-then-externally-score pattern. They clear $10^3$–$10^5$ poses per CPU-hour, which makes them appropriate for virtual-screening triage, but the returned scores are shape-and-complementarity heuristics calibrated for rigid biomolecular pockets, and the absolute scores are not interpretable as quantitative interaction energies for generic, chemically diverse probe–target dimers with halogens, σ-holes, or charged fragments.

**Generative ML pose and conformer models.** RDKit's ETKDG[85] produces intramolecular conformers of a single molecule from knowledge-based torsion priors in milliseconds, while GeoDiff,[86] torsional diffusion,[87] and NeuralPLexer[14] extend generative diffusion or flow models to dimer and complex geometries at seconds per sample on a single GPU. These models are excellent candidate generators, but they sample from a learned geometric prior and do not emit calibrated electronic binding scores; quantitative ranking still requires an external energy evaluator downstream.

Across all three families the accurate option (CREST-style search on a quantum surface) spends nearly all of its wall-clock inside semiempirical or DFT single-points, while the fast options either return uncalibrated scores or no energies at all. An interatomic-potential surrogate with DFT-level accuracy for the electronic binding score would replace that inner-loop evaluator without forcing a redesign of the outer sampling logic; Section A.3 reviews why pretrained universal MLIPs are beginning to offer the accuracy and coverage needed to test this replacement directly.

## A.3 Universal machine learning interatomic potentials as a binding-score surrogate

RAPIDS predicts a pose-ranked relaxed vacuum electronic binding score

$$\Delta E_{\text{bind}} = E_{PT}(R_{PT}^{\text{opt}}) - E_P(R_P^{\text{opt}}) - E_T(R_T^{\text{opt}})$$

on the ωB97M-V/def2-TZVPD surface exposed by the FAIRChem UMA OMol head, with an optional ALPB implicit-water correction; this is neither a binding free energy nor a docking score. The question for this subsection is narrow: has the universal machine learning interatomic potential (MLIP) literature (architectures, training data, and pretrained checkpoints) reached the point where using a high-level binding-score evaluator inside an outer pose-search loop is *plausible enough to test empirically*, and what must a screening pipeline do on top of the MLIP to make its outputs trustworthy? The evidence below is proxy evidence (sibling models, covered-chemistry benchmarks, related training sets); whether it carries over to the specific UMA OMol head on our 5,567-dimer probe–target regime is the empirical question answered by the benchmarks in the results section of this paper.

**From invariant regressors to equivariant message passing.** MLIPs evolved from Gaussian kernels and invariant neural networks over hand-crafted descriptors[88,89] to E(3)-equivariant message-passing architectures that carry spherical tensors through every layer. NequIP,[20] MACE,[2] and Allegro[21] are representative: by embedding rotational symmetry into the representation rather than learning it, they reach sub-kcal/mol force-field accuracy with roughly an order of magnitude less training data than invariant predecessors. EquiformerV2 later extended this family to higher angular-momentum channels at transformer scale.[90]

**Molecular-chemistry data at training scale.** The second enabler was training data that actually covers drug-like, probe–target-relevant chemistry. SPICE provides >1.1 M conformations of small molecules, dimers, dipeptides, and solvated amino acids at ωB97M-D3(BJ)/def2-TZVPPD;[15] OMol25 releases over 100 M ωB97M-V/def2-TZVPD single-points across 83 elements with explicit solvation, variable charge, and variable spin, up to 350 atoms per system;[1] and DES370K supplies CCSD(T)-level dimer interaction energies for external charged validation.[19] Materials-side corpora such as OC20, MPtrj, and OMat24 broadened periodic-table coverage for joint pretraining[91–93] but are peripheral to the dimer interaction energies of interest here.

**Universal pretrained models.** Several pretrained universal MLIPs now ship as off-the-shelf energy/force evaluators, including MACE-MP-0 for atomistic materials chemistry,[94] MACE-OFF23 for organic molecules,[95] and UMA, an eSEN backbone with a Mixture of Linear Experts (MoLE) jointly pretrained on OMol25, OC20, OMat24, ODAC23, and OMC25.[3] RAPIDS invokes the UMA OMol head specifically because it is trained at ωB97M-V/def2-TZVPD with VV10 dispersion, a high-performing NCI DFT level and the OMol25/UMA OMol training level, so its native predicted quantity lives at the same electronic-structure level as the relaxed score RAPIDS wants to rank.

**What the pretrained model supplies.** Four properties matter for inner-loop use. *Accuracy*: MACE-OFF23 reports low errors on neutral closed-shell organic-molecule test sets derived from SPICE and DES370K dimers relative to its training theory,[95] and ANI-1ccx approaches CCSD(T)/CBS conformer rankings at neural-network cost.[96] *Coverage*: UMA and MACE-MP span most of the periodic table with a single checkpoint,[3,94] and OMol25 extends to charged and open-shell species and to explicit-solvent environments.[1] *Speed*: for a typical probe–target dimer the dominant cost per energy/force call drops by several orders of magnitude relative to a hybrid-DFT/def2-TZVPD reference on comparable hardware; concrete wall-clock comparisons against CREST+DFT are reported in the results section of this paper. *Differentiability*: analytic forces come from one backward pass, which is what lets RAPIDS drive LBFGS geometry optimization on the MLIP surface without finite differences.

**Known failure modes that the pipeline must own.** Within covered chemistry a pretrained MLIP can be a promising inner-loop surrogate for DFT single-point evaluation inside a relaxed-pose molecular screening loop when evaluated relative to the same reference functional, but it still comes with important caveats. Three failure modes persist: (i) out-of-distribution extrapolation on unusual oxidation states, rare element combinations, or reactive geometries; (ii) absence of explicit long-range electrostatics in the finite-cutoff local architectures shared by most current backbones, which is consequential for highly polar or charged dimers; and (iii) inherited systematic error of the training functional, so that a UMA OMol prediction is at best as accurate as ωB97M-V/def2-TZVPD, and any residual deviation from CCSD(T) on a given motif propagates into every score.

**Fine-tuning as an alternative path, and its costs.** A natural way to address these failure modes is to fine-tune the foundation MLIP on task-specific data, and a growing literature reports clear gains from this route, including correcting systematic potential-energy-surface softening,[16] reaching sub-kJ/mol sublimation enthalpies on molecular crystals with limited high-level data,[17] and frozen-weight transfer learning for materials chemistry.[18] Fine-tuning is, however, expensive on two axes that compound. First, it requires curating a downstream-relevant labeled dataset at the target electronic-structure level, which for non-covalent dimers means CCSD(T)/CBS-style or high-level DFT references that are themselves the bottleneck the surrogate was meant to bypass. Second, training and adapting modern equivariant message-passing architectures is itself compute-heavy: EquiformerV2 reports on the order of 1,368 GPU-days for a single OC20 run,[90] and dedicated systems work has been needed to bring equivariant-MLIP training and inference down to practical cost.[97] RAPIDS therefore complements rather than replaces this fine-tuning literature: it asks how far one can push a pretrained universal MLIP at inference time, with no additional training or labeled data beyond a small per-benchmark calibration slice, before fine-tuning becomes worth its data and compute cost.

**What RAPIDS adds.** A universal pretrained MLIP provides an inner-loop energy surrogate, but it does not by itself perform the pose search, guard against the ML-specific failure modes it introduces, or calibrate against a gold-standard reference. RAPIDS supplies these pieces. The remainder of the paper lays out the tiered orientation sampling, LBFGS with smart continuation, four per-configuration guards (topology/geometry/energy/convergence) together with a scan-level Energy Consistency Guard (ECG), implicit-solvation post-correction, and benchmark harness that together wrap the UMA OMol head as a probe–target binding-score predictor for screening, benchmarked against CREST+DFT on 5,567 dimers drawn from 18 NCI datasets.

## A.4 Multi-fidelity decision-making and offline policy evaluation

A benchmark in which each retained system has been scored by every available method is, by construction, an offline multi-fidelity decision log: each dimer is a context, each fidelity is an arm, and each successfully computed entry of the 5,567 × 9 result table is a realized (reward, cost) pair. Cheap arms (the RAPIDS MLIP-relaxed score, plain DFT single-points on the MLIP geometry) are fast and sometimes inaccurate; expensive arms (CREST_xTB_DFT) are slow and not uniformly more accurate. As cataloged in §A.2, no single arm dominates the others across the 18 datasets, which motivates treating method selection as an adaptive decision problem rather than a fixed pipeline.

### A.4.1 Multi-armed and multi-fidelity bandits

The stochastic multi-armed bandit problem formalizes sequential choice under uncertainty with a logarithmic regret lower bound,[98] matched by upper-confidence-bound strategies[23] and by posterior-sampling strategies whose modern regret analyses[99] revived an idea of Thompson.[57] Contextual extensions such as LinUCB[24] condition arm choice on a feature vector, which is the object most similar to the per-system descriptors RAPIDS already computes. Multi-fidelity bandit variants make arm cost and resolution explicit: MF-GP-UCB[25] and its continuous-approximation successor BOCA[26] use cheap low-fidelity queries to localize the optimum before committing expensive evaluations, and multi-information-source knowledge-gradient acquisition generalizes this to biased surrogates.[100,101] The hyperparameter-optimization line (successive halving,[102] Hyperband,[103] BOHB,[104] and freeze-thaw Bayesian optimization[105]) treats partial training budget as a fidelity axis, and entropy-based acquisition[106] quantifies information gain per unit cost. Multi-fidelity Bayesian optimization has been used for molecular and materials design, where expensive DFT or experimental measurements are paired with cheaper force-field or semi-empirical surrogates.[107]

#### A.4.2 Meta-Bayesian optimization

In meta-Bayesian optimization (meta-BO), evaluations from related tasks improve BO on a new task.[27] One common approach learns a surrogate that captures features shared across tasks, either by combining surrogates fitted to each task[108–110] or by fitting a joint model or shared prior.[111–118] Meta-acquisition methods instead recast BO as a Markov decision process and meta-train a reinforcement-learning agent on related tasks.[28,119–121] RAPIDS-Select shares the high-level idea that past evaluations inform the choice on a new task: its regressors are fitted on the execution log of other systems and then select a fidelity for a new one. It commits one fidelity per system, with no surrogate update or acquisition loop.

#### A.4.3 Offline policy evaluation and bandit replay

When new online experiments are costly, a logged dataset of (context, action, reward) tuples can be used to estimate the value of counterfactual policies. Importance-sampling estimators reweight logged rewards by the target/logging propensity ratio;[122] doubly-robust estimators combine importance sampling with a learned reward model to reduce variance while retaining consistency;[29,123] self-normalized variants and counterfactual risk minimization bound generalization when propensities are small.[124] For exchangeable logs with a uniform logging policy, Li et al.[125] showed that a simple "replay" estimator (step through logged events and accept only those whose logged action matches the target policy's choice) is unbiased and has become the standard offline evaluation protocol in recommendation and advertising. These estimators are the machinery a future *online* RAPIDS deployment would need, where only the arm actually selected per system is logged; on the precomputed benchmark used here the factorial (context × arm) design means they collapse to direct estimation on the subset of cells that completed successfully. The current paper goes one step further and replays deployable rule-based/frozen policies together with retrospective full-information and online-feedback policies, so the method section treats this matrix as an offline replay object first and as a conceptual bandit analogue second.

#### A.4.4 Why a replay-ready benchmark matters

The 5,567 × 9 precomputed reward–cost table released with RAPIDS is, operationally, a replay log over probe–target binding-score evaluation (remaining SCF-failure cells are exposed as NaN and handled by a declared fallback in the replay protocol of §3.4). Any allocation policy that maps an incoming dimer to one fidelity (or to a sequence of fidelities with an early-exit rule) can be evaluated on this log without running a single new DFT job, and comparisons across policies inherit the same benchmark references and wall-clock costs. This subsection provides the vocabulary; the method and results sections instantiate specific policies, from an oracle lower bound through contextual and cost-aware variants, and compare their cost–accuracy curves against that lower bound rather than reporting an explicit cumulative-regret trajectory.

### A.5 Large language model agents for autonomous chemistry

Large language models can orchestrate tool-mediated workflows: rather than emitting a final answer in one pass, they decide when to call an external API, how to parse its output, and how to chain results into a longer reasoning trajectory.[126,127] Chemistry has been one of the earliest natural-science domains where this pattern took hold, because many chemistry sub-problems (structure retrieval, reaction condition lookup, property prediction, simulation submission) already exist as callable tools behind stable interfaces. What the agent layer adds is a way to compose those tools from a natural-language goal.

#### A.5.1 Autonomous-chemistry agent frameworks

ChemCrow couples GPT-4 to eighteen expert-designed tools spanning literature search, reaction prediction, retrosynthesis, and safety checks, and closes end-to-end loops for target synthesis and chromophore design.[33] Coscientist pairs an LLM planner with documentation search, Python execution, and a cloud-lab API to carry out palladium-catalyzed cross-coupling optimization on physical hardware.[34] Chemist-X targets reaction-condition recommendation with retrieval-augmented generation over molecular databases,[35] while CACTUS connects open-source LLMs to cheminformatics tools for molecular-property queries and drug-likeness assessment.[128] ChemToolAgent quantifies when tool augmentation helps versus hurts on general chemistry problems,[129] and ChemAgent introduces a self-updating memory library that improves multi-step chemical reasoning.[130] Materials-discovery agents follow a similar recipe: ChatMOF couples an LLM to MOF databases and a generator for inverse design;[36] HoneyComb retrieves from a curated materials knowledge base and a tool hub;[37] LLMatDesign prompts a reasoning model through iterative materials-property optimization.[131] Adjacent scientific-reasoning agents such as SciAgent and drug-pipeline agents such as PharmAgents extend the pattern to broader domains.[132,133] Verification strategies from general agent research (self-consistency voting, tree search over reasoning paths, and search-guided fine-tuning) are increasingly imported to reduce tool-call errors and unfounded claims.[134–136]

### A.5.2 LLMs for chemical optimization

Some work places LLMs inside the optimization loop instead of using them as tool-calling planners. BO-ICL uses frozen LLMs for in-context regression with uncertainty, encoding each material as its synthesis and testing procedure. It matches or outperforms Gaussian processes on aqueous-solubility and oxidative-coupling-of-methane benchmarks, and in live reverse water–gas shift experiments it found multimetallic catalysts approaching equilibrium CO yield within 6 and 10 iterations, from pools of 3,700 and 360,000 candidates.[137] LLAMBO frames BO in natural language so that an LLM proposes and evaluates candidates conditioned on past evaluations, and it is effective at zero-shot warm-starting.[138]

Other approaches use the LLM as a representation for a conventional surrogate. BoChemian feeds fixed LLM embeddings into BO of chemical reactions.[139] Ranković et al. train the LLM through the uncertainty-aware objective of the optimizer, raising the discovery rate of high-yielding Buchwald–Hartwig conditions in 50 iterations from 24% with static embeddings to 43%.[140]

The LLM can also serve directly as the surrogate or the search operator. LICO adds embedding and prediction layers to a base LLM and trains it for in-context prediction across molecular functions, reaching state-of-the-art performance on the low-budget PMO-1K benchmark.[141] MolLEO replaces the random crossover and mutation operators of an evolutionary algorithm with chemistry-aware LLMs.[142] Kristiadi et al. find that LLMs help BO over molecules only when pretrained or fine-tuned on domain-specific data.[143]

### A.5.3 The open gap for non-covalent binding

The frameworks above concentrate on synthesis and retrosynthesis, materials inverse design, or literature-grounded QA. Non-covalent binding, the dominant physics in sensor design, lead optimization, and supramolecular chemistry, remains underserved because every available energy oracle fails one agent-level requirement. A DFT single-point takes minutes to hours per dimer, which is incompatible with the many tool calls a tree-search or self-consistency agent emits per task (LATS, tree-of-thoughts and related orchestrators grow this budget with tree width and sample count); empirical force fields and docking scores are fast but cannot rank the sub-kcal/mol differences that separate competing probes or competing poses for a given target. Commercial AI-assisted discovery platforms expose docking, FEP+, and predictive-model pipelines to chemists, but we are not aware of a published, open-protocol, MLIP-level binding-score oracle callable through an agent interface. RAPIDS addresses that gap: it exposes MLIP-speed binding-score predictions (§A.3), wrapped in validation guards and served through a Model Context Protocol endpoint whose tool surface is described in §A.6.[39] The main contribution here is the oracle, the benchmark matrix, and the agent-ready interface; a worked agent case study is documented in the RAPIDS GitHub repository.

## A.6 Tool exposure for scientific agents: Model Context Protocol (MCP) and progressive-disclosure skills

Large language models now treat tool use as a first-class capability: a prompt declares external functions, and the model emits structured calls that the host executes and returns to the context window.[144,145] Raw function-calling scales badly, however: every added tool inflates the prompt, schemas drift across vendors, and invocations are rarely captured in a replayable form. The Model Context Protocol (MCP), together with Anthropic's Agent Skills pattern, which operationalizes *progressive disclosure* for LLM tool exposure, address these failure modes, and together shape how a scientific tool such as RAPIDS is surfaced to an agent.

### A.6.1 From function calling to the Model Context Protocol

Early tool-augmented LLMs interleaved reasoning with external actions[127] and learned to self-select APIs.[126] Scaling past a handful of tools motivated retrieval-over-APIs benchmarks such as ToolLLM/ToolBench[146] and Gorilla,[147] and orchestration frameworks including LangGraph,[148] AutoGen,[149] CrewAI,[150] and Semantic

Kernel.[151] Tool exposure nonetheless remained ad hoc: each client re-implemented schema plumbing, authentication, and logging. MCP, announced by Anthropic in November 2024,[39,40] standardizes this interface: an MCP *server* advertises three core primitives (tools, resources, and prompts) over JavaScript Object Notation Remote Procedure Call (JSON-RPC), and any MCP *client* can consume them. Current client implementations include VS Code's agent mode[152] and the clients cataloged in the MCP specification,[40] making a single MCP server reachable from the growing ecosystem of scientific and developer agents.

### A.6.2 Progressive disclosure and the Skills pattern

Even with a clean transport, advertising dozens of tools with full schemas saturates the context window. Anthropic's Agent Skills pattern,[41,42] described later within the broader Anthropic agent-tooling ecosystem that also includes Computer Use,[153] operationalizes *progressive disclosure*: only a short skill name and one-sentence description are loaded up front, while full instructions, schemas, and reference files are fetched on demand. For a scientific package the consequences are concrete. The RAPIDS tool surface would otherwise displace task-relevant context, whereas a progressively disclosed summary remains near-free until invoked. Progressive disclosure is also compatible with client-side audit logging: hosts such as Claude Desktop and Claude Code log each tool invocation's typed inputs and outputs, so that together with server-reported versioning a reviewer can reconstruct which tool version produced which number, a deployment property the MCP transport makes straightforward but does not itself guarantee.

### A.6.3 Consequences for RAPIDS

MLIP-based binding prediction is only useful to an agent if it is callable with a clean interface, a small prompt footprint, and logs that downstream clients can audit. MCP provides the transport; progressive disclosure, whether via a full Skills wrapper or via MCP tools whose surface area is deliberately minimized, keeps the prompt footprint small. RAPIDS packages its guarded pose search, validation guards, fidelity catalog, and offline-replay artifacts as a small Skill-level surface over the underlying MCP server, so an agent can call the live oracle, inspect the nine-fidelity ladder, and request replay-based allocation advice through the same interface. Agent-side orchestration is delegated to the frameworks and clients surveyed in §A.5, and the concrete tool inventory is documented in the RAPIDS GitHub repository.

# B Detailed Methods

This section records the exact quantity definitions, initialization choices, thresholds, pseudocode, and protocol details that support the compact main-text Methods section. Both pipelines evaluated in this paper target the same predicted quantity: a relaxed vacuum electronic binding energy

$$\Delta E_{\text{bind}} = E_{PT}(R_{PT}^{\text{opt}}) - E_P(R_P^{\text{opt}}) - E_T(R_T^{\text{opt}})$$

for a probe–target dimer on the ωB97M-V/def2-TZVPD[56,154,155] high-level functional surface. Because both RAPIDS and CREST+DFT relax the complex and the isolated monomers before subtraction, this is *not* a strict frozen-geometry interaction energy. Throughout the paper the benchmark target remains the published interaction-energy reference

$$\Delta E_{\text{int}}^{\star} = E_{PT}(R_{PT}^{\star}) - E_P(R_P^{PT,\star}) - E_T(R_T^{PT,\star}),$$

so every reported MAE should be read as a benchmark-reference deviation after method-specific pose relaxation. The sections below expand the compact main-text summaries in §3.2–§3.4.

For benchmark evaluation on the 5,567-system completed split[1], PubChem retrieval[43] is bypassed: monomer coordinates, charges, and the benchmark reference geometry are taken from the dataset-provided dimer records, the two monomers are separated from that record, and each pipeline then reinitializes the complex from those monomers according to its own search rule. For a fixed system and fidelity, the isolated probe and target monomers are evaluated or optimized once with that method's charge/spin assignment and reused when scoring candidate poses of the same system.

[1] The underlying collection contains two additional unsupported A24 systems, for which at least one lower-level engine in the pipeline stack does not support the chemistry.

**Table S1** Benchmark map for the 5,567 completed systems used in the paper. Counts are completed-system counts. The last column records the published benchmark level treated as ground truth in this work together with the original dataset citation.

| Benchmark | *N* | Charge | Representative chemistry | Published ground truth |
|---|---|---|---|---|
| A24 | 22 | Neutral | Canonical small-molecule dimers. | CCSD(T)/CBS-style interaction energies; A24 benchmark.[44] |
| S66 | 66 | Neutral | Balanced H-bond, dispersion, and mixed organic dimers. | CCSD(T)/CBS interaction energies; S66 benchmark.[45] |
| X40 | 40 | Neutral | Halogen-bond dimers. | CCSD(T)/CBS interaction energies; X40 benchmark.[46] |
| HB300SPX | 300 | Neutral | Extended hydrogen-bond set with heavier donor and acceptor chemistry. | NCIAtlas CCSD(T)/CBS-style interaction energies; HB300SPX.[48] |
| HB375 | 375 | Neutral | Diverse neutral hydrogen-bond dimers. | NCIAtlas CCSD(T)/CBS-style interaction energies; HB375.[47] |
| SH250 | 250 | Neutral | σ-hole interactions across halogen, chalcogen, and pnictogen donors. | NCIAtlas CCSD(T)/CBS-style interaction energies; SH250.[49] |
| D1200_HBCNO | 308 | Neutral | London-dispersion subset containing only H, B, C, N, and O. | NCIAtlas CCSD(T)/CBS-style interaction energies; D1200.[50] |
| D1200_Halogens | 310 | Neutral | Halogen-containing London-dispersion subset. | NCIAtlas CCSD(T)/CBS-style interaction energies; D1200.[50] |
| D1200_PS | 293 | Neutral | P/S-containing London-dispersion subset. | NCIAtlas CCSD(T)/CBS-style interaction energies; D1200.[50] |
| BFDb_BBI | 100 | Neutral | Protein backbone–backbone fragments. | BFDb CCSD(T)-F12 silver-standard interaction energies.[51] |
| BFDb_HSG | 21 | Neutral | Indinavir–residue fragments from the HIV-II protease complex (PDB 1HSG). | BFDb CCSD(T)-F12 silver-standard interaction energies.[51] |
| BFDb_NBC1 | 10 | Neutral | Nucleobase stacking fragments. | BFDb CCSD(T)-F12 silver-standard interaction energies.[51] |
| BFDb_SSI_dispersion | 1325 | Neutral | Sidechain pairs dominated by dispersion. | BFDb CCSD(T)-F12 silver-standard interaction energies.[51] |
| BFDb_SSI_mixed | 765 | Neutral | Sidechain pairs with mixed polar and dispersion character. | BFDb CCSD(T)-F12 silver-standard interaction energies.[51] |
| BFDb_SSI_other | 240 | Neutral | Sidechain pairs outside the polar, mixed, and dispersion buckets. | BFDb CCSD(T)-F12 silver-standard interaction energies.[51] |
| BFDb_SSI_polar | 266 | Neutral | Polar and electrostatic sidechain pairs. | BFDb CCSD(T)-F12 silver-standard interaction energies.[51] |
| IHB100 | 100 | Charged | Ionic hydrogen bonds with formally charged fragments. | NCIAtlas CCSD(T)/CBS-style interaction energies; IHB100 set in the HB/IHB benchmark family.[47] |
| BFDb_SSI_charged | 776 | Charged | Salt bridges, charged sidechains, and same-charge repulsion pairs. | BFDb CCSD(T)-F12 silver-standard interaction energies.[51] |

## B.1 Detailed RAPIDS Implementation

RAPIDS (Rapid Atomistic Probe–target Interaction Discovery Scaffold) builds on the externally pretrained UMA OMol head.[1,3] The work here contributes the orientation sampler, the optimizer with smart continuation, the validation stack, the solvation post-correction, the offline-replay harness, and the application-layer MCP/Skills packaging. To the best of our knowledge, this is the first systematic use of a pretrained universal MLIP as an interaction-system scaffold benchmarked at this scale and checked against wet-lab sensing selectivity. The implementation supports probe–target–substrate assemblies, but the benchmarked pathway in this paper is the vacuum probe–target dimer case.

**Model provenance.** Table S2 pins the exact checkpoints of the three MLIP backbones, used as released with no task-specific fine-tuning; reproducibility is fixed by package version and, for UMA, the Hugging Face snapshot hash.

**Table S2** Model provenance for the three MLIP backbones (used as released, no fine-tuning).

| | **UMA**[3] | **MACE-OMol**[2] | **ORB-OMol**[60] |
|---|---|---|---|
| Checkpoint | `uma-s-1p2` | `mh-1` (omol) | `orb_v3_conservative_omol` |
| Package | fairchem-core 2.16.0 | mace-torch 0.3.16 | orb-models 0.5.5 |
| Parameters | 291 M | 6.4 M | 25.8 M |
| Cutoff | 6.0 Å[a] | 6.0 Å | 6.0 Å (max. 120 nbrs) |
| Force type | conservative[b] | conservative | conservative |
| Precision | float32 | float64 | float32 |
| Checkpoint hash | HF snapshot[c] | ckpt md5 `140206029`… | orb-models 0.5.5 |

[a] From the UMA-s design (`max_radius` in the `uma-s-1p2` config); the released averaged-model wrapper does not expose the attribute. [b] Conservative (energy-gradient) forces per the UMA-s design; the wrapper does not expose a `direct_forces` flag, and the residual-force and reference-relaxation analyses (Fig. 4) are consistent with conservative gradients. [c] Hugging Face snapshot `7210de6fe86ad94854b21b881fefbcfdfeab373b` (`checkpoints/uma-s-1p2.pt`).

**Execution scope.** The benchmarked RAPIDS path is the guarded screening workflow: structure building, tiered search, LBFGS relaxation, per-configuration validation, scan-level ECG aggregation, and optional solvation.

**Energy surface.** Every energy or force call is a forward (forward–backward for gradients) pass through the UMA OMol head, an eSEN backbone with a Mixture of Linear Experts (MoLE) pretrained at ωB97M-V/def2-TZVPD with VV10 dispersion on OMol25.[1,3] The head is used here only as a surrogate energy/force evaluator; it does not run explicit DFT at inference time. No self-consistent-field (SCF) iterations, no basis-set selection, and no finite-difference gradients are performed at inference time.

**Tiered orientation sampling.** The pose search is a three-tier structured scan rather than a stochastic metadynamics. Tier 1 always runs and enumerates $n_{\text{anchors}} \times n_{\text{orientations}} = 3 \times 3 = 9$ probe placements using three anchor offsets and three coarse orientations. Tier 2 adds six basin-hopping perturbations of the best Tier-1 configuration, with randomized rotations and translations around that candidate. Tier 3 adds nine random refinements near the running best. Tier 1→2 upgrades fire if any of four criteria hold on the eligible Tier-1 set: $\sigma_E > 0.3$ eV (≈6.9 kcal/mol), top-two gap < 0.02 eV (≈0.46 kcal/mol), eligible fraction < 80%, or system size above 50 atoms; Tier 2→3 fires if Tier 2 improves the Tier-1 minimum by more than 0.1 eV (≈2.3 kcal/mol). Caller-forced minimum tiers (min_tier ≥ 2 or ≥ 3) override these triggers, whereas an empty eligible Tier-1 set blocks any upgrade.

Formally, letting $\mathcal{R}_1$ denote the Tier-1 result set, define

$$\mathcal{R}_1^{\text{elig}} = \{r \in \mathcal{R}_1 : c(r) \in \{\text{high}, \text{medium}\}\}, \tag{1}$$

$$\sigma_1 = \sigma\left(\{E(r) : r \in \mathcal{R}_1^{\text{elig}}\}\right), \tag{2}$$

$$\eta_1 = \frac{|\mathcal{R}_1^{\text{elig}}|}{|\mathcal{R}_1|}, \tag{3}$$

$$\Delta E_{1,2} = E_2^{(1)} - E_{\text{min}}^{(1)}. \tag{4}$$

We use the conventions $\sigma_1 = 0$ if $|\mathcal{R}_1^{\text{elig}}| = 0$, $\Delta E_{1,2} = +\infty$ if $|\mathcal{R}_1^{\text{elig}}| < 2$, $\eta_1 = 0$ if $|\mathcal{R}_1| = 0$, and $E_{\text{min}}^{(2)} = +\infty$ if the eligible Tier-2 set $\mathcal{R}_2^{\text{elig}}$ is empty. The two upgrade gates are then

$$U_{1\to 2} = \begin{cases} 1, & \text{if max_tier} \geq 2,\ |\mathcal{R}_1^{\text{elig}}| > 0, \text{ and at least one of the following} \\ & \text{holds: } \sigma_1 > 0.3 \text{ eV};\ \Delta E_{1,2} < 0.02 \text{ eV};\ \eta_1 < 0.8;\ N_{\text{atoms}} > 50;\ \text{min_tier} \geq 2, \\ 0, & \text{otherwise,} \end{cases}$$

$$U_{2\to 3} = \begin{cases} 1, & \text{if max_tier} \geq 3 \text{ and either min_tier} \geq 3 \text{ or} \\ & \left(|\mathcal{R}_2^{\text{elig}}| > 0 \wedge E_{\text{min}}^{(2)} < E_{\text{min}}^{(1)} - 0.1 \text{ eV}\right), \\ 0, & \text{otherwise.} \end{cases}$$

The second gate uses the best eligible Tier-2 energy only when such a candidate exists. These definitions are implementation-faithful summaries of the prose trigger rules above.

**Local optimization.** Each configuration is relaxed with limited-memory BFGS on the MLIP surface to $f_{\text{max}} = 0.05$ eV Å$^{-1}$, capped at 100 steps per run with up to three *smart continuations* that re-enter optimization from the best previously seen geometry.

**Validation guards.** RAPIDS separates per-configuration guards from scan-level aggregation. The per-configuration guard stack contains topology, geometry, energy, and convergence checks. The *topology guard* compares covalent bond graphs of the initial and final structures (ase.neighborlist[156] with element-pair thresholds and a 15% bond-change tolerance). If an intramolecular bond change is detected on a neutral system, the worker re-optimizes the same geometry with fallback UMA heads and downgrades confidence by majority vote; for systems with any nonzero fragment or complex charge ($q_P \neq 0$, $q_T \neq 0$, or $q_{PT} \neq 0$), the worker downgrades confidence directly and returns early without executing the later guards. When topology is preserved, the *geometry guard* flags severe non-bonded contacts via the MolProbity criterion[157] $d_{ij} < r_i^{\mathrm{vdW}} + r_j^{\mathrm{vdW}} - 0.4$ Å together with strained intramolecular bonds. The *energy guard* writes $E_{\mathrm{bind}} \equiv \Delta E_{\mathrm{bind}}$ and applies two thresholds using the smaller fragment atom count $N_{\mathrm{small}} = \min(|P|, |T|)$ rather than the full complex size: $|E_{\mathrm{bind}}|/N_{\mathrm{small}} \leq$ 0.65 eV/atom and $|E_{\mathrm{bind}}| \leq 2.2$ eV (≈50.7 kcal/mol). The benchmarked flow uses this two-threshold rule. The absolute 2.2 eV ceiling is an empirical catastrophic-failure threshold aligned with the 50 kcal/mol clipping scale used in evaluation. That clipping scale lies well above even very strong monovalent host–guest binding free energies, which are on the order of ~20–24 kcal/mol for benchmark CB[7] complexes,[158,159] and it is also consistent with chemistry-ML practice that treats catastrophic thresholds pragmatically[160] and regards energy deviations of roughly 40–56 kcal/mol as unreasonably large.[161] Finally, the *convergence guard* downgrades only topology-preserved runs, assigning low to diverged smart-continuation sequences and capping step-limit exits at medium. These guard thresholds live in the worker's native eV units and act as catastrophic-failure sentinels rather than as ranking thresholds in the 1–2 kcal/mol regime discussed in §A.1.

At the level of abstraction used in the paper, let $b_{\mathrm{early}}(r) = 1$ denote the broken-topology branch that returns before the later guards, and $b_{\mathrm{early}}(r) = 0$ otherwise. Each candidate $r$ then determines the executed-guard index set

$$\mathcal{G}_{\mathrm{exec}}(r) = \begin{cases} \{\mathrm{topo}\}, & b_{\mathrm{early}}(r) = 1, \\ \{\mathrm{topo}, \mathrm{geom}, \mathrm{en}, \mathrm{conv}\}, & b_{\mathrm{early}}(r) = 0. \end{cases}$$

The emitted alert set is then

$$\mathcal{A}(r) = \{\, g_k(r) \colon k \in \mathcal{G}_{\mathrm{exec}}(r) \,\}.$$

The energy alert itself is

$$\begin{aligned} g_{\mathrm{en}}(r) &= \mathbf{1}\left[\frac{|E_{\mathrm{bind}}(r)|}{N_{\mathrm{small}}(r)} > 0.65 \text{ eV/atom} \lor |E_{\mathrm{bind}}(r)| > 2.2 \text{ eV}\right], \\ E_{\mathrm{bind}}(r) &\equiv \Delta E_{\mathrm{bind}}(r), \\ N_{\mathrm{small}}(r) &= \min(|P|, |T|). \end{aligned}$$

The reported confidence is an order-sensitive mapping

$$c(r) = \phi(\mathcal{A}(r)),$$

where unrecoverable topology breaks and non-physical energies force low, while clash-only or step-limit events cap confidence at medium. In the broken-topology branch, the worker returns immediately after fallback-task consensus and the later guards are not evaluated.

**Energy Consistency Guard.** The scan-level *Energy Consistency Guard* (ECG) is executed only in the MCP scan_orientations path, not in standalone CLI runs. It operates on the high/medium-confidence candidate set

$$\mathcal{R}^{\mathrm{elig}} = \{r \colon c(r) \in \{\mathrm{high}, \mathrm{medium}\}\},$$

using the geometric-similarity predicate

$$S(r_i, r_j) = \mathbf{1}\left[\Delta\mathrm{COM}(r_i, r_j) \leq 1.5 \text{ Å} \land \Delta d_{\min}(r_i, r_j) \leq 0.5 \text{ Å}\right],$$

and the similarity-restricted energy span

$$\Delta_{\mathrm{span}} = \max_{i,j \colon S(r_i, r_j) = 1} |E_{\mathrm{bind}}(r_i) - E_{\mathrm{bind}}(r_j)|,$$

with the convention $\Delta_{\mathrm{span}} = 0$ when no pair satisfies the similarity predicate, and

$$E_{\min} = \min_{r \in \mathcal{R}^{\mathrm{elig}}} E_{\mathrm{bind}}(r).$$

Let

$$b_{\mathrm{sim}} = \mathbf{1}\left[\exists\, i, j \colon\ S(r_i, r_j) = 1\right].$$

The ECG mode is then

$$m_{\mathrm{ECG}} = \begin{cases} \text{manual_review}, & \Delta_{\mathrm{span}} > 10 \text{ eV} \ \land\ b_{\mathrm{sim}} = 1, \\ \text{median}, & \Delta_{\mathrm{span}} > 3 \text{ eV} \ \land\ b_{\mathrm{sim}} = 1, \\ \text{median}, & E_{\min} < Q_1 - 3\,\mathrm{IQR}, \\ \text{minimum}, & \text{otherwise}. \end{cases}$$

with committed energy

$$E^{\star} = \begin{cases} \min\limits_{r \in \mathcal{R}^{\text{elig}}} E_{\text{bind}}(r), & m_{\text{ECG}} = \text{minimum}, \\ \operatorname*{median}\limits_{r \in \mathcal{R}^{\text{elig}}} E_{\text{bind}}(r), & m_{\text{ECG}} = \text{median}. \end{cases}$$

When $m_{\text{ECG}}$ = manual_review, the run is flagged for DFT follow-up and the worker reports both the minimum and median summaries rather than forcing a single committed scalar. The ECG thresholds (3 and 10 eV, i.e. ≈69 and 231 kcal/mol) are deliberately far above the ranking scale of interest and serve only as catastrophic-instability sentinels.

**Implicit solvation.** A vacuum-mode run can additionally attach a single-point GFN2-xTB[9] calculation with the ALPB implicit-water model[162] on the optimized probe, target, and complex geometries. The solvation correction is

$$\Delta\Delta G_{\text{solv}} = \left(G_{PT}^{\text{aq}} - G_{PT}^{\text{gas}}\right) - \left(G_{P}^{\text{aq}} - G_{P}^{\text{gas}}\right) - \left(G_{T}^{\text{aq}} - G_{T}^{\text{gas}}\right),$$

and the reported ALPB-solvation-corrected electronic score is

$$\Delta G_{\text{sol}} = \Delta E_{\text{bind}}^{\text{vac}} + \Delta\Delta G_{\text{solv}}.$$

We write the three-body correction as $\Delta\Delta G_{\text{solv}}$ to distinguish it from the final score $\Delta G_{\text{sol}}$. This quantity omits entropy, thermal, and standard-state corrections and should not be read as a full solution-phase binding free energy. The benchmark metrics in this paper remain vacuum electronic binding scores unless solvation is explicitly requested.

**Output.** Each simulation records the optimized geometries, trajectories, binding energies, solvation corrections when requested, convergence status, per-guard validation outcomes, and a confidence-annotated best-pose $\Delta E_{\text{bind}}$.

**Algorithmic summary.** Algorithms 1 and 2 summarize the guarded MCP execution path used throughout the paper.

**Algorithm 1** RAPIDS guarded scan

Retrieve molecules and build the initial probe–target(-substrate) complex.
Run Tier 1 grid search (3 × 3 = 9 placements), optimizing and validating each candidate with Algorithm 2.
**if** the eligible Tier 1 set is empty **then**
    Return the current result set without upgrading.
**end if**
**if** $\sigma_E > 0.3$ eV **or** top-two gap < 0.02 eV **or** valid fraction < 80% **or** $N_{\text{atoms}} > 50$ **or** min_tier≥ 2 **then**
    Run Tier 2 basin hopping (6 perturbed configurations around the Tier 1 best).
**end if**
**if** Tier 2 improves the Tier 1 minimum by more than 0.1 eV **or** min_tier≥ 3 **then**
    Run Tier 3 random refinement.
**end if**
Apply the ECG to the high/medium-confidence set and choose the final scan selection mode.
**if** vacuum mode with solvation requested **then**
    Attach the xTB+ALPB correction on the optimized probe, target, and complex geometries.
**end if**
Return the ranked result set, committed energy, confidence annotations, and run artifacts.

**Algorithm 2** Worker-side optimize and validate

Relax the candidate on the UMA OMol surface with LBFGS and up to three smart continuations.
Initialize confidence to high and the alert set to empty.
Run the topology guard on the initial and final bond graphs.
**if** topology is not preserved **then**
    **if** the system is neutral and fallback UMA heads are available **then**
        Re-optimize with the fallback heads and downgrade confidence by the majority verdict.
    **else**
        Downgrade confidence directly and mark the run as DFT-recommended.
    **end if**
    Return immediately; later guards are skipped in the broken-topology branch.
**end if**
Run the geometry guard and flag severe clashes or strained bonds.
Run the energy guard using $|E_{\text{bind}}|/N_{\text{small}} \leq 0.65$ eV/atom and $|E_{\text{bind}}| \leq 2.2$ eV.
Apply convergence downgrades only on topology-preserved runs.
Return the optimized geometry, predicted energy, confidence, and alert record.

## B.2 Detailed CREST + Composite-DFT Baseline Settings

The baseline is a cost-controlled CREST + composite-DFT high-fidelity pipeline inspired by the three-stage recipe of Bursch et al.:[10] semiempirical conformer sampling, tight semiempirical reoptimization, and a composite-DFT two-step treatment.

**Stage 1: CREST NCI sampling.** Initial probe–target dimers are built by translating the monomers along the $z$ axis so that their centers of mass are separated by 4.0 Å. CREST[8] then runs GFN2-xTB[9] metadynamics inside a soft logfermi wall that confines the two partners, giving conformers within a 6 kcal mol$^{-1}$ window above the putative global minimum without imposing a reference topology. Duplicates are removed internally by energy and RMSD. If CREST produces no conformers (e.g. convergence failure), the 4.0 Å starting geometry is forwarded to Stage 2 as a fallback.

**Stage 2: GFN2-xTB tight reoptimization.** The ten lowest-energy CREST conformers and each isolated monomer are reoptimized with tight GFN2-xTB settings. The lowest-energy xTB-optimized complex is retained as the Stage-3 input, and an xTB-level binding energy $\Delta E_{\mathrm{xTB}}$ is recorded as an intermediate.

**Stage 3: ORCA composite DFT.** The retained probe, target, and complex geometries are each treated with a two-step ORCA 6.1.1 workflow. The first step applies the composite r$^2$SCAN-3c rung with the modified def2-mTZVPP basis together with D4 and gCP corrections.[163–165] The second step evaluates the final single-point energy at the reference ωB97M-V/def2-TZVPD level in ORCA. For hybrid and meta-GGA evaluations we use the RIJCOSX approximation (RI-J together with seminumerical exchange) with the def2/J auxiliary basis. The parsed species energies are combined as a relaxed binding energy $\Delta E_{\mathrm{bind}}{}^{\mathrm{DFT}} = E_{PT} - E_P - E_T$.

**Intermediate DFT arms used in the 9-arm matrix.** The six DFT intermediate arms all start from the RAPIDS-committed complex geometry and are all executed in ORCA. The three single-point (SP) arms evaluate the RAPIDS-optimized complex together with the corresponding RAPIDS-optimized isolated monomers at one explicit single-point level: PBE-D3BJ/def2-TZVP SP, ωB97X-D3BJ/def2-TZVP SP, and ωB97M-V/def2-TZVP SP.[52–56] The three geometry-reoptimization plus single-point (GeoSP) arms first reoptimize the complex and the isolated probe and target monomers once at the same functional with def2-TZVP and TightOpt/TightSCF, then evaluate a same-functional def2-TZVPD single-point on that optimized geometry: PBE-D3BJ GeoSP, ωB97X-D3BJ GeoSP, and ωB97M-V GeoSP. In all six arms the fragment charges $q_P$, $q_T$, and $q_{PT}$ are carried explicitly, closed-shell singlets are assumed unless the dataset requires otherwise, and no counterpoise/BSSE correction is applied.

**Charge, spin, and resources.** Complex and monomer charges are set independently: the workflow carries $q_P$, $q_T$, and $q_{PT}$ separately, so a salt bridge can have $q_{PT} = 0$ but charged fragments. Unless otherwise noted systems are closed-shell singlets. No symmetry, thermal, or solvation corrections are applied; reported baseline energies are relaxed vacuum electronic binding energies rather than strict frozen-geometry interaction energies. RAPIDS inference and every explicit DFT evaluation in this paper are run locally, while CREST and xTB remain CPU-side preprocessing. All reported costs are wall-clock times for this implementation and are not normalized to energy draw or cloud-dollar cost.

## B.3 Detailed Pipeline Correspondence

Table S3 aligns the two pipelines stage by stage.

**Table S3** Stage-by-stage correspondence between RAPIDS and the CREST + DFT high-fidelity baseline pipeline. Both produce relaxed vacuum electronic binding energies, but RAPIDS does so with a UMA surrogate trained at the ωB97M-V/def2-TZVPD level while the baseline evaluates explicit DFT through ORCA 6.1.1. The gold reference used in the results is instead the published benchmark interaction energy at the benchmark geometry.

| Stage | RAPIDS | CREST + DFT |
|---|---|---|
| Evaluator used in search | UMA OMol (single MLIP pass) | GFN2-xTB (Stages 1–2), ORCA DFT (Stage 3) |
| Pose search | Tiered 9 + 6 + 9 scan with basin-hopping and trigger/min-tier upgrades | GFN2-xTB metadynamics, 6 kcal/mol window |
| Geometry relax | LBFGS on MLIP, smart continuation | xTB tight → assembled r$^2$SCAN-3c TightOpt |
| Reported electronic score | UMA OMol surrogate trained at ωB97M-V/def2-TZVPD | Explicit ωB97M-V/def2-TZVPD via ORCA |
| Solvation | Optional xTB+ALPB post-correction in vacuum mode | Vacuum only |
| Failure checks | Per-config guards + ECG (MCP path) | SCF + optimization convergence |

The two pipelines agree on two points and differ on four. Both target the same high-level functional rung, ωB97M-V/def2-TZVPD, and both restrict themselves to vacuum electronic energies unless solvation is explicitly requested. They differ on (i) the evaluator at inference (UMA OMol surrogate inference against explicit SCF plus geometry optimization), which drives wall-clock cost; (ii) the sampler (a structured, cached, upgrade-triggered tiered scan against metadynamics at a fixed energy window), which trades exploration for reproducibility; (iii) the committed geometry (a minimum on the UMA OMol surface against a minimum on the r$^2$SCAN-3c surface), so that an observed score gap on common systems combines two distinct contributions, the MLIP's fixed-geometry surrogate error against the explicit SP rung used in the matrix and the geometric deviation between UMA-optimized and r$^2$SCAN-3c-optimized complexes; and (iv) the failure-mode model, namely ML-specific guards with multi-head consensus against SCF and optimization convergence diagnostics. The published benchmark target against which errors are reported is neither pipeline: it is the dataset-provided interaction-energy reference at the benchmark geometry. The results section therefore reports benchmark-reference deviations of method-specific relaxed binding scores, and uses intermediate fidelities that apply the reference DFT (as a single-point or as a geometry-reoptimization plus single-point) on top of RAPIDS-committed geometries to separate the surface and geometry contributions. Section 3.4 additionally describes how the full 5,567 × 9 precomputed result matrix is treated as an offline multi-fidelity replay log for comparing allocation heuristics and replay policies.

## B.4 Telescoping decomposition: full per-system definitions

The compact decomposition stated in §4.3 expands as follows. For each system $s$, the four scores are

$$\hat{y}_{\mathrm{R}}(s) = \widehat{\Delta E}_{\mathrm{bind}}^{\mathrm{RAPIDS}}(s), \tag{5}$$

$$\hat{y}_{\mathrm{SP}}(s) = \widehat{\Delta E}_{\mathrm{bind}}^{\mathrm{RAPIDS_wB97M-V_SP}}(s), \tag{6}$$

$$\hat{y}_{\mathrm{Geo}}(s) = \widehat{\Delta E}_{\mathrm{bind}}^{\mathrm{RAPIDS_wB97M-V_GeoSP}}(s), \tag{7}$$

$$\hat{y}_{\mathrm{C}}(s) = \widehat{\Delta E}_{\mathrm{bind}}^{\mathrm{CREST_xTB_DFT}}(s). \tag{8}$$

The exact telescoping identity at the system level is

$$\hat{y}_{\mathrm{R}}(s) - \hat{y}_{\mathrm{C}}(s) = \delta_{\mathrm{surf}}(s) + \delta_{\mathrm{geom}}(s) + \delta_{\mathrm{samp}}(s),$$

with the three axis-aligned increments

$$\delta_{\mathrm{surf}}(s) = \hat{y}_{\mathrm{R}}(s) - \hat{y}_{\mathrm{SP}}(s), \tag{9}$$

$$\delta_{\mathrm{geom}}(s) = \hat{y}_{\mathrm{SP}}(s) - \hat{y}_{\mathrm{Geo}}(s), \tag{10}$$

$$\delta_{\mathrm{samp}}(s) = \hat{y}_{\mathrm{Geo}}(s) - \hat{y}_{\mathrm{C}}(s). \tag{11}$$

## B.5 Frozen-geometry method-only benchmark

To isolate method (energy-surface) bias from geometry effects, we evaluate a companion benchmark in which geometry is held fixed at the published reference structure for every system, so no method performs any geometry optimization. We take the gold-standard reference geometries that produced the benchmark CCSD(T)/CBS-style and CCSD(T)-F12 interaction energies (the same 5,567 neutral and charged systems used throughout, provided as a complex and two separated monomers per system) and compute a single interaction energy $E_{\mathrm{int}} = E(\mathrm{complex}) - E(\mathrm{probe}) - E(\mathrm{target})$ with each of ten methods. The ten methods are GFN2-xTB;[9] the three MLIP backbones UMA,[3] MACE-OMol,[2] and ORB-OMol,[60] each as a single-point energy on the reference geometry; and six explicit DFT single points formed by three functionals (PBE-D3BJ, ωB97X-D3BJ, ωB97M-V) crossed with two basis sets (def2-TZVP and def2-TZVPD). All DFT single points use ORCA 6.1.1 with TightSCF, the RIJCOSX approximation for the hybrid and range-separated functionals, and the def2/J auxiliary basis. Because geometry is identical across methods and identical to the geometry underlying the reference energy, each per-method error is a pure surface/method bias with no geometry-search or relaxation contribution. This is the complement to the multi-fidelity benchmark of §3.1, where geometry is varied per arm. Aggregate per-method statistics are in Table S13 and the result is discussed in §4.7.

## B.6 Detailed Offline-Replay Protocol

**Purpose.** Sections 3.2–3.3 fix two full pipelines at one operating point each. Here we replay the resulting 5,567 × 9 prediction/time matrix as an offline multi-fidelity decision log, so arbitrary allocation policies over the same nine arms are scored on the same benchmark stream without issuing new DFT jobs (forward reference §A.4).

**Setup.** A context $s_t$ is one of 5,567 probe–target systems, drawn in permutation $\pi_k$ (seed $k$) from the 18 benchmarks. The arm set $\mathcal{A}$ is the nine fidelities of §3.3: RAPIDS, three DFT_SP on RAPIDS geometry, three DFT_GeoSP, CREST_xTB, CREST_xTB_DFT. Each log cell holds a predicted relaxed binding score $\Delta\hat{E}_{\text{bind}}(s_t, a)$, wall-clock time $\tau(s_t, a)$, and a status flag; $\Delta E_{\text{int}}^{\star}(s_t)$ is the published benchmark interaction-energy reference (CCSD(T)/CBS or equivalent). Reward is capped negative absolute benchmark-reference deviation,

$$r(s_t, a) = -\min\left(|\widehat{\Delta E}_{\text{bind}}(s_t, a) - \Delta E_{\text{int}}^{\star}(s_t)|, E_{\max}\right),$$

with $E_{\max}$ = 50 kcal/mol, and cost is $c(s_t, a) = \tau(s_t, a)$ in seconds. Reference values are never exposed at arm-selection time, but some replayed policy families do consume them during retrospective updates as described below.

**Policy formalization.** At each step $t$, a replay policy $\pi$ maps the current context to an ordered arm sequence,

$$\mathcal{M}_t = \pi(s_t) \in \mathcal{A}^*,$$

where $\mathcal{A}^*$ denotes the set of all finite ordered sequences over the arm set $\mathcal{A}$. Writing

$$\mathcal{M}_t = (a_{t,1}, \dots, a_{t,m_t}),$$

with committed arm

$$a_t^{\dagger} = \text{last_nonNaN}(\mathcal{M}_t),$$

committed prediction

$$\hat{y}_t^{\pi} = \widehat{\Delta E}_{\text{bind}}(s_t, a_t^{\dagger}),$$

and paid cost

$$c_t^{\pi} = \sum_{j=1}^{m_t} \tau(s_t, a_{t,j}).$$

The replay reward is

$$r_t^{\pi} = -\min(|\hat{y}_t^{\pi} - \Delta E_{\text{int}}^{\star}(s_t)|, E_{\max}),$$
$$E_{\max} = 50 \text{ kcal/mol},$$

and the corresponding capped error objective is

$$\text{MAE}_{\text{cap}}(\pi) = \frac{1}{T}\sum_{t=1}^{T} \min(|\hat{y}_t^{\pi} - \Delta E_{\text{int}}^{\star}(s_t)|, E_{\max}).$$

We therefore interpret replay as a budgeted multi-fidelity allocation problem,

$$\min_{\pi} \quad \text{MAE}_{\text{cap}}(\pi)$$
$$\text{s.t.} \quad \mathbb{E}\left[\sum_{t=1}^{T} c_t^{\pi}\right] \le B,$$

while reporting the empirical Pareto frontier across policies rather than committing to a single budget $B$. This optimization view is the underlying decision problem that replay approximates; the paper's reported outputs remain empirical cost–MAE/ρ comparisons among concrete policies.

**Policies replayed.** Four families are implemented. *Static*: nine ALWAYS-*X* policies and an ORACLE lower bound that, for each $s_t$, chooses $\arg\min_a |\Delta\hat{E}_{\text{bind}} - \Delta E_{\text{int}}^{\star}|$ while paying only that arm's cost. *Context-free selectors*: RANDOM; UCB1[23] on (RAPIDS-magnitude bin, arm) pairs over five quantile bins on $|\Delta\hat{E}_{\text{bind}}^{\text{RAPIDS}}|$, with and without a cost-normalized score (UCB statistic divided by each method's median cost); Thompson sampling[57] on the same bins; an explore-then-exploit COST-AWARE policy that pulls all nine arms on the first 10% of systems and then greedily selects arms according to predicted benefit per unit cost; and MFMI-GREEDY.[166] *Escalation / stacking*: DISAGREEMENT, PROGRESSIVE LADDER (RAPIDS → PBE-D3BJ_SP → ωB97M-V_SP → ωB97M-V_GeoSP → CREST_xTB_DFT), online STACKING, STACKING ML, CHEAP ENSEMBLE, BIAS CORRECTION. *Context-aware / meta*: LEARNED SELECTOR (trained on RAPIDS-value features only), CHEMUCB on eight chem-feature bins, CHEM LEARNED SELECTOR (the chem-feature twin of LEARNED SELECTOR), ALORS/CHEMALORS,[58] LINUCB[24] on chemical plus cheap-prediction features, leave-one-benchmark-out BUCKET PRIOR (adaptive and BUCKET LEARNED variants), and a gradient-boosted-machine[59] family (GBM STACKING, META-GBM, META-GBM ADAPTIVE, GBM META-WARM, HYBRID SELECT+STACK). Priors and meta-models are trained leave-one-benchmark-out. The reported study compares 40 replay policies spanning these families.

**Per-system selection.** Single-shot and escalating policies share one protocol: on each $s_t$ a policy returns an ordered sequence $\mathcal{M}(s_t) \in \mathcal{A}^*$; the committed prediction is the last non-NaN value in $\mathcal{M}(s_t)$ and the paid cost is $\sum_{j=1}^{m_t} \tau(s_t, a_{t,j})$ for $\mathcal{M}(s_t) = (a_{t,1}, \ldots, a_{t,m_t})$. Static and context-free selectors use $m_t = 1$ after any warm-up phase. Escalation policies extend $\mathcal{M}$ only when an explicit disagreement trigger fires; PROGRESSIVE LADDER, for example, applies three sequential disagreement thresholds on consecutive-prediction gaps at successive ladder stages and sweeps four threshold triples. Stacking policies instead run a fixed cheap-arm prefix on every system: STACKING always executes a short list of cheap arms and only then escalates on disagreement, while the STACKING ML / GBM STACKING / META-GBM meta-learners are non-triggered: they always call the same cheap-feature prefix and fuse its outputs through a learned combiner, paying the full prefix cost on every $s_t$. For the full-information selector families, posterior or weight updates use the all-arm error vector available in the precomputed matrix; this is retrospective full-information replay, not selected-arm bandit feedback.

**Feedback regimes and deployability.** The replay suite mixes four feedback assumptions. Rule-based policies (ALWAYS-*X*, DISAGREEMENT, PROGRESSIVE LADDER, CHEAP ENSEMBLE) never consume benchmark references during replay-time updates and are directly deployable as written. META-GBM is a frozen leave-one-benchmark-out meta-model and is also directly deployable. Online regression / bias-correction policies (STACKING, STACKING ML, GBM STACKING, BIAS CORRECTION, META-GBM ADAPTIVE) update after each processed system using that system's benchmark reference and therefore require a delayed ground-truth feedback loop to deploy. UCB/Thompson/ChemUCB/BucketPrior/Learned-Selector-style policies receive full-information retrospective updates from the errors of all nine arms on the current system because the factorial log contains every arm; they should be read as retrospective selectors rather than as causal online bandits. We keep all four classes because together they map the achievable cost–accuracy frontier on the same matrix, but only the rule-based and frozen-model subsets are directly deployable without an external feedback channel.

**Cost model.** Each $\tau(s_t, a)$ is the standalone wall-clock time recorded for that arm's independently executed job in the benchmark log. Summing these costs along $\mathcal{M}(s_t)$ therefore approximates an uncached escalation workflow and does not assume artifact reuse between, for example, SP and GeoSP arms. The replay comparisons are internally consistent across policies, but the absolute cumulative cost of a cached production workflow could be lower than the standalone sums reported here. For RAPIDS-Select specifically, the reported per-system cost is the committed arm's logged chemistry wall-clock time; constructing the ~156 cheap features, including the three-backbone MLIP scan summaries used for the cross-backbone disagreement feature, adds two further MLIP scans (58 and 39 s per system on average) that are small relative to the DFT and CREST escalation arms and are not included in the committed-arm cost. Counting the three MLIP scans for RAPIDS-Select and the RAPIDS scan that seeds every DFT arm, the in-distribution saving relative to Always-PBE-D3BJ GeoSP is 33%.

**Evaluation metrics.** Per seed and benchmark the stream of committed predictions is sampled at $N_{\text{budget}} = 30$ checkpoints uniformly spaced in cumulative cost; at each we record the capped MAE and the Spearman rank correlation ρ between committed predictions and references. Per-benchmark values are seed medians; "neutral" (16 benchmarks) and "charged" (2 benchmarks) aggregates are medians across benchmarks of those seed medians. ORACLE is reported as a per-stream lower bound, not as an achievable deployment policy; no explicit cumulative-regret curve is reported.

**Reproducibility.** Each policy is replayed under $N_{\text{seeds}} = 10$ stream permutations (numpy.random.default_rng(seed), *seed* ∈ {0, …, 9}). Uncertainty bands are non-parametric $[p_{10}, p_{90}]$ intervals over the per-seed benchmark-aggregated curves. Priors and Meta-GBMs are refit leave-one-benchmark-out per target.

**Artifacts.** Per-seed curves, seed-aggregated policy summaries, and the figures underlying §4 are provided in the Zenodo deposit and the RAPIDS GitHub repository (see Data and code availability).

**Reference-relaxation drift and residual-force probes.** The two diagnostics behind Fig. 4 are defined as follows. *Reference-relaxation drift* seeds each backbone (UMA, MACE-OMol, ORB-OMol) at the benchmark reference geometry and relaxes the complex on the MLIP surface (ASE LBFGS, $f_{\max} = 0.05$ eV/Å, 30 Å vacuum); the drift is the root-mean-square deviation (RMSD) between the relaxed and reference complex coordinates after removing rigid-body translation and rotation by Kabsch superposition, with atoms matched by index. *Residual force* is the single-point DFT gradient (ωB97M-V/def2-TZVP, ORCA 6.1.1) evaluated at the frozen RAPIDS-committed geometry; we report the per-atom root-mean-square force magnitude (Hartree/Bohr converted to eV/Å). To test whether this residual force is a genuine geometric signal rather than noise, we compute, for each system, the cosine between the residual-force vector at the committed geometry and the displacement from that geometry to the corresponding ωB97M-V GeoSP-optimized geometry, again after Kabsch rigid-body removal; a positive cosine means the residual force points toward the DFT minimum, which holds for ~97% of systems (98% neutral, 96% charged). These are diagnostic interventions on the committed geometry rather than a formal causal identification.

# C Supplementary results figures

## C.1 Neutral offline-replay trajectories

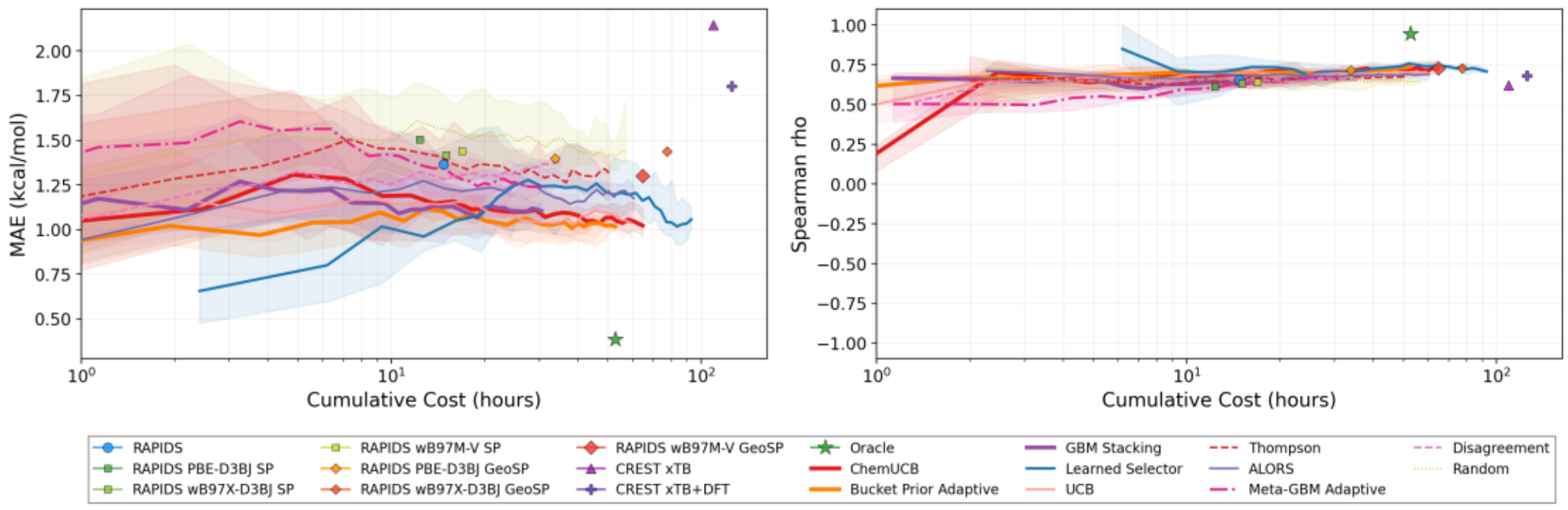

**Fig. S1** Neutral-only offline-replay trajectories for the 16 neutral benchmarks (the deployable one-shot selector is in §4.8). The left subplot shows MAE versus cumulative cost, in hours on a log scale; the right subplot shows Spearman ρ versus cumulative cost. Dashed lines are fixed single-fidelity baselines; solid lines are representative allocation heuristics and replay policies; shaded bands show the non-parametric $[p_{10}, p_{90}]$ range across ten stream permutations. The best-performing replayed selector, ChemUCB-8, reaches 65.1 h / 1.01 kcal/mol, which is 52% of the Always-CREST xTB+DFT cumulative cost (125.3 h) while lowering MAE from 1.80 to 1.01 kcal/mol. These policies use retrospective all-arm feedback, so the curves are an offline upper bound rather than a one-shot deployable procedure.

## C.2 Full 18-benchmark fidelity overview

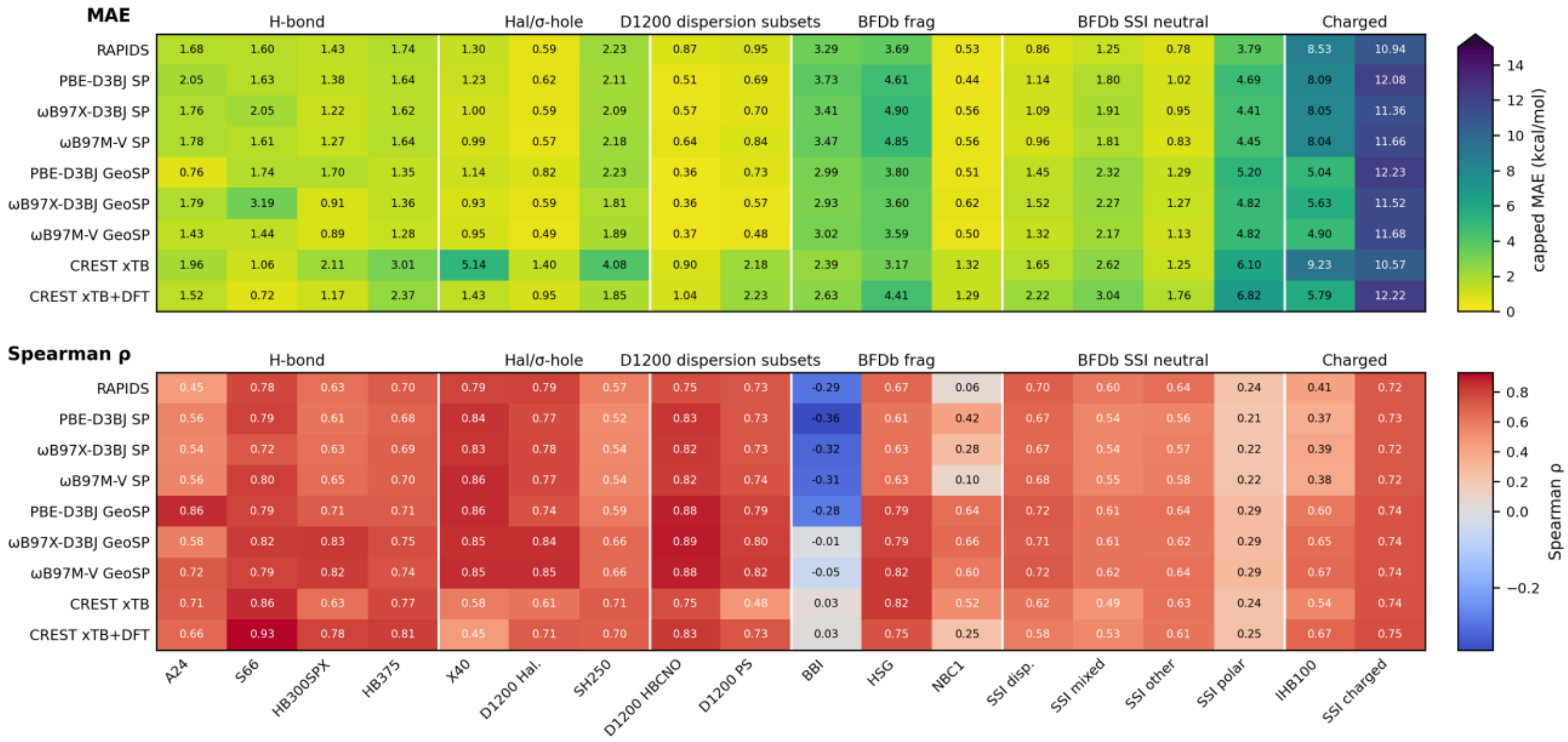

**Fig. S2** Full benchmark-relative summary of the nine fidelity methods across all 18 benchmarks. Top panel: capped MAE (kcal/mol). Bottom panel: Spearman ρ. The main text concentrates on the neutral story and defers these full per-benchmark heatmaps, including the two charged benchmarks, to this ESI.

## C.3 Charged-system diagnostics

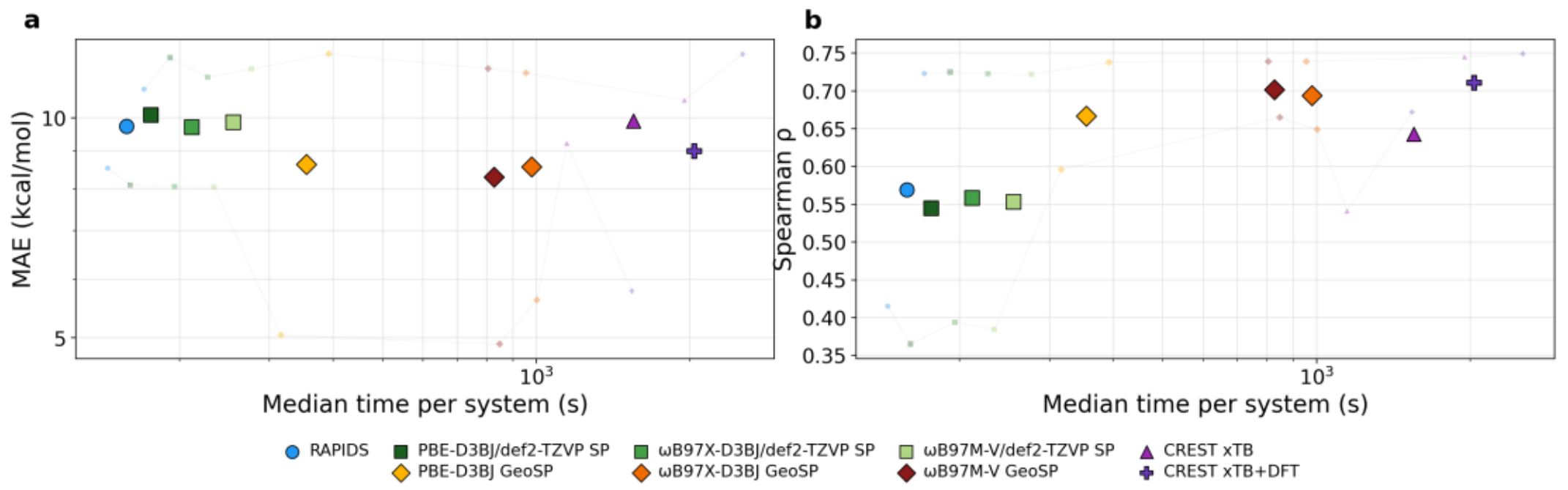

**Fig. S3** Charged static-fidelity overview across the two charged benchmarks (IHB100 and BFDb_SSI_charged), the charged counterpart of the neutral overview in Fig. 3. Left panel: MAE versus median wall-clock time per system, in seconds on a log scale. Right panel: Spearman ρ versus median wall-clock time per system. Faint markers and light gray connectors show benchmark-level values for the nine static fidelities; the larger colored markers show the median across the charged benchmarks for each fidelity. In contrast to the neutral ladder, RAPIDS and the fixed-geometry DFT single points sit high on the error axis (MAE 8.0–12.1 kcal/mol); DFT geometry reoptimization and CREST xTB+DFT recover accuracy on IHB100, visualizing the trust boundary of §4.5.

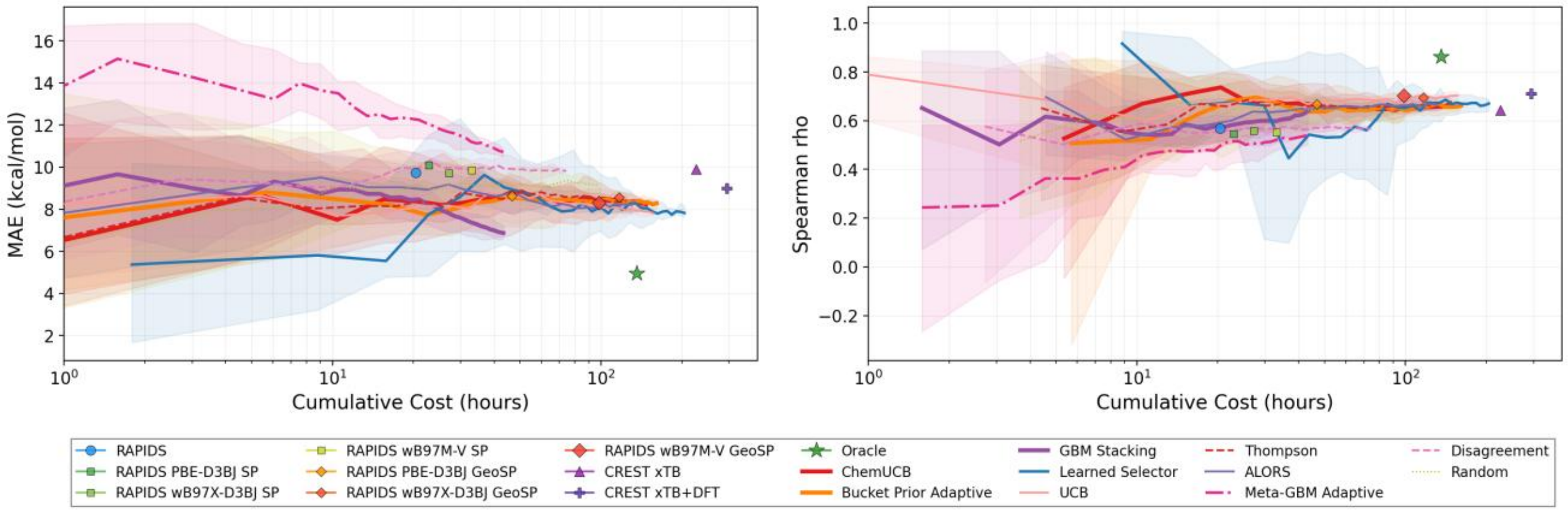

**Fig. S4** Charged-only offline-replay trajectories deferred from the main text. Left panel: capped MAE versus cumulative cost, in hours on a log scale. Right panel: Spearman ρ versus cumulative cost. This is the charged counterpart of the neutral replay figure (Fig. S1) and shows the charged cost–accuracy ladder: RAPIDS direct output (9.7 kcal/mol, ρ = 0.57) sits with the fixed-geometry DFT single points (9.7–10.1 kcal/mol) at the low-cost end, DFT geometry reoptimization gives the most accurate static arms (8.3–8.6 kcal/mol, ρ = 0.67–0.70), and the stacking policy over the two cheapest arms (GBM-Stack-2cheap) reaches 6.9 kcal/mol at 43 h, below every static arm.

## C.4 Error-distribution diagnostics

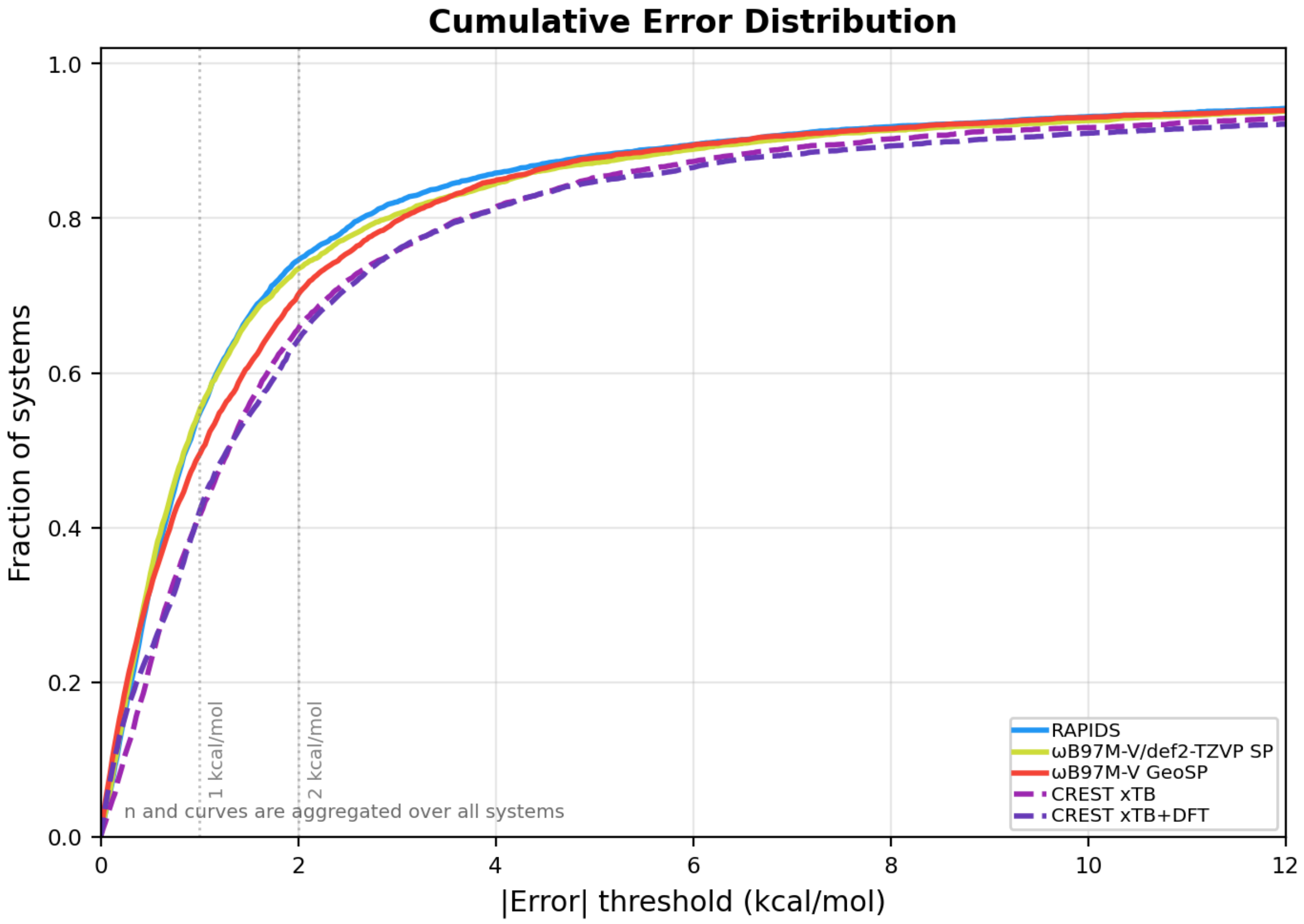

**Fig. S5** Distribution-level diagnostics deferred from the main text. Cumulative distribution of absolute errors for RAPIDS, ωB97M-V/def2-TZVP SP, ωB97M-V GeoSP, CREST xTB, and CREST xTB+DFT, aggregated over all available systems. RAPIDS tracks or leads the DFT and CREST curves across the whole range: 55% of its errors are within 1 kcal/mol and 93% within 10 kcal/mol.

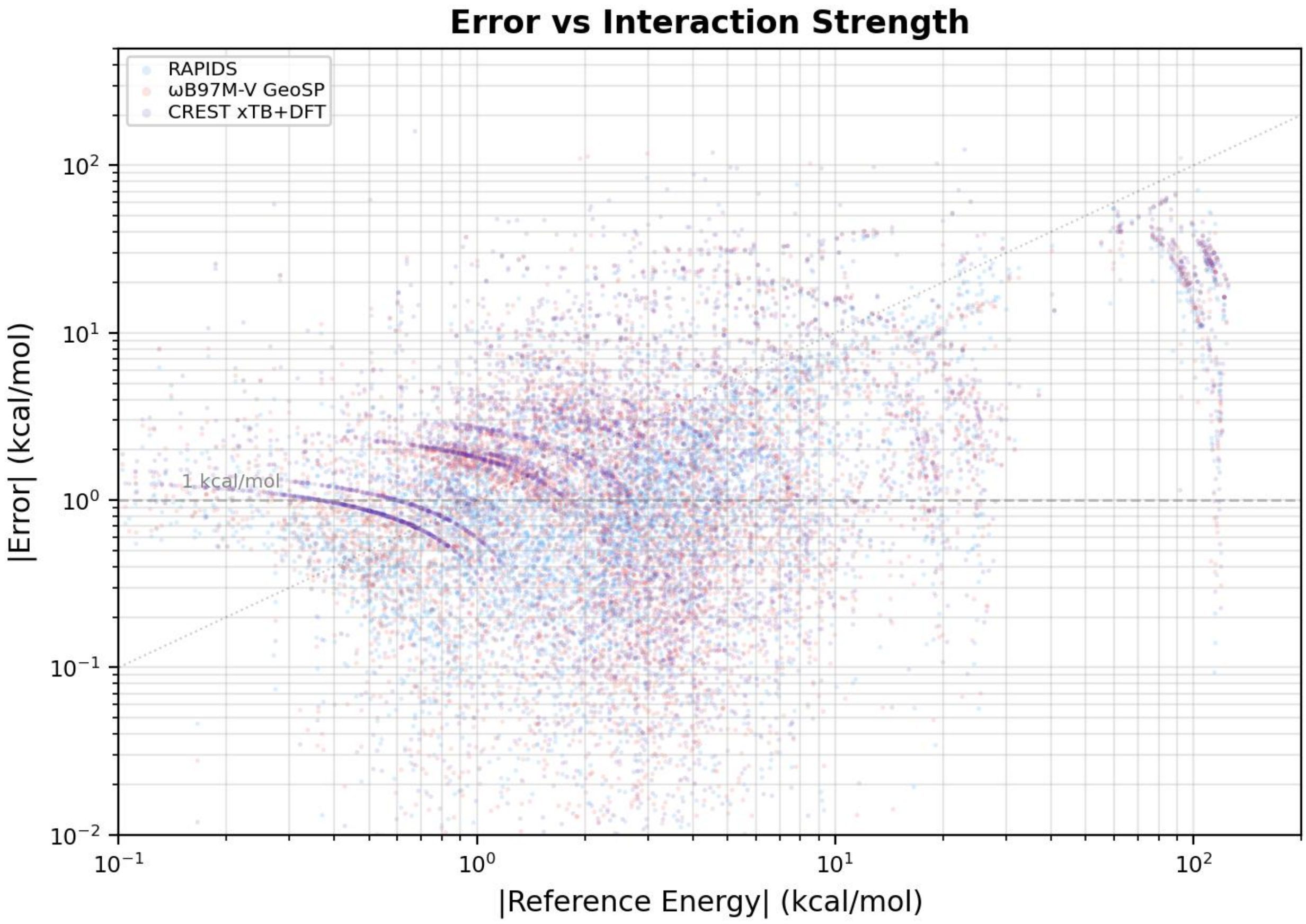

**Fig. S6** Absolute error versus absolute reference interaction energy for RAPIDS, ωB97M-V GeoSP, and CREST xTB+DFT. This panel complements Fig. S5: the large errors of all three methods concentrate on strongly bound and charged systems.

## C.5 Static aggregate summaries

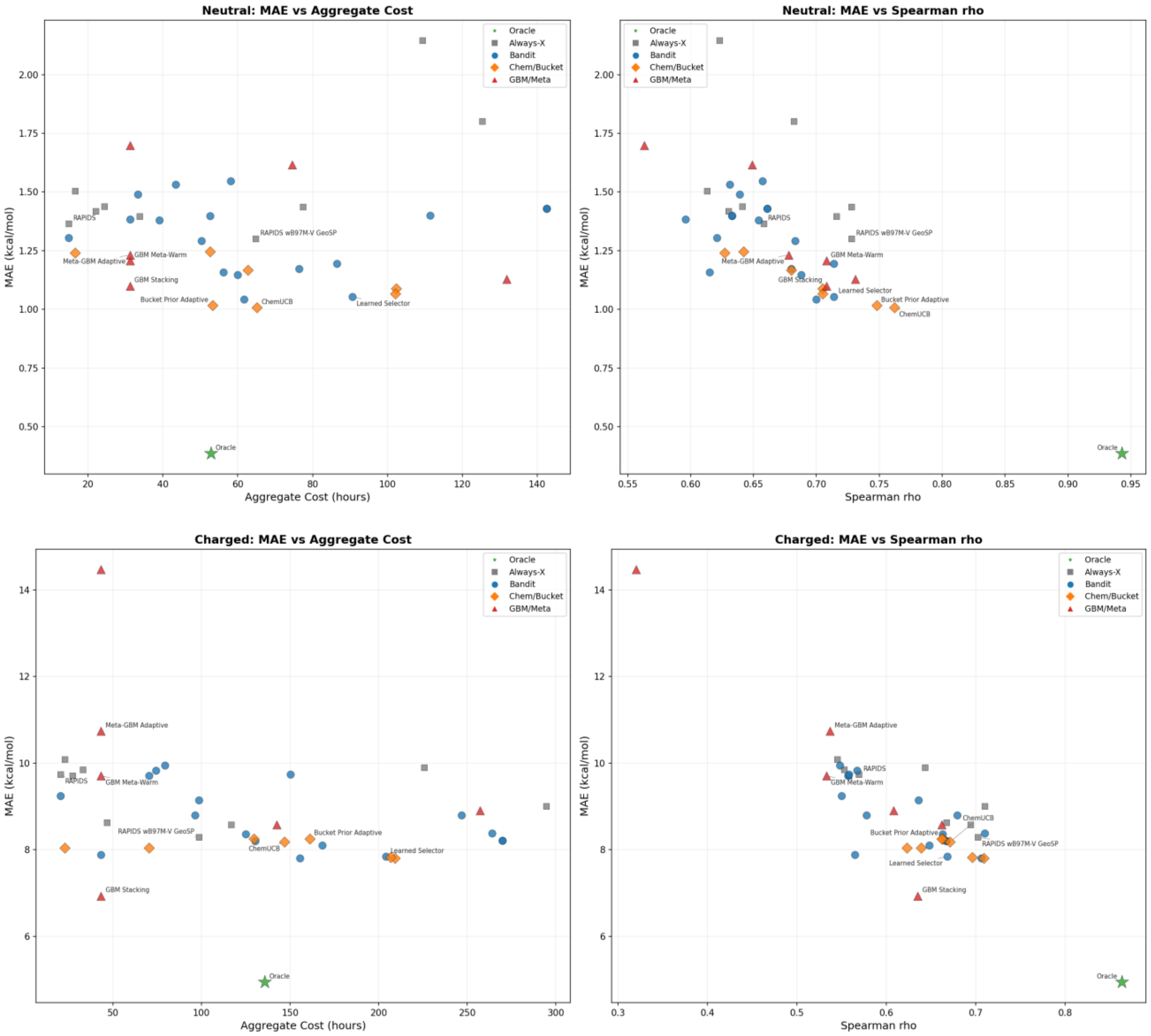

**Fig. S7** Aggregate cost–accuracy summaries deferred from the main text. Top: neutral aggregate. Bottom: charged aggregate. In each panel, the left subplot shows capped MAE versus aggregate cost and the right subplot shows Spearman ρ versus MAE. Markers show the full 40-policy inventory colored by policy family, with the fixed single-fidelity Always-$X$ arms as gray squares and ORACLE as the per-system lower bound. These summaries complement the main-text neutral static-fidelity overview by retaining the aggregate view for both charge classes.

## C.6 Policy-family rankings

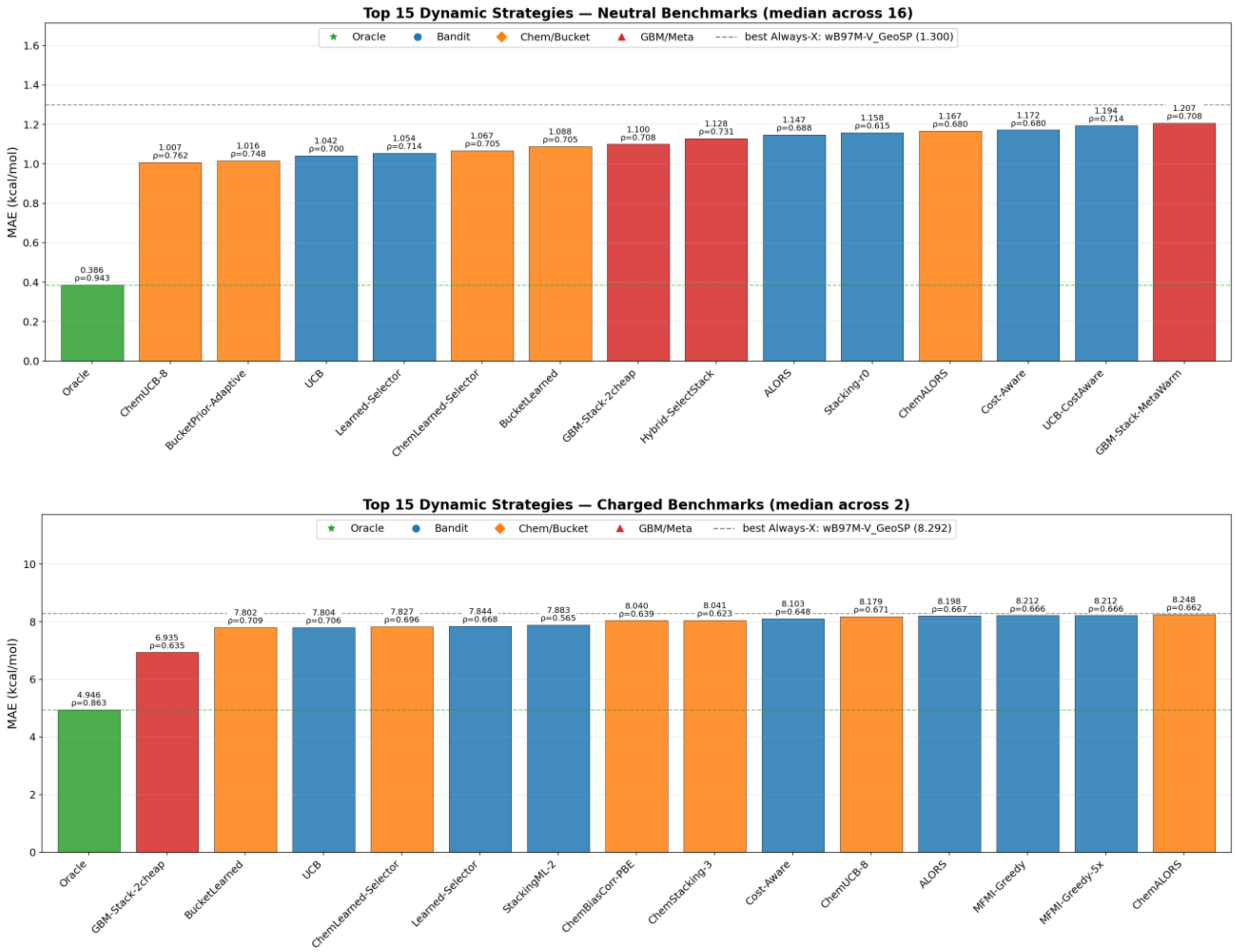

**Fig. S8** Top-15 replay policies by final capped MAE, neutral aggregate (top) and charged aggregate (bottom). The best static ALWAYS-*X* policy (dashed line) and ORACLE are shown for reference; values are those of Table S7. The policy families that rank near the top differ between the two aggregates.

## C.7 RAPIDS-Select: baselines, ablation, and cost–risk frontier

This section details the one-shot cost-aware selector of §4.8. For each system $x$, RAPIDS-SELECT commits the arm $\arg\min_a [\,\mathrm{cost}(a) + \lambda\,\widehat{\mathrm{err}}_a(x)\,]$, where $\widehat{\mathrm{err}}_a$ is a per-arm gradient-boosted estimate of arm $a$'s absolute error from ∼156 cheap features and λ = 800; the pool is the five-arm subset RAPIDS, PBE-D3BJ SP, PBE-D3BJ GeoSP, CREST xTB, CREST xTB+DFT, trained with charged systems included and no hard charge short-circuit. All external charged results use the 981-system DES370K benchmark with CCSD(T)/CBS-quality references.

Against literature one-shot selectors re-evaluated on the identical caches, benchmarks, and per-system cost accounting (Fig. S9, gray crosses), RAPIDS-Select matches or beats each in distribution at a lower cost: the cost-blind SATzilla-style empirical-hardness rule[30,31] reaches the same accuracy at 2.6× the cost (removing the cost term costs +611 s [+383, +852] at statistically equal error, paired bootstrap over 18 benchmarks); the FrugalML-style budgeted cascade[61] sits on the same frontier at higher cost and needs a tuned risk threshold, whereas RAPIDS-Select is a single closed-form rule; a one-shot learning-to-defer rule[62,63] and ALORS with a cost term[58] need 23% more and 2.1× the cost for comparable error; and a multiclass Rice-style classifier[30] that predicts the oracle-best arm directly needs 2.8× the cost for statistically indistinguishable in-distribution error and, on the external charged set, reaches lower error only at 2.7× the cost, because it discards the error magnitudes the cost rule needs. Under a stricter leave-one-chemistry-family-out split the cost advantage holds (−31% cost versus Always-PBE-D3BJ GeoSP at lower MAE), with the held-out charged family the sole hard case (MAE 11.1 kcal/mol, 38% catastrophic), consistent with the trust boundary (Table S4). An ablation (Fig. S10) shows the routing signal is redundant across feature groups, while the single dominant factor is charged-aware training data.

**Table S4** Leave-one-group-out evaluation of RAPIDS-Select. Each chemistry group is held out in turn (all of its benchmarks); the per-arm regressors are retrained on all remaining systems and the held-out group is routed at λ = 800. The last two rows hold out each charged benchmark on its own. Groups: small-molecule mixed (A24, S66), hydrogen bond (HB300SPX, HB375), halogen (X40, D1200_Halogens), σ-hole (SH250), dispersion (D1200_HBCNO, D1200_PS), protein fragments (BFDb BBI, HSG, NBC1, SSI-mixed, SSI-other), sidechain dispersion (BFDb SSI-dispersion), sidechain polar (BFDb SSI-polar), and charged (IHB100, BFDb SSI-charged). Cost is mean per-system wall-clock seconds, MAE is in kcal/mol, and Cat. is the fraction with |error| > 10 kcal/mol. The static arms need no training and are shown on the same systems for reference.

| | | RAPIDS-Select | | | Always-PBE-D3BJ GeoSP | | | Always-RAPIDS | |
|---|---|---|---|---|---|---|---|---|---|
| Held-out group | *N* | Cost | MAE | Cat. (%) | Cost | MAE | Cat. (%) | MAE | Cat. (%) |
| Small-molecule mixed | 88 | 211 | 1.69 | 3.4 | 351 | 1.49 | 3.4 | 1.62 | 2.3 |
| Hydrogen bond | 675 | 261 | 1.52 | 0.9 | 382 | 1.51 | 1.0 | 1.60 | 0.9 |
| Halogen | 350 | 274 | 0.65 | 0.3 | 840 | 0.86 | 1.1 | 0.67 | 0.0 |
| σ-hole | 250 | 350 | 1.95 | 1.6 | 415 | 2.23 | 1.2 | 2.23 | 1.2 |
| Dispersion | 601 | 364 | 0.73 | 0.2 | 807 | 0.54 | 0.3 | 0.91 | 0.2 |
| Protein fragments | 1136 | 351 | 1.42 | 0.7 | 807 | 2.17 | 1.8 | 1.37 | 0.7 |
| Sidechain dispersion | 1325 | 242 | 0.87 | 0.2 | 542 | 1.45 | 0.4 | 0.86 | 0.2 |
| Sidechain polar | 266 | 227 | 4.01 | 6.0 | 540 | 5.20 | 12.4 | 3.79 | 5.3 |
| Charged | 876 | 1232 | 11.06 | 37.7 | 384 | 11.41 | 38.9 | 10.67 | 39.5 |
| Charged, IHB100 only | 100 | 377 | 6.81 | 24.0 | 316 | 5.04 | 17.0 | 8.53 | 36.0 |
| Charged, SSI-charged only | 776 | 1079 | 12.04 | 40.9 | 392 | 12.23 | 41.8 | 10.94 | 39.9 |

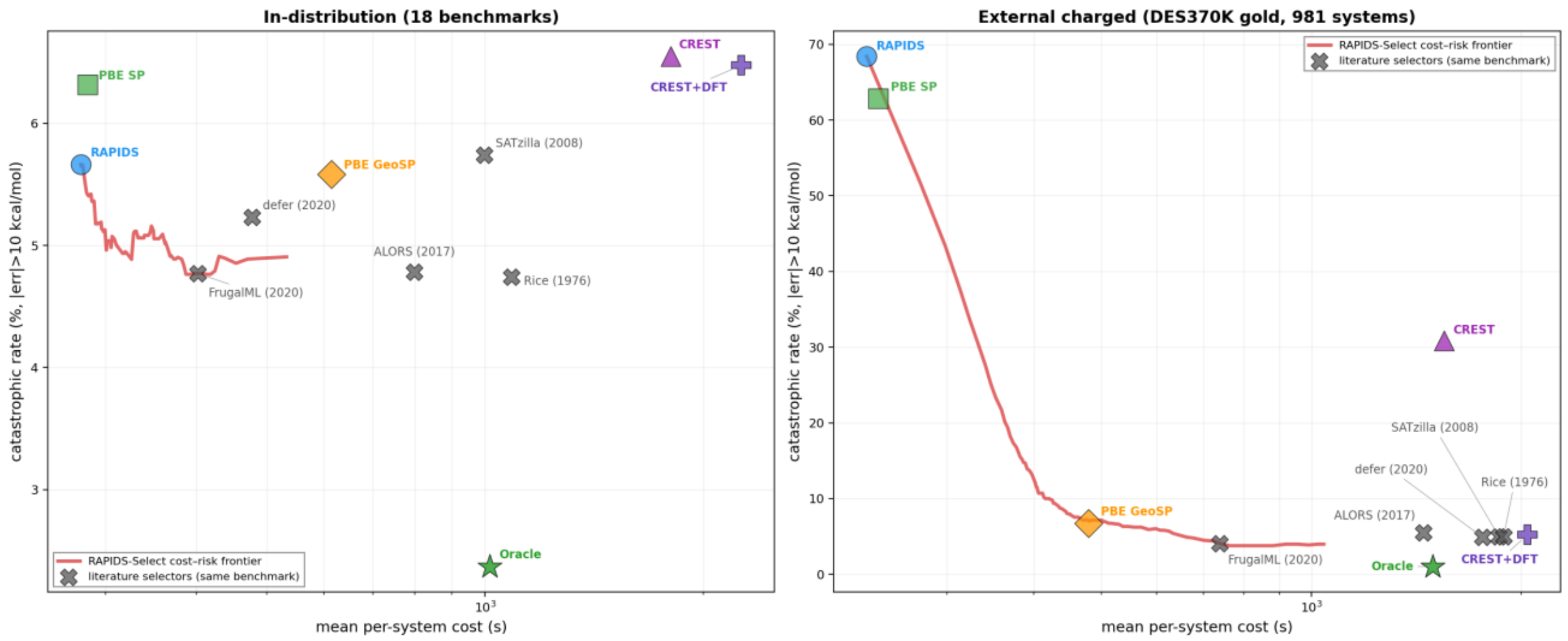

**Fig. S9** RAPIDS-Select cost–risk frontier. Catastrophic rate (fraction with |error| > 10 kcal/mol) versus mean per-system cost, in-distribution (left) and charged on the 981-system DES370K benchmark (right). The red curve is the RAPIDS-Select operating frontier traced by the threshold on its predicted catastrophe probability: a practitioner sets a catastrophic-risk target and reads off the minimum-cost operating point. Static single-fidelity arms (colored), literature selectors (gray crosses), and ORACLE (per-system best) are shown for reference; every marker in the left panel is a macro average over the 18 benchmarks. The frontier starts at Always-RAPIDS and lies below every other static arm in distribution; on charged it passes through the PBE-D3BJ GeoSP arm and continues to lower catastrophic rates. The $\lambda = 800$ operating point of Table 1 is set by the cost rule and is not on this threshold curve.

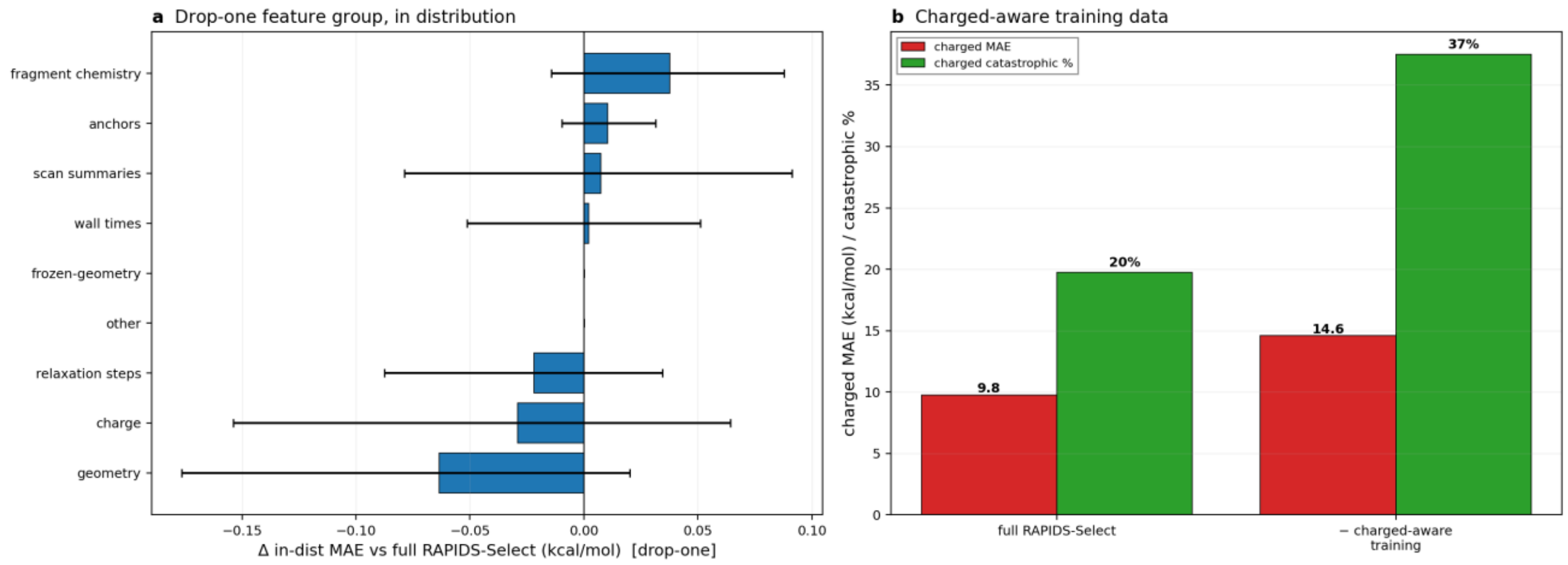

**Fig. S10** RAPIDS-Select ablation. **Left:** drop-one feature-group importance for in-distribution routing; no single feature group is critical (every paired-bootstrap 95% CI includes zero; the largest effect is 0.06 kcal/mol), indicating a redundant routing signal. **Right:** the dominant factor is charged-aware training data; removing charged systems from the training set sharply increases charged MAE and catastrophic rate (on the 981 DES370K systems, 3.01 → 8.07 kcal/mol and 5.7% → 25.4%; the panel plots the full 1,155-system charged training pool), whereas dropping any single feature group does not.

## C.8 Representative benchmark panels

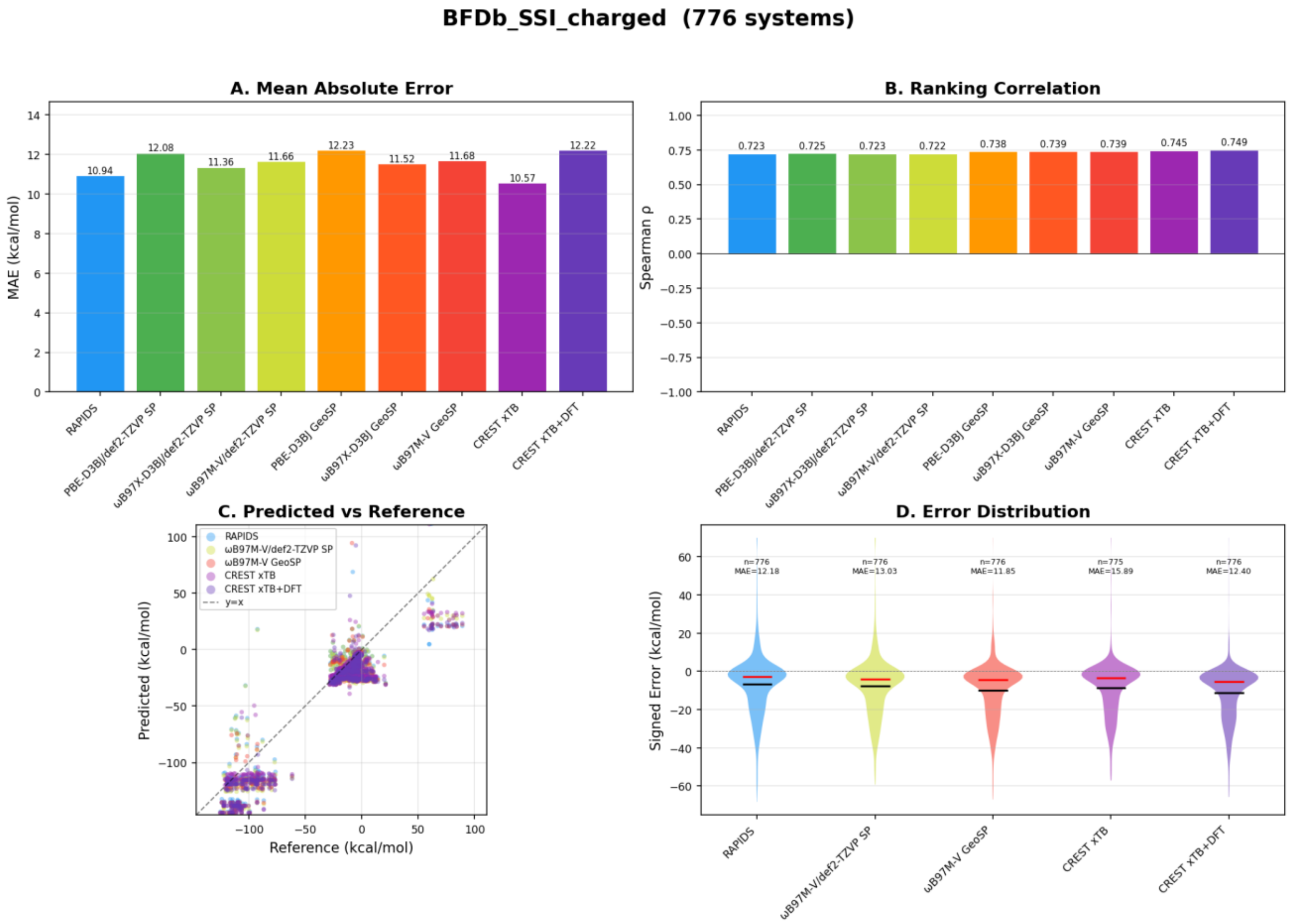

**Fig. S11** Representative charged benchmark: BFDb SSI charged. The four subpanels report method-wise MAE, method-wise Spearman ρ, predicted-versus-reference scatter for representative methods, and signed-error distributions. Every relaxing method lands at a similar error here (10.6–12.2 kcal/mol): the reference is a fixed-geometry interaction energy of crystal-derived sidechain pairs, and vacuum relaxation moves all methods away from it.

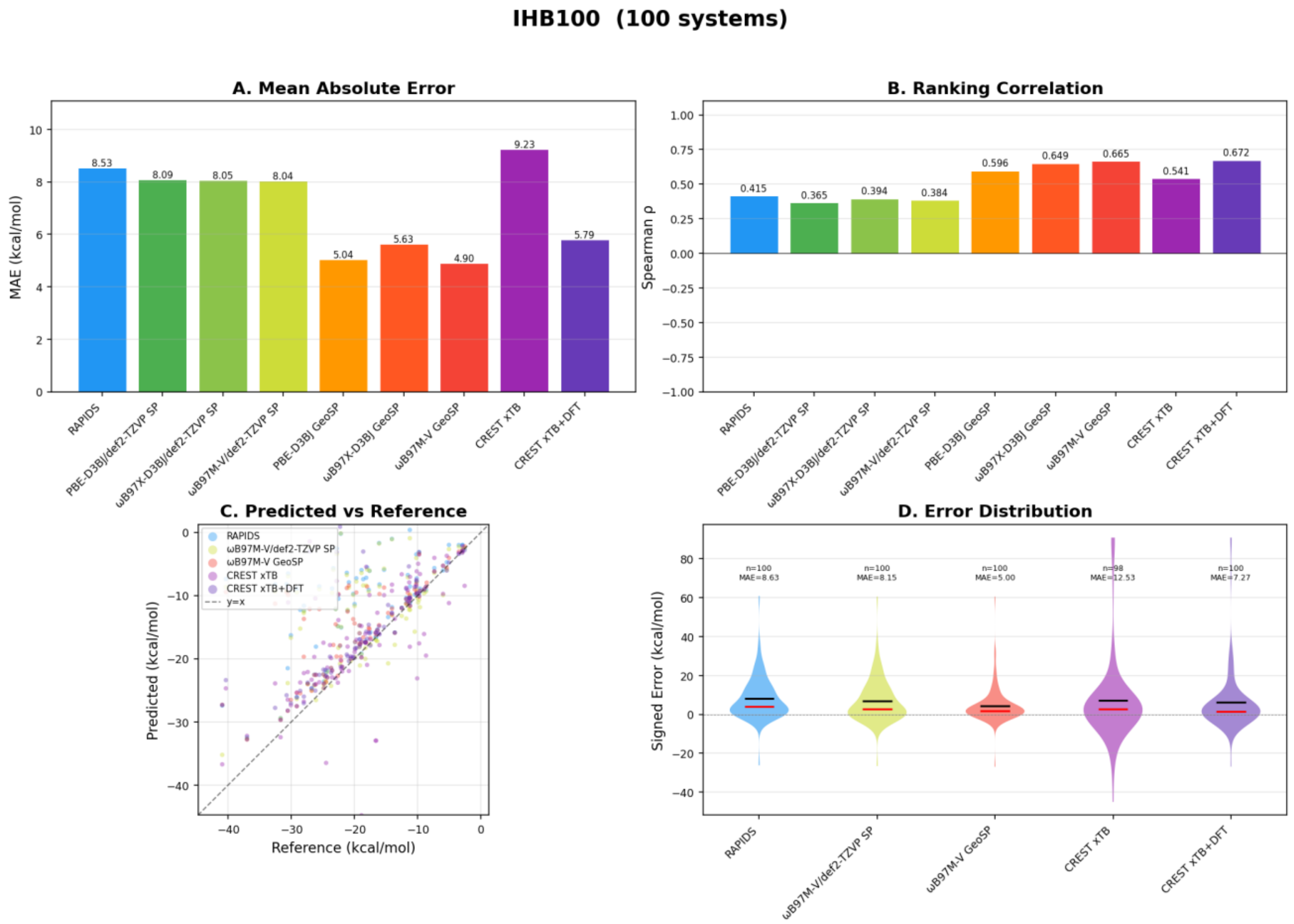

**Fig. S12** Representative charged benchmark: IHB100. The four subpanels report method-wise MAE, method-wise Spearman ρ, predicted-versus-reference scatter for representative methods, and signed-error distributions. RAPIDS and the DFT single points on its geometry sit near 8 kcal/mol, and the error drops to 4.9–5.6 kcal/mol once the geometry is reoptimized at the DFT level.

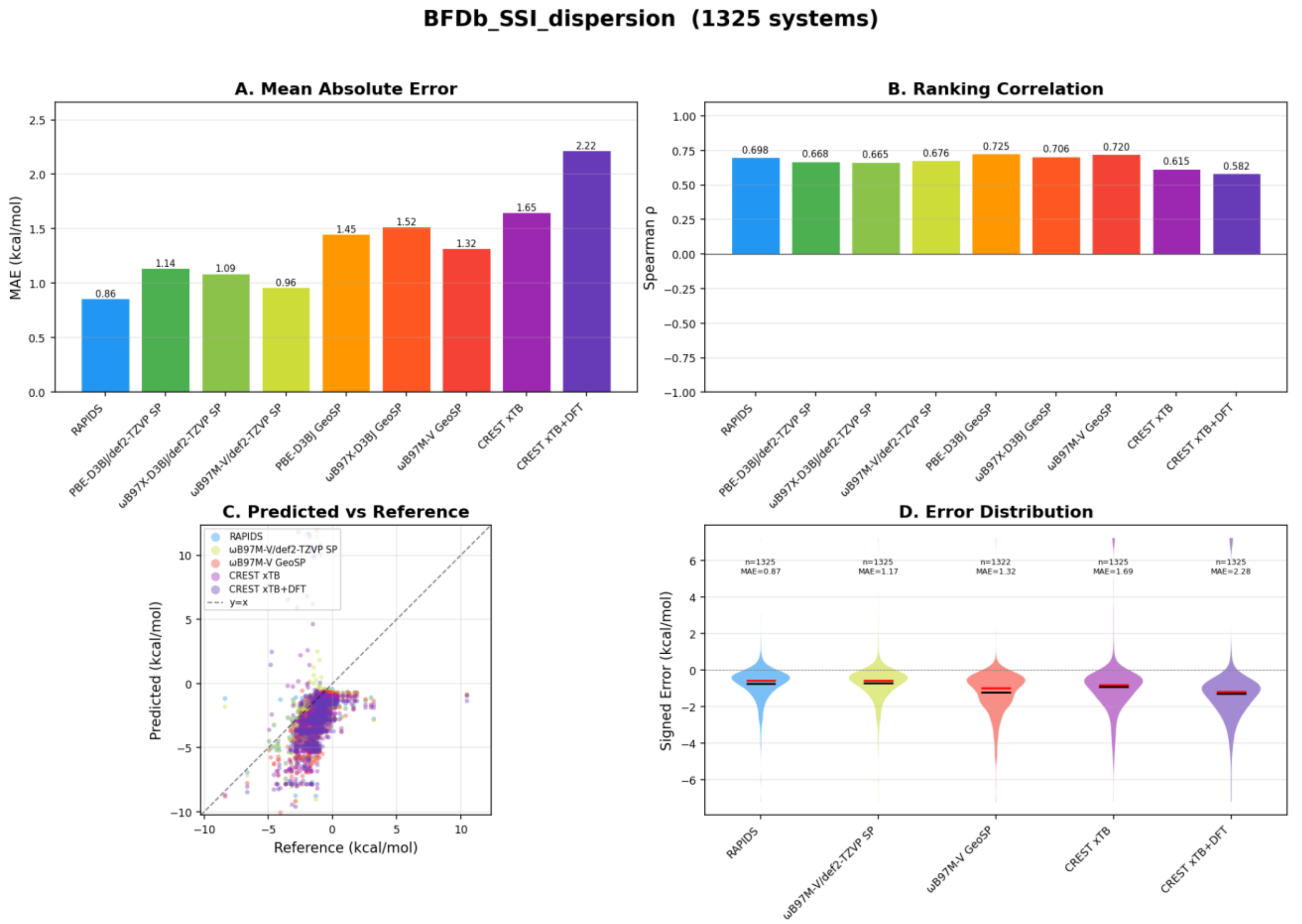

**Fig. S13** Representative neutral benchmark: BFDb SSI dispersion. The four subpanels report method-wise MAE, method-wise Spearman ρ, predicted-versus-reference scatter for representative methods, and signed-error distributions. This is the benchmark-relative neutral setting in which RAPIDS is strongest in the main text.

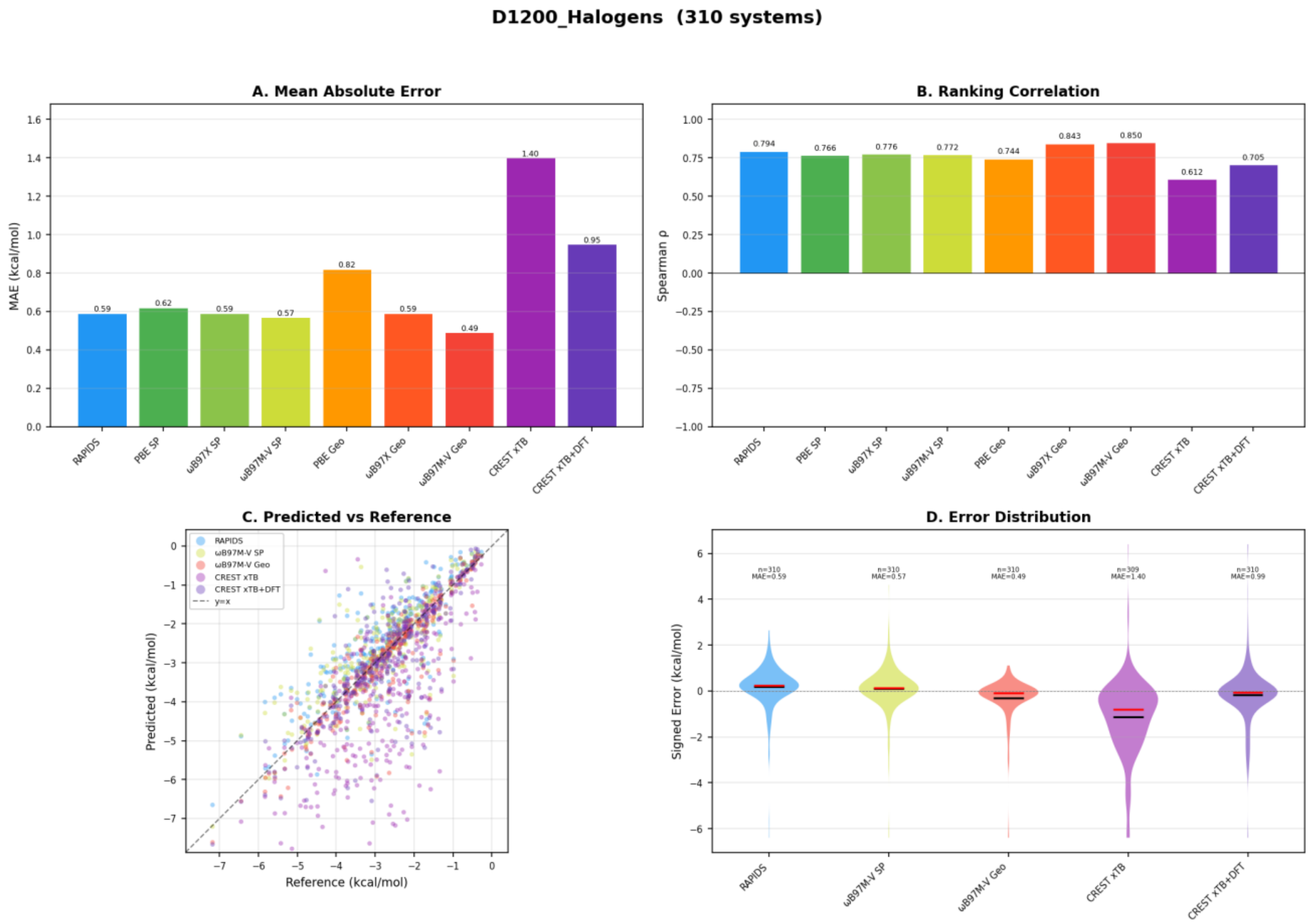

**Fig. S14** Representative neutral benchmark: D1200 halogens. The four subpanels report method-wise MAE, method-wise Spearman ρ, predicted-versus-reference scatter for representative methods, and signed-error distributions. This panel shows a complementary neutral case in which explicit DFT and CREST variants outperform RAPIDS.

## C.9 Cross-backbone diagnostic

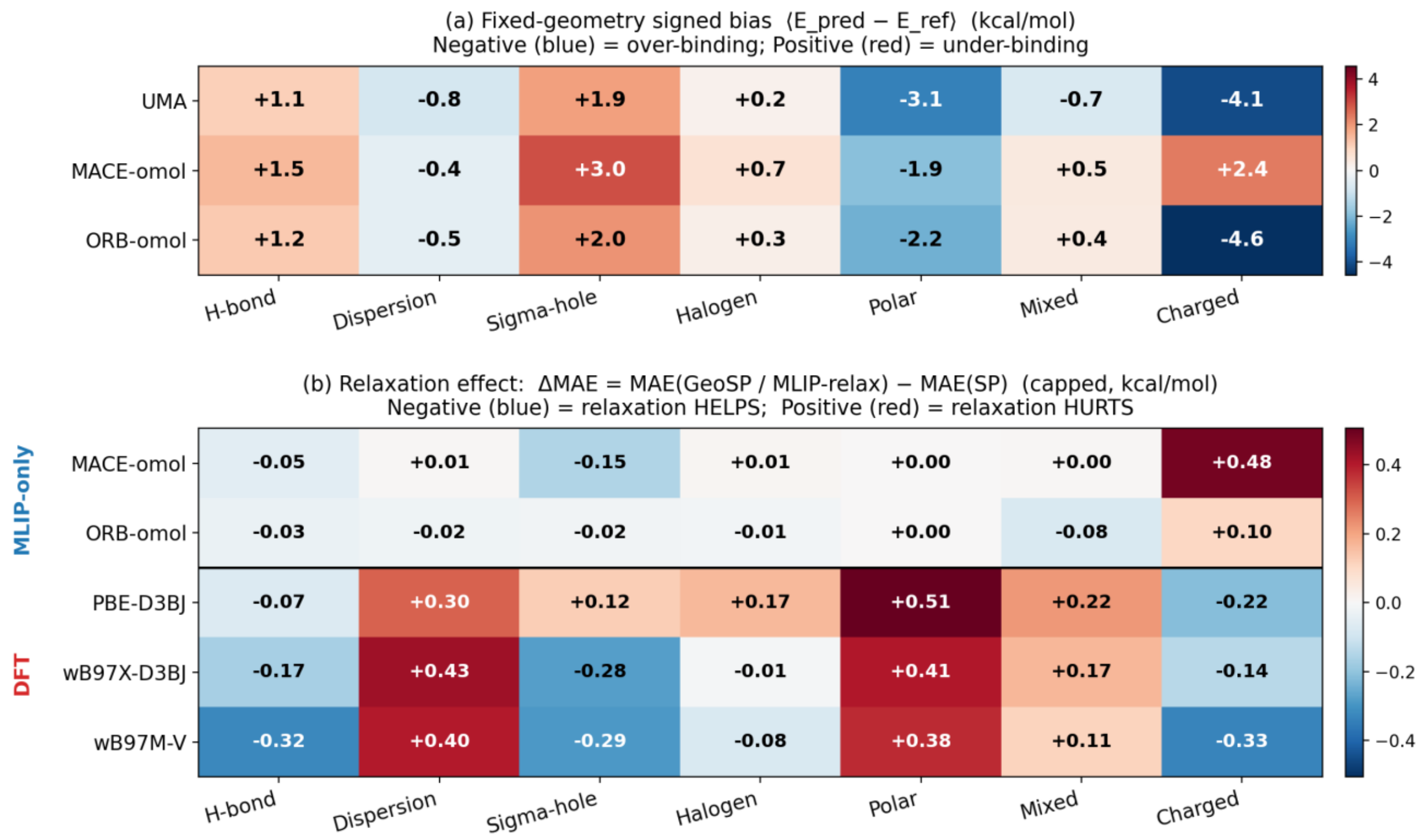

**Fig. S15** Cross-backbone diagnostic on RAPIDS-committed geometries. (a) Signed bias ⟨$E_{pred} - E_{ref}$⟩ by chemistry family for UMA, MACE-OMol, and ORB-OMol fixed-geometry single points (red = under-binding, blue = over-binding). (b) Relaxation effect ΔMAE = MAE(GeoSP/relax) − MAE(SP) per backbone and family (red = relaxation hurts, blue = helps). The Charged column pools IHB100 and BFDb_SSI_charged (876 systems). Interpretation in §4.6.

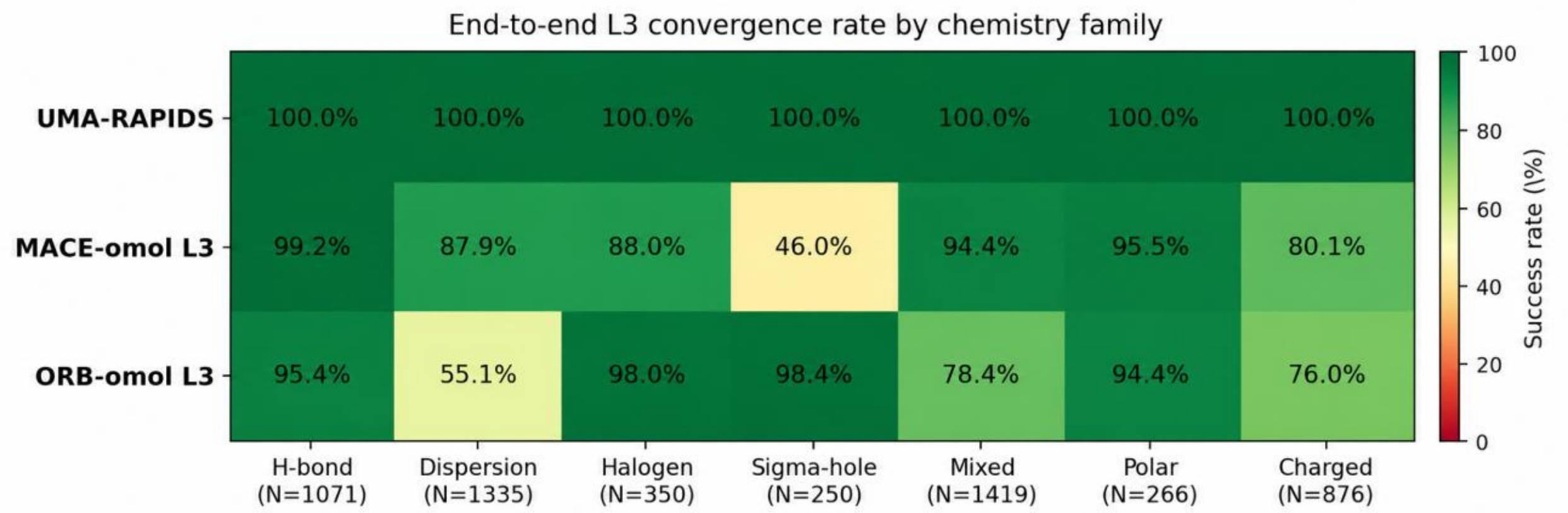

**Fig. S16** L3 end-to-end convergence rate by chemistry family. UMA-RAPIDS is the universal-coverage baseline; MACE-OMol L3 and ORB-OMol L3 have complementary failure pockets (details below).

**End-to-end L3 results.** On the fair subset where both alternative L3 runs converge ($N$ = 4,116), ORB-OMol end-to-end L3 has the lowest MAE of all methods at 2.43 kcal/mol, followed by UMA-RAPIDS at 2.46, ORB-OMol relaxation (L2) at 2.49 and ORB-OMol fixed-geometry SP at 2.50; all four are ahead of every DFT arm (best: ωB97X-D3BJ SP 2.53, ωB97M-V SP 2.54, ωB97M-V GeoSP 2.56) at a fraction of the cost, while MACE-OMol L1/L2/L3 cluster at 2.91–3.06. On the same subset UMA-RAPIDS (2.46) outperforms five of the six alternative-backbone arms, every DFT arm, and both CREST arms, trailing only ORB-OMol L3 by 0.03 kcal/mol. Coverage is asymmetric (Fig. S16): MACE-OMol converges on 0% of X40 and 46% of SH250 (heavy-halogen and σ-hole motifs), whereas ORB-OMol converges on only 55% of BFDb_SSI_dispersion. Across the full 5,567-system stream, MACE-OMol succeeds where ORB-OMol fails on 837 systems and ORB-OMol succeeds where MACE-OMol fails on 260, with 354 systems on which both alternative backbones fail. ORB-OMol's lead over DFT-GeoSP therefore holds only where it converges: UMA-RAPIDS remains the main pipeline because it covers the full stream, while MACE-OMol and ORB-OMol are usable as alternative-backbone arms with explicit chemistry-family exclusions. The 837/260 convergence-outcome asymmetry also provides a deployment-side complementary signal to the prediction-level disagreement signal of §4.6: escalation can be triggered by which backbone converges, in addition to the magnitude of prediction disagreement.

**Cross-backbone bias correction.** The chemistry-family-structured bias of finding (i) is learnable across backbones, but the calibration groups have to follow chemistry family rather than benchmark identity. A per-benchmark bias offset does not transfer cleanly across backbones (it helps only MACE-OMol L3, by roughly 9%, and is neutral-to-harmful at L1/L2), mirroring its weakened effect on UMA-RAPIDS itself after the geometry correction (§4.4). A chemistry-family-grouped variant (ChemBiasCorr) does transfer: it reduces neutral MAE by roughly 9–13% when applied to UMA-RAPIDS, MACE-OMol L3, and ORB-OMol L3, bringing all three to 1.23–1.33 kcal/mol and approaching ωB97M-V GeoSP. The effect is strongest at the end-to-end L3 level (MACE-OMol L3 −13%, ORB-OMol L3 −10%), confirming that the residual surface bias is shared across the OMol-trained backbones and structured by chemistry family.

## C.10 WetSenseBench: wet-lab-grounded selectivity validation

This section documents the WetSenseBench evaluation summarized in §4.9. The benchmark tests whether RAPIDS, with no DFT follow-up, reproduces the qualitative host–guest selectivity calls reported in published wet-lab sensing studies.

**Task and terminology.** Each evaluation unit is one pair (probe, target, interferent). The *probe* is the sensor/receptor molecule (fixed within a study); the *target* is the analyte the study reports the probe binds selectively; the *interferent* is a competing analyte the study reports is bound more weakly. The wet-lab gold label is binary (`target_favored / interferent_favored`). RAPIDS runs two independent de-novo conformer searches, one for the probe–target complex and one for the probe–interferent complex, takes the most negative converged interaction energy $E_{\text{int}} = E(\text{probe} + \text{analyte}) - E(\text{probe}) - E(\text{analyte})$ on each side, and predicts from the sign of $\Delta E = E_{\text{int}}(\text{probe} + \text{target}) - E_{\text{int}}(\text{probe} + \text{interferent})$ ($\Delta E < 0 \Rightarrow$ predict `target_favored`). A prediction is correct when its sign matches the gold label. Wet-lab papers report only qualitative selectivity calls, not unified binding constants, so all metrics are binary at the pair level. We report two: *pair-level accuracy* (fraction of pairs called correctly) and a *DOI-level top-binder* check (does RAPIDS' lowest-energy complex across the whole study sit on the same analyte the study names as its design target).

**Corpus construction.** Fourteen wet-lab papers were extracted from the open chemistry literature by a multi-agent pipeline, each agent emitting probe/target/interferent SMILES, conditions, readout, and a qualitative gold label per triple. Six of the fourteen were excluded from the paper-facing table for documented reasons (Table S6): three are UMA OMol failures or scope exclusions (unphysical energies for a Pt complex and for one receptor giving all-positive energies, and TTF chalcogens), one yields a single evaluable pair after harvest, and two have extraction/data issues. The 8 retained studies (Table S5) give 92 evaluated pairs. We note a sampling bias: four of the eight studies (#1, #2, #3, #7) use picric acid (2,4,6-trinitrophenol) as the design target, reflecting the prominence of nitroaromatic explosive sensing in the host–guest literature rather than any model preference. All geometries are vacuum-optimized; there is no explicit solvent, counterion, or ionic-strength modeling.

**Table S5** The 8 paper-facing WetSenseBench studies. "Class" is the structural design: rank-series (1 target, *N* interferents), matrix (multiple targets and interferents), or the CB[6] hydrocarbon panel. Top-binder is the DOI-level check (§4.9). All predictions are RAPIDS with no DFT follow-up.

| # | DOI | Probe (sensor) | Class | $n_{\text{pairs}}$ | acc. | top-binder |
|---|---|---|---|---|---|---|
| 1 | 10.1021/acs.joc.6b00267 | tetraphenylethene–pyrone | rank-series | 5 | 100% | ✓ |
| 2 | 10.1021/acsomega.6b00288 | bisthiocarbonohydrazone | rank-series | 9 | 78% | ✓ |
| 3 | 10.1039/C4CC06225F | reduced [3+2] organic cage | rank-series | 9 | 100% | ✓ |
| 4 | 10.1016/j.talanta.2016.03.026 | amino/imidazole calix[4]arene | rank-series | 12 | 67% | × |
| 5 | 10.1252/jcej.28.267 | α-/β-cyclodextrin | matrix | 9 | 89% | ✓ |
| 6 | 10.1021/acsomega.3c02349 | aza-BODIPY | matrix | 9 | 89% | ✓ |
| 7 | 10.1021/jo502281b | tris(benzimidazole) tripod | matrix | 17 | 76% | ✓ |
| 8 | 10.1002/anie.201104119 | cucurbit[6]uril (CB[6]) | CB[6] | 22 | 32% | × |
| All 8 (pair-level) | | | | 92 | 70.7% | 6/8 |
| Excluding CB[6] (#8) | | | | 70 | **82.9%** | |

**Table S6** The six studies excluded from WetSenseBench during the 14 → 8 filter: three are UMA OMol failures or scope exclusions, three have insufficient evaluable data; none was excluded because of its ranking.

| DOI | Reason for exclusion |
|---|---|
| 10.1021/acs.joc.5b00348 | Pt complex; UMA OMol energies +88, +69 kcal/mol (unphysical). |
| 10.1007/s00216-009-2819-4 | TTF chalcogens out of UMA scope; receptor SMILES fragmented. |
| 10.1021/am505636f | All energies positive (+2 to +87 kcal/mol); UMA fails on this receptor. |
| 10.1016/j.dyepig.2017.05.002 | Only 1 evaluable pair after harvest. |
| 10.1002/anie.200802762 | Out of evaluation scope (data/extraction issues). |
| 10.1002/elan.201800199 | Out of evaluation scope (data/extraction issues). |

**Rank-series and matrix studies (Figs. S17–S18).** The rank-series studies (#1–#4: one target against *N* interferents) show the expected confidence structure: chemistry-matched wins carry large negative $\Delta E$ (down to −26.1 kcal/mol for #1), while near-ties cluster near zero. The two errors in #2 are 2,4-dinitrotoluene (+2.4) and 2,4,6-trinitrotoluene (+12.7), where RAPIDS slightly overbinds TNT/DNT relative to picric acid for a small Schiff-base probe. The matrix studies (#5–#7: multiple targets × interferents) localize their errors *within* chemical families rather than across them: RAPIDS rarely confuses a strong nitroaromatic target with an unrelated interferent, and the residual misses are intra-class orderings (which of several nitroaromatics binds strongest; whether α-cyclodextrin prefers p-nitrophenol or o-nitrophenol, $\Delta E = +10.3$).

**Failure mode #4 (calix[4]arene), a near-miss.** The DOI-level top-binder miss in #4 is RAPIDS picking chloroform ($E_{\text{int}}$ = −11.69 kcal/mol) over the wet-lab target dichloromethane (−9.64), a gap of 2.05 kcal/mol; dichloromethane still ranks above 8 of the 12 analytes. The four wrong pairs (carbon tetrachloride +0.78, acetonitrile +0.94, methanol +0.91, chloroform +2.05) are all near-ties; chloroform–calixarene host–guest binding is a documented strong interaction in solution, so RAPIDS returns a chemically reasonable ordering that disagrees with sensor-design intent.

**Failure mode #8 (CB[6]), the precision-limit case (Fig. S19).** The 32% CB[6] score has three separate causes. (i) *Precision floor (dominant)*: 14 of 22 labels have $|\Delta E| \leq 1$ kcal/mol, inside the model's intrinsic noise, so these binary calls are at chance level. (ii) *Scope mismatch*: the (n-butane, propane) pair appears as two labels with *opposite* wet-lab gold (high-confidence salt-free `target_favored`; medium-confidence 50 mM NaOAc `interferent_favored`), because salt destabilizes the n-butane complex; gas-phase RAPIDS returns the same $\Delta E \approx -0.7$ on both, so one label is necessarily wrong by construction. (iii) *Geometric artifact*: two guests (neopentane and isopentane) yielded non-physical ≈−32 kcal/mol binding energies from strained conformer-scan poses (real CB[6]–hydrocarbon affinities are 2–7 kcal/mol[4,71]); these two outliers drive the DOI-level top-binder miss. CB[6] is retained as the benchmark's boundary case: it marks where MLIP-based selectivity prediction stops being reliable, namely when selectivity-determining gaps fall below the ~1 kcal/mol floor or when the wet-lab readout depends on solvation/ionic-strength physics outside a gas-phase model's scope.

**Caveats.** (i) All wet-lab labels are qualitative; we make no quantitative binding-energy claim. (ii) The corpus over-represents explosive (picric-acid) sensing. (iii) All geometries are gas-phase, vacuum-optimized. (iv) Per-study $N$ is small (5–22 pairs), so we omit per-study error bars and quote the aggregate 7-study / 70-pair number as the reliable figure. (v) The structure–identity assignment of every rendered complex in Fig. 6 was cross-verified: the simulation directory encodes the expected (label, side) and the rendered atom composition matches the combined element-count multiset of the probe and analyte SMILES via hydrogen-explicit OpenBabel conversion;[167] renderings use PyMOL.[168]

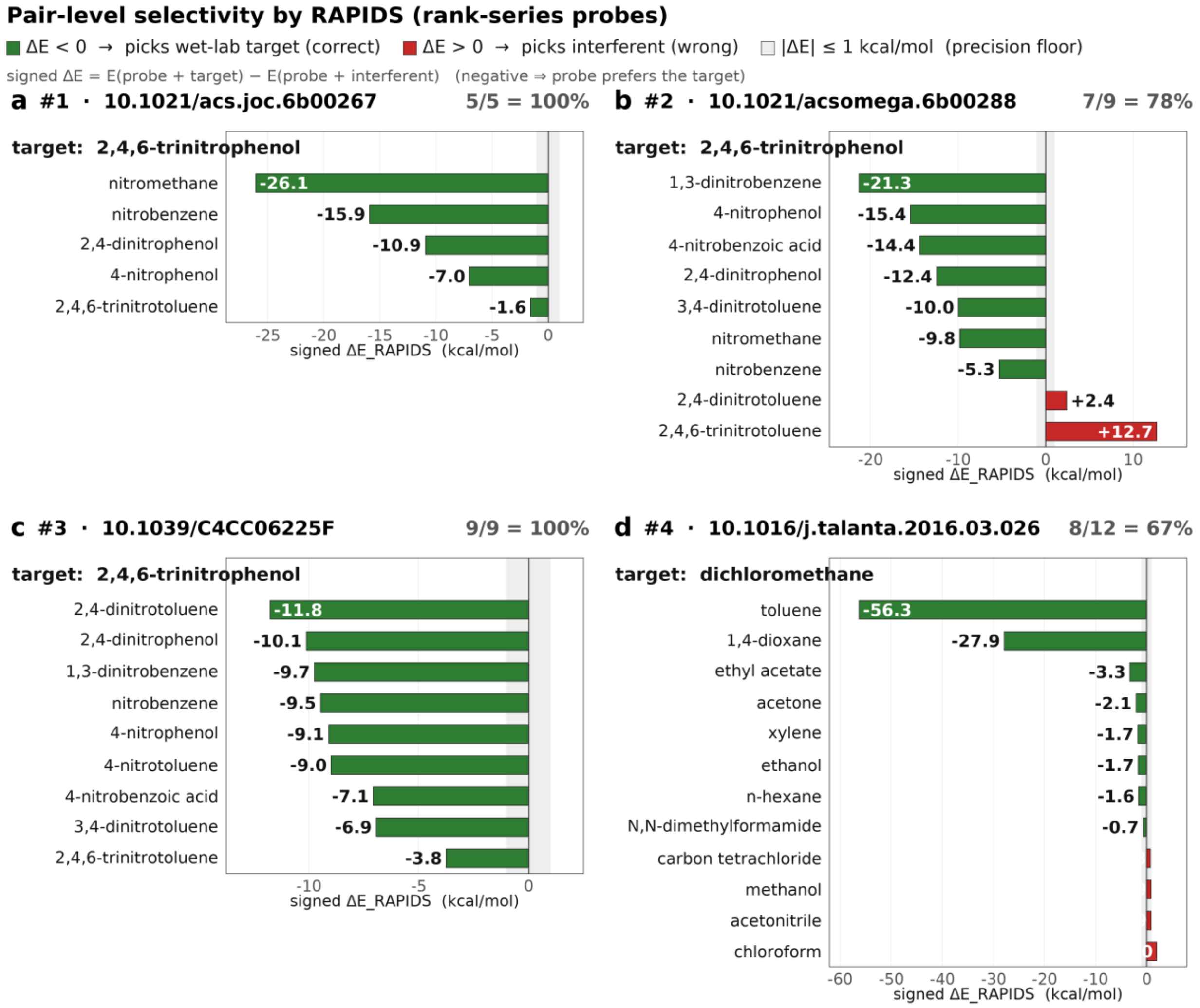

**Fig. S17** WetSenseBench rank-series studies (#1–#4). Each sub-panel lists, for one study, the signed $\Delta E_{\mathrm{RAPIDS}} = E_{\mathrm{int}}(\text{probe + target}) - E_{\mathrm{int}}(\text{probe + interferent})$ per interferent (kcal/mol), where $E_{\mathrm{int}}$ is the RAPIDS interaction (binding) energy. Green bars ($\Delta E < 0$) are correct target-favored calls; red bars ($\Delta E > 0$) are misses. The light gray band at $|\Delta E| \leq 1$ kcal/mol marks the MLIP precision floor; rows are sorted by $\Delta E$.

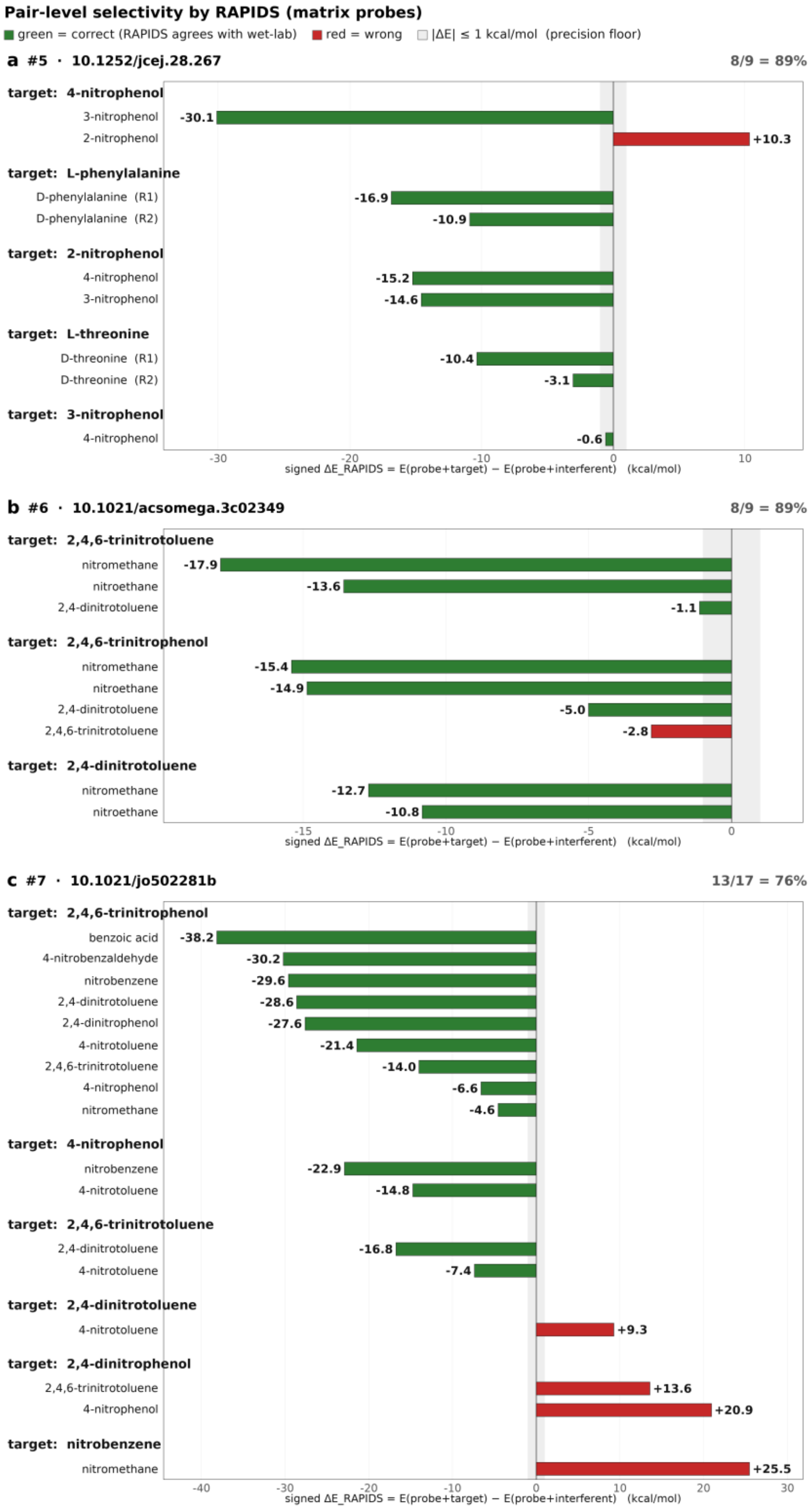

**Fig. S18** WetSenseBench matrix studies (#5–#7), grouped by target within each study. Bar conventions as in Fig. S17. Errors concentrate on intra-class orderings (closely related nitroaromatics and nitrophenol isomers) rather than target-versus-unrelated-interferent confusions.

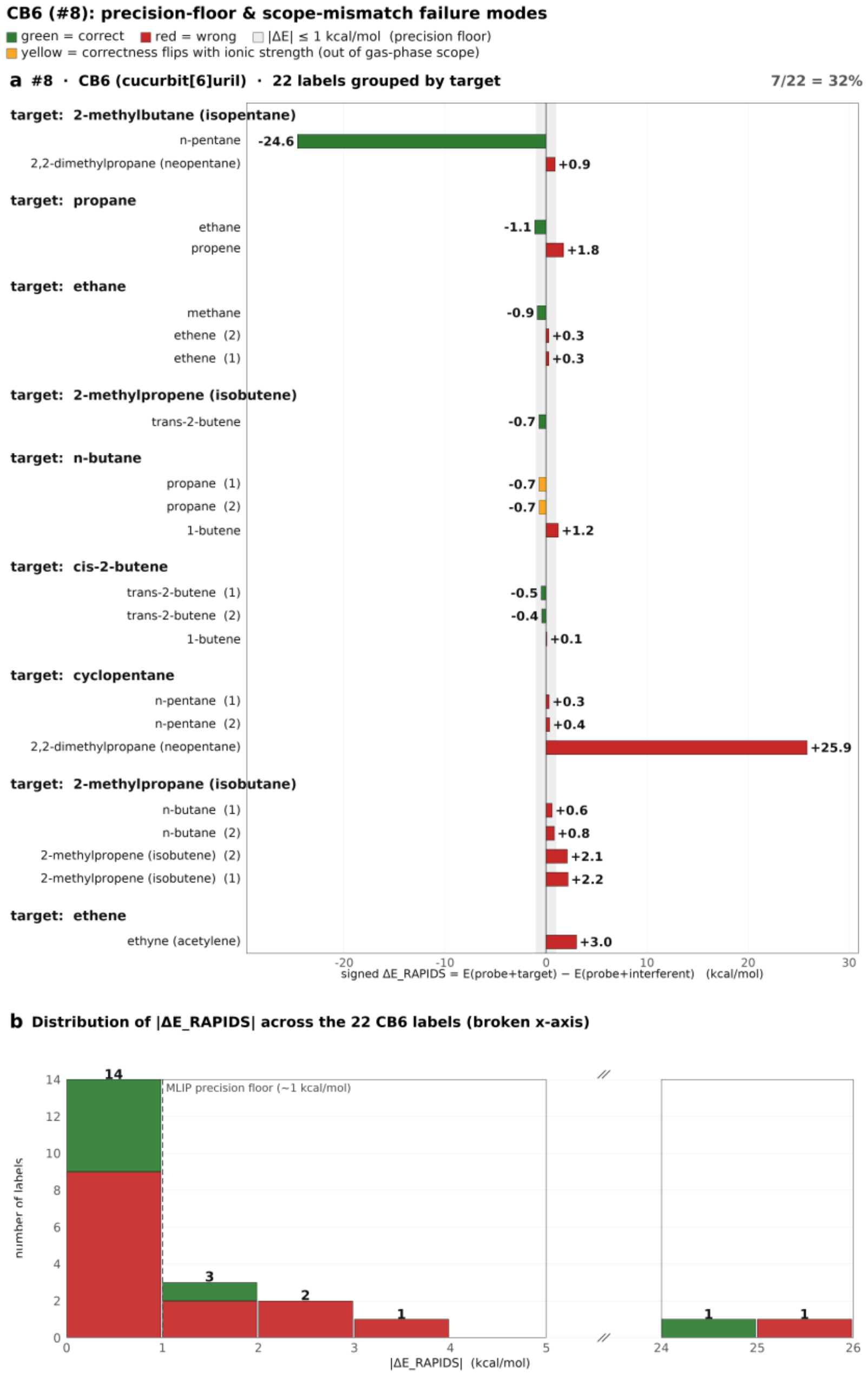

**Fig. S19** WetSenseBench CB[6] study (#8), the precision-limit case. (a) Per-pair $\Delta E$ for all 22 labels; the condition-dependent (n-butane, propane) pair is drawn as two yellow bars with the same RAPIDS $\Delta E$ but opposite wet-lab gold. (b) Histogram of $|\Delta E|$ (broken axis), stacked correct/wrong, with the 1 kcal/mol precision-floor line; 14 of 22 labels fall inside the floor and the two 24–26 kcal/mol entries are conformer-scan pose artifacts (each the $|\Delta E|$ of a non-physical $\approx -32$ kcal/mol guest pose against a normal $\approx -7$ kcal/mol counterpart).

# D Supplementary tables

## D.1 Replay-policy inventory

**Table S7** Inventory of the 40 replay policies reported from the offline-replay study. Deployability labels follow Section B.6: *Direct* denotes no replay-time benchmark-reference updates, *Delayed-GT* requires per-system ground-truth feedback, *Retrospective* denotes selectors replayed with full-information or post hoc feedback, and *Oracle* is the unattainable lower bound. Neutral and charged aggregates are medians across the corresponding benchmark groups.

| Strategy | Category | Deployability | Neutral MAE | Neutral ρ | Neutral h | Charged MAE | Charged ρ | Charged h |
|---|---|---|---|---|---|---|---|---|
| Always-CREST_xTB | Always-X | Direct | 2.145 | 0.623 | 109.4 | 9.896 | 0.643 | 225.7 |
| Always-CREST_xTB_DFT | Always-X | Direct | 1.802 | 0.682 | 125.3 | 9.007 | 0.710 | 294.5 |
| Always-PBE-D3BJ_GeoSP | Always-X | Direct | 1.397 | 0.716 | 33.8 | 8.633 | 0.667 | 46.7 |
| Always-PBE-D3BJ_SP | Always-X | Direct | 1.504 | 0.613 | 16.5 | 10.085 | 0.545 | 22.9 |
| Always-RAPIDS | Always-X | Direct | 1.365 | 0.658 | 14.8 | 9.735 | 0.569 | 20.4 |
| Always-wB97M-V_GeoSP | Always-X | Direct | 1.300 | 0.728 | 64.8 | 8.292 | 0.702 | 98.5 |
| Always-wB97M-V_SP | Always-X | Direct | 1.439 | 0.641 | 24.4 | 9.849 | 0.553 | 33.1 |
| Always-wB97X-D3BJ_GeoSP | Always-X | Direct | 1.437 | 0.728 | 77.4 | 8.574 | 0.694 | 116.8 |
| Always-wB97X-D3BJ_SP | Always-X | Direct | 1.418 | 0.630 | 22.0 | 9.706 | 0.558 | 27.2 |
| Oracle | Oracle | Oracle | 0.386 | 0.943 | 52.8 | 4.946 | 0.863 | 135.7 |
| ALORS | Bandit | Retrospective | 1.147 | 0.688 | 60.0 | 8.198 | 0.667 | 130.4 |
| BiasCorr-RAPIDS | Bandit | Delayed-GT | 1.304 | 0.621 | 14.8 | 9.245 | 0.550 | 20.4 |
| CheapEnsemble-3 | Bandit | Direct | 1.398 | 0.633 | 52.6 | 9.712 | 0.557 | 70.5 |
| CheapEnsemble-5 | Bandit | Direct | 1.399 | 0.633 | 111.4 | 9.739 | 0.558 | 150.2 |
| Cost-Aware | Bandit | Retrospective | 1.172 | 0.680 | 76.4 | 8.103 | 0.648 | 168.1 |
| Disagreement-t1 | Bandit | Direct | 1.379 | 0.654 | 39.1 | 9.831 | 0.567 | 74.3 |
| Ladder-t1 | Bandit | Direct | 1.489 | 0.639 | 33.3 | 9.952 | 0.548 | 79.4 |
| Learned-Selector | Bandit | Retrospective | 1.054 | 0.714 | 90.6 | 7.844 | 0.668 | 204.1 |
| MFMI-Greedy | Bandit | Retrospective | 1.429 | 0.661 | 142.5 | 8.212 | 0.666 | 269.9 |
| MFMI-Greedy-5x | Bandit | Retrospective | 1.429 | 0.661 | 142.5 | 8.212 | 0.666 | 269.9 |
| Random | Bandit | Direct | 1.546 | 0.657 | 58.1 | 9.143 | 0.636 | 98.5 |
| Stacking-r0 | Bandit | Delayed-GT | 1.158 | 0.615 | 56.2 | 8.799 | 0.578 | 96.4 |
| StackingML-2 | Bandit | Delayed-GT | 1.384 | 0.596 | 31.2 | 7.883 | 0.565 | 43.3 |
| Thompson | Bandit | Retrospective | 1.292 | 0.683 | 50.3 | 8.358 | 0.663 | 124.9 |
| Thompson-CostAware | Bandit | Retrospective | 1.531 | 0.631 | 43.4 | 8.799 | 0.679 | 246.7 |
| UCB | Bandit | Retrospective | 1.042 | 0.700 | 61.7 | 7.804 | 0.706 | 155.5 |
| UCB-CostAware | Bandit | Retrospective | 1.194 | 0.714 | 86.5 | 8.378 | 0.710 | 263.9 |
| BucketLearned | Chem/Bucket | Retrospective | 1.088 | 0.705 | 102.3 | 7.802 | 0.709 | 209.2 |
| BucketPrior-Adaptive | Chem/Bucket | Retrospective | 1.016 | 0.748 | 53.3 | 8.255 | 0.663 | 161.1 |
| ChemALORS | Chem/Bucket | Retrospective | 1.167 | 0.680 | 62.7 | 8.248 | 0.662 | 129.7 |
| ChemBiasCorr-PBE | Chem/Bucket | Delayed-GT | 1.240 | 0.627 | 16.5 | 8.040 | 0.639 | 22.9 |
| ChemLearned-Selector | Chem/Bucket | Retrospective | 1.067 | 0.705 | 102.1 | 7.827 | 0.696 | 206.7 |
| ChemStacking-3 | Chem/Bucket | Delayed-GT | 1.246 | 0.642 | 52.6 | 8.041 | 0.623 | 70.5 |
| ChemUCB-8 | Chem/Bucket | Retrospective | 1.007 | 0.762 | 65.1 | 8.179 | 0.671 | 146.8 |
| GBM-Stack-2cheap | GBM/Meta | Delayed-GT | 1.100 | 0.708 | 31.2 | 6.935 | 0.635 | 43.3 |
| GBM-Stack-MetaWarm | GBM/Meta | Delayed-GT | 1.207 | 0.708 | 31.2 | 9.709 | 0.533 | 43.3 |
| Hybrid-SelectStack | GBM/Meta | Delayed-GT | 1.128 | 0.731 | 131.9 | 8.908 | 0.608 | 257.1 |
| LinUCB | GBM/Meta | Retrospective | 1.617 | 0.649 | 74.5 | 8.583 | 0.662 | 142.4 |
| Meta-GBM | GBM/Meta | Direct | 1.699 | 0.563 | 31.2 | 14.468 | 0.320 | 43.3 |
| Meta-GBM-Adaptive | GBM/Meta | Delayed-GT | 1.232 | 0.678 | 31.2 | 10.742 | 0.537 | 43.3 |

## D.2 Full 18 × 9 per-benchmark MAE matrix

Table S8 and Table S9 give the per-benchmark capped MAE (kcal/mol) of every fidelity method on every benchmark, where each method's relaxed binding score is compared against the published benchmark interaction-energy reference. These are the numbers summarized by the full heatmap in Fig. S2.

**Table S8** Per-benchmark capped MAE (kcal/mol), part A (benchmarks 1–9), computed between each method's relaxed binding score and the published benchmark interaction-energy reference.

| Method | A24 | S66 | HB300SPX | HB375 | X40 | D1200_Hal. | SH250 | D1200_HBCNO | D1200_PS |
|---|---|---|---|---|---|---|---|---|---|
| RAPIDS | 1.68 | 1.60 | 1.43 | 1.74 | 1.30 | 0.59 | 2.23 | 0.87 | 0.95 |
| PBE-D3BJ SP | 2.05 | 1.63 | 1.38 | 1.64 | 1.23 | 0.62 | 2.11 | 0.51 | 0.69 |
| ωB97X-D3BJ SP | 1.76 | 2.05 | 1.22 | 1.62 | 1.00 | 0.59 | 2.09 | 0.57 | 0.70 |
| ωB97M-V SP | 1.78 | 1.61 | 1.27 | 1.64 | 0.99 | 0.57 | 2.18 | 0.64 | 0.84 |
| PBE-D3BJ GeoSP | 0.76 | 1.74 | 1.70 | 1.35 | 1.14 | 0.82 | 2.23 | 0.36 | 0.73 |
| ωB97X-D3BJ GeoSP | 1.79 | 3.19 | 0.91 | 1.36 | 0.93 | 0.59 | 1.81 | 0.36 | 0.57 |
| ωB97M-V GeoSP | 1.43 | 1.44 | 0.89 | 1.28 | 0.95 | 0.49 | 1.89 | 0.37 | 0.48 |
| CREST xTB | 1.96 | 1.06 | 2.11 | 3.01 | 5.14 | 1.40 | 4.08 | 0.90 | 2.18 |
| CREST xTB+DFT | 1.52 | 0.72 | 1.17 | 2.37 | 1.43 | 0.95 | 1.85 | 1.04 | 2.23 |

**Table S9** Per-benchmark capped MAE (kcal/mol), part B (benchmarks 10–18), computed between each method's relaxed binding score and the published benchmark interaction-energy reference.

| Method | BBI | HSG | NBC1 | SSI-disp. | SSI-mixed | SSI-other | SSI-polar | IHB100 | SSI-charged |
|---|---|---|---|---|---|---|---|---|---|
| RAPIDS | 3.29 | 3.69 | 0.53 | 0.86 | 1.25 | 0.78 | 3.79 | 8.53 | 10.94 |
| PBE-D3BJ SP | 3.73 | 4.61 | 0.44 | 1.14 | 1.80 | 1.02 | 4.69 | 8.09 | 12.08 |
| ωB97X-D3BJ SP | 3.41 | 4.90 | 0.56 | 1.09 | 1.91 | 0.95 | 4.41 | 8.05 | 11.36 |
| ωB97M-V SP | 3.47 | 4.85 | 0.56 | 0.96 | 1.81 | 0.83 | 4.45 | 8.04 | 11.66 |
| PBE-D3BJ GeoSP | 2.99 | 3.80 | 0.51 | 1.45 | 2.32 | 1.29 | 5.20 | 5.04 | 12.23 |
| ωB97X-D3BJ GeoSP | 2.93 | 3.60 | 0.62 | 1.52 | 2.27 | 1.27 | 4.82 | 5.63 | 11.52 |
| ωB97M-V GeoSP | 3.02 | 3.59 | 0.50 | 1.32 | 2.17 | 1.13 | 4.82 | 4.90 | 11.68 |
| CREST xTB | 2.39 | 3.17 | 1.32 | 1.65 | 2.62 | 1.25 | 6.10 | 9.23 | 10.57 |
| CREST xTB+DFT | 2.63 | 4.41 | 1.29 | 2.22 | 3.04 | 1.76 | 6.82 | 5.79 | 12.22 |

## D.3 Per-benchmark MAE ladder for the decomposition of §4.3

Table S10 gives the per-benchmark values of the benchmark-relative MAE ladder used in §4.3. This ladder tracks how the benchmark-reference MAE changes when one rung of the fidelity stack is swapped for the next, but it is not itself a direct estimate of MAE($|\delta_{surf}|$), MAE($|\delta_{geom}|$), or MAE($|\delta_{samp}|$) from the system-level decomposition. To avoid colliding with the system-level decomposition symbols $\delta_{surf}(s)$, $\delta_{geom}(s)$, and $\delta_{samp}(s)$ used in the main text, the table reports the corresponding *adjacent MAE differences* as $\Delta MAE_{surf}$, $\Delta MAE_{geom}$, and $\Delta MAE_{samp}$. Entries are capped MAE (kcal/mol) of the four ladder rungs and their pairwise differences.

**Table S10** Per-benchmark MAE ladder and adjacent MAE differences, all benchmark-relative. Columns: capped MAE of RAPIDS, ωB97M-V/def2-TZVP SP on RAPIDS geometry, ωB97M-V GeoSP on RAPIDS geometry, and CREST xTB+DFT, each compared against the published benchmark interaction-energy reference; then the three adjacent differences $\Delta MAE_{surf} = MAE_{RAPIDS} - MAE_{wB97M\text{-}V_SP}$, $\Delta MAE_{geom} = MAE_{wB97M\text{-}V_SP} - MAE_{wB97M\text{-}V_GeoSP}$, $\Delta MAE_{samp} = MAE_{wB97M\text{-}V_GeoSP} - MAE_{CREST_xTB_DFT}$. Negative $\Delta MAE$ indicates the higher-fidelity rung has a *worse* MAE against the benchmark reference, which is the benchmark-relative pattern identified on the BFDb SSI neutral subsets (§4.2).

| Benchmark | RAPIDS | ωB97M-V/ def2-TZVP SP | ωB97M-V GeoSP | CREST xTB+DFT | $\Delta MAE_{surf}$ | $\Delta MAE_{geom}$ | $\Delta MAE_{samp}$ |
|---|---|---|---|---|---|---|---|
| A24 | 1.68 | 1.78 | 1.43 | 1.52 | −0.10 | +0.35 | −0.09 |
| S66 | 1.60 | 1.61 | 1.44 | 0.72 | −0.01 | +0.17 | +0.72 |
| HB300SPX | 1.43 | 1.27 | 0.89 | 1.17 | +0.16 | +0.37 | −0.28 |
| HB375 | 1.74 | 1.64 | 1.28 | 2.37 | +0.10 | +0.36 | −1.09 |
| X40 | 1.30 | 0.99 | 0.95 | 1.43 | +0.31 | +0.04 | −0.48 |
| D1200_Halogens | 0.59 | 0.57 | 0.49 | 0.95 | +0.02 | +0.08 | −0.46 |
| SH250 | 2.23 | 2.18 | 1.89 | 1.85 | +0.04 | +0.29 | +0.04 |
| D1200_HBCNO | 0.87 | 0.64 | 0.37 | 1.04 | +0.24 | +0.26 | −0.67 |
| D1200_PS | 0.95 | 0.84 | 0.48 | 2.23 | +0.10 | +0.37 | −1.76 |
| BFDb_BBI | 3.29 | 3.47 | 3.02 | 2.63 | −0.19 | +0.45 | +0.39 |
| BFDb_HSG | 3.69 | 4.85 | 3.59 | 4.41 | −1.16 | +1.26 | −0.82 |
| BFDb_NBC1 | 0.53 | 0.56 | 0.50 | 1.29 | −0.04 | +0.06 | −0.79 |
| BFDb_SSI_dispersion | 0.86 | 0.96 | 1.32 | 2.22 | −0.09 | −0.36 | −0.90 |
| BFDb_SSI_mixed | 1.25 | 1.81 | 2.17 | 3.04 | −0.56 | −0.35 | −0.87 |
| BFDb_SSI_other | 0.78 | 0.83 | 1.13 | 1.76 | −0.05 | −0.30 | −0.62 |
| BFDb_SSI_polar | 3.79 | 4.45 | 4.82 | 6.82 | −0.65 | −0.38 | −2.00 |
| IHB100 | 8.53 | 8.04 | 4.90 | 5.79 | +0.49 | +3.14 | −0.89 |
| BFDb_SSI_charged | 10.94 | 11.66 | 11.68 | 12.22 | −0.72 | −0.02 | −0.54 |

## D.4 Per-benchmark validation-guard trigger rates

**Table S11** RAPIDS validation-guard trigger rates per benchmark. A system is counted as flagged when at least one element of the validation stack downgrades the confidence from high, namely one of the four per-configuration guards (topology / geometry / energy / convergence) or the scan-level Energy Consistency Guard (ECG). $N_{missing}$ counts systems with no committed RAPIDS prediction because at least one required lower-level engine in the benchmarked pipelines (UMA or CREST/xTB preprocessing) does not support the chemistry. The trigger rate is computed on completed predictions, i.e. *flagged*/$N_{completed}$.

| **Benchmark** | $N_{completed}$ | $N_{missing}$ | **flagged** | **rate** |
|---|---|---|---|---|
| A24 | 22 | 2 | 5 | 22.7% |
| S66 | 66 | 0 | 0 | 0.0% |
| X40 | 40 | 0 | 4 | 10.0% |
| HB300SPX | 300 | 0 | 7 | 2.3% |
| HB375 | 375 | 0 | 0 | 0.0% |
| SH250 | 250 | 0 | 23 | 9.2% |
| D1200_HBCNO | 308 | 0 | 10 | 3.2% |
| D1200_PS | 293 | 0 | 56 | 19.1% |
| D1200_Halogens | 310 | 0 | 10 | 3.2% |
| BFDb_BBI | 100 | 0 | 17 | 17.0% |
| BFDb_HSG | 21 | 0 | 4 | 19.0% |
| BFDb_NBC1 | 10 | 0 | 2 | 20.0% |
| BFDb_SSI_dispersion | 1,325 | 0 | 218 | 16.5% |
| BFDb_SSI_mixed | 765 | 0 | 129 | 16.9% |
| BFDb_SSI_other | 240 | 0 | 33 | 13.8% |
| BFDb_SSI_polar | 266 | 0 | 47 | 17.7% |
| IHB100 | 100 | 0 | 4 | 4.0% |
| BFDb_SSI_charged | 776 | 0 | 429 | 55.3% |
| **Total** | **5,567** | **2** | **998** | **17.9%** |

## D.5 Per-benchmark guard flagged-vs-unflagged capped-MAE

Table S12 reports the per-benchmark capped-MAE of the RAPIDS predictions broken out by whether the validation stack (§3.2) emits a flagged label. These are the numbers summarized at the aggregate level in the main-text reliability note at the end of §4.5.

**Table S12** RAPIDS capped MAE on the unflagged (ok) subset versus the flagged subset, with an explicit missing count for systems that yield no committed RAPIDS prediction because at least one required lower-level engine (UMA or CREST/xTB preprocessing in the benchmarked pipelines) does not support the chemistry. Benchmarks with zero flagged predictions (S66, HB375) have no ratio defined; ratio < 1 indicates the flagged subset carries a *lower* MAE than the unflagged subset, which we observe on two benchmarks (BFDb_HSG, BFDb_SSI_polar).

| Benchmark | $N_{ok}$ | $N_{flagged}$ | $N_{missing}$ | $MAE_{ok}$ | $MAE_{flagged}$ | ratio |
|---|---|---|---|---|---|---|
| A24 | 17 | 5 | 2 | 0.77 | 4.78 | 6.22 |
| S66 | 66 | 0 | 0 | 1.60 | — | — |
| X40 | 36 | 4 | 0 | 1.00 | 4.01 | 4.01 |
| HB300SPX | 293 | 7 | 0 | 1.42 | 2.01 | 1.42 |
| HB375 | 375 | 0 | 0 | 1.74 | — | — |
| SH250 | 227 | 23 | 0 | 2.01 | 4.33 | 2.15 |
| D1200_HBCNO | 298 | 10 | 0 | 0.86 | 1.19 | 1.38 |
| D1200_PS | 237 | 56 | 0 | 0.93 | 1.01 | 1.09 |
| D1200_Halogens | 300 | 10 | 0 | 0.58 | 0.89 | 1.52 |
| BFDb_BBI | 83 | 17 | 0 | 3.08 | 4.30 | 1.40 |
| BFDb_HSG | 17 | 4 | 0 | 3.69 | 3.66 | 0.99 |
| BFDb_NBC1 | 8 | 2 | 0 | 0.43 | 0.94 | 2.19 |
| BFDb_SSI_dispersion | 1107 | 218 | 0 | 0.83 | 1.04 | 1.25 |
| BFDb_SSI_mixed | 636 | 129 | 0 | 1.16 | 1.71 | 1.47 |
| BFDb_SSI_other | 207 | 33 | 0 | 0.75 | 0.96 | 1.28 |
| BFDb_SSI_polar | 219 | 47 | 0 | 3.93 | 3.14 | 0.80 |
| IHB100 | 96 | 4 | 0 | 7.74 | 27.31 | 3.53 |
| BFDb_SSI_charged | 347 | 429 | 0 | 9.16 | 12.38 | 1.35 |
| Total (all 18 bm) | 4569 | 998 | 2 | 2.02 | 6.38 | 3.15 |

## D.6 Frozen-geometry method-only benchmark

**Table S13** Frozen-geometry method-only benchmark (§4.7, Section B.5): all ten methods evaluated as single points on the identical published reference geometries, so the per-method error is a pure surface/method bias with no geometry contribution. Neutral and charged columns are the median capped MAE (per-system error clipped at 50 kcal/mol) across the 16 neutral and 2 charged benchmarks, following the aggregation convention of §3.4; bias and Spearman ρ are computed over all 5,567 systems. Rows sorted by neutral MAE. On the reference geometry the UMA surface reaches charged MAE 0.14 kcal/mol on BFDb_SSI_charged, statistically tied with the best DFT method and 75× below its 10.94 kcal/mol error on the MLIP-relaxed geometry of §4.5.

| Method | Group | Neut. MAE | Chg. MAE | Bias | ρ |
|---|---|---|---|---|---|
| ωB97M-V/def2-TZVPD | DFT | 0.12 | 0.21 | −0.01 | 0.9986 |
| UMA | MLIP | 0.13 | 0.24 | −0.02 | 0.9926 |
| ORB-OMol | MLIP | 0.14 | 3.90 | −0.11 | 0.9934 |
| ωB97X-D3BJ/def2-TZVPD | DFT | 0.21 | 0.35 | −0.09 | 0.9977 |
| PBE-D3BJ/def2-TZVPD | DFT | 0.28 | 0.76 | −0.32 | 0.9957 |
| ωB97M-V/def2-TZVP | DFT | 0.30 | 1.33 | −0.28 | 0.9967 |
| MACE-OMol | MLIP | 0.31 | 8.42 | +1.14 | 0.9487 |
| ωB97X-D3BJ/def2-TZVP | DFT | 0.31 | 1.07 | −0.35 | 0.9979 |
| PBE-D3BJ/def2-TZVP | DFT | 0.52 | 2.22 | −0.64 | 0.9941 |
| GFN2-xTB | semi-emp. | 0.67 | 8.98 | −2.03 | 0.9091 |